\pdfoutput=1

\documentclass[11pt,twoside,a4paper,cmspaper,final,collab]{cms-tdr}
\def\svnVersion{330683}\def\cmsCernNoTag{CERN-PH-EP/2013-037}\def\cmsCernDate{\today}\def\cmsMessage{Published in Physical Review D as \href{http://dx.doi.org/10.1103/PhysRevD.93.032004}{\doi{10.1103/PhysRevD.93.032004.}}}

\begin{document}\cmsNoteHeader{EXO-12-041}

\hyphenation{had-ron-i-za-tion}
\hyphenation{cal-or-i-me-ter}
\hyphenation{de-vices}
\RCS$Revision: 310811 $
\RCS$HeadURL: svn+ssh://svn.cern.ch/reps/tdr2/papers/EXO-12-041/trunk/EXO-12-041.tex $
\RCS$Id: EXO-12-041.tex 310811 2015-11-16 21:18:14Z dmorse $

\providecommand{\cPj}{\ensuremath{\cmsSymbolFace{j}}\xspace}
\newcommand{\eejj}{\ensuremath{\Pe\Pe\cPj\cPj}}
\newcommand{\enujj}{\ensuremath{\Pe\cPgn\cPj\cPj}}
\newcommand{\mumujj}{\ensuremath{\Pgm\Pgm\cPj\cPj}}
\newcommand{\munujj}{\ensuremath{\Pgm\cPgn\cPj\cPj}}
\newcommand{\emujj}{\ensuremath{\Pe\Pgm\cPj\cPj}}
\newcommand{\zjets}{\ensuremath{\cPZ^{\mathrm{0}}}\mathrm{+jets}}
\newcommand{\wjets}{\ensuremath{\cPW^{\pm}}\mathrm{+jets}}
\newcommand{\lljj}{\ensuremath{\mathrm{\ell\ell}\cPj\cPj}}
\newcommand{\lnujj}{\ensuremath{\mathrm{\ell}\cPgn\cPj\cPj}}

\newcommand{\st}{\ensuremath{S_{\mathrm{T}}}}
\newcommand{\mee}{\ensuremath{m_{\Pe\Pe}}}
\newcommand{\mej}{\ensuremath{m_{\Pe\cPj}}}
\newcommand{\mejmin}{\ensuremath{m_{\Pe\cPj}^{\text{min}}}}
\newcommand{\mejavg}{\ensuremath{m_{\Pe\cPj}^{\text{average}}}}
\newcommand{\mtjnu}{\ensuremath{m_{\text{T, j}\nu}}}

\newcommand{\mt}{\ensuremath{m_{\text{T, e}\nu}}}

\newcommand{\enujjWJetsMonteCarloScaleFactor}{0.85 \pm 0.01\stat \pm 0.01\syst}
\newcommand{\enujjTTBarMonteCarloScaleFactor}{0.97 \pm 0.02\stat \pm 0.01\syst}
\newcommand{\eejjZJetsMonteCarloScaleFactor} {0.97 \pm 0.01\stat}

\newcommand{\enujjWJetsMonteCarloScaleFactorMETRescaled}{0.95 \pm 0.02\stat \pm 0.01\syst}
\newcommand{\enujjTTBarMonteCarloScaleFactorMETRescaled}{1.07 \pm 0.03\stat \pm 0.01\syst}

\newcommand{\enujjWJetsMonteCarloScaleFactorMETandMTRescaled}{0.97 \pm 0.02\stat \pm 0.01\syst}
\newcommand{\enujjTTBarMonteCarloScaleFactorMETandMTRescaled}{1.08 \pm 0.03\stat \pm 0.01\syst}

\newcommand{\eejjZControlRegionContamination}{4\%}

\newcommand{\electronRecoDataMCScaleFactor}{0.98}
\newcommand{\electronRecoDataMCScaleFactorRelUnc}{1.5}
\newcommand{\electronRecoDataMCScaleFactorSqr}{0.96}

\newcommand{\wjetsXSection}{37509.0\unit{pb}}
\newcommand{\zjetsXSection}{3503.71\unit{pb}}
\newcommand{\ttbarXSection}{234\unit{pb}}
\newcommand{\stopSChannelXSection}{5.55\unit{pb}}
\newcommand{\stopTChannelXSection}{87.1\unit{pb}}
\newcommand{\stopTWChannelXSection}{22.2\unit{pb}}
\newcommand{\wwXSection}{57.1\unit{pb}}
\newcommand{\wzXSection}{32.3\unit{pb}}
\newcommand{\zzXSection}{8.26\unit{pb}}

\newcommand{\percentQCDatEEJJLimit}{1\%}
\newcommand{\percentQCDatENuJJLimit}{3\%}

\newcommand{\percentContaminationClosureTest}{5\%}
\newcommand{\percentContaminationClosureTestFinal}{55\%}

\newcommand{\closureTestLowSTPredicted}{13100}
\newcommand{\closureTestLowSTPredictedUnc}{400}
\newcommand{\closureTestLowSTObserved}{12100}
\newcommand{\closureTestLowSTObservedUnc}{400}
\newcommand{\closureTestLowSTRatio}{1.08}
\newcommand{\closureTestLowSTRatioUnc}{0.05}

\newcommand{\closureTestMidSTPredicted}{877}
\newcommand{\closureTestMidSTPredictedUnc}{46.7}
\newcommand{\closureTestMidSTObserved}{600}
\newcommand{\closureTestMidSTObservedUnc}{53}
\newcommand{\closureTestMidSTRatio}{1.46}
\newcommand{\closureTestMidSTRatioUnc}{0.15}

\newcommand{\qcdSystematicUncertaintyPerEle}{30\%}
\newcommand{\qcdSystematicUncertaintyTwoEle}{60\%}

\newcommand{\emujjContamination}{2\%}
\newcommand{\emujjRecoScaleFactor}{0.974  \pm 0.011\stat}

\newcommand{\mumujjRecoScaleFactor}{97.5 \pm 0.4\stat}
\newcommand{\munujjRecoScaleFactor}{97.2 \pm 0.5\stat}

\newcommand{\enujjWJetsShapeUncertainty}{5.92\%}
\newcommand{\enujjTTBarShapeUncertainty}{8.17\%}
\newcommand{\eejjZJetsShapeUncertainty}{8.70\%}

\newcommand{\electronEnergyScaleUncBarrel}{0.4\%}
\newcommand{\electronEnergyScaleUncEndcap}{4.1\%}

\newcommand{\electronEnergyResolutionUncBarrel}{1.006}
\newcommand{\electronEnergyResolutionUncEndcap}{1.015}

\newcommand{\lumiUncertainty}{2.6\%}

\newcommand{\eejjObservedLimit}{1005}
\newcommand{\eejjExpectedLimit}{1030}
\newcommand{\enujjObservedLimit}{845}
\newcommand{\enujjExpectedLimit}{890}

\newcommand{\enujjObservedLimitCombined}{845}
\newcommand{\enujjExpectedLimitCombined}{932}

\newcommand{\eejjObservedLimitNoSyst}{1010}
\newcommand{\eejjExpectedLimitNoSyst}{1030}
\newcommand{\enujjObservedLimitNoSyst}{850}
\newcommand{\enujjExpectedLimitNoSyst}{895}

\newcommand{\eejjObservedLimitMuon}{1015}
\newcommand{\eejjExpectedLimitMuon}{980}
\newcommand{\enujjObservedLimitMuon}{825}
\newcommand{\enujjExpectedLimitMuon}{890}

\newcommand{\PV}{\ensuremath{\cmsSymbolFace{V}\xspace}}
\newcommand{\PVV}{\ensuremath{\cmsSymbolFace{VV}\xspace}}

\newlength\cmsFigWidth
\ifthenelse{\boolean{cms@external}}{\setlength\cmsFigWidth{0.85\columnwidth}}{\setlength\cmsFigWidth{0.4\textwidth}}
\ifthenelse{\boolean{cms@external}}{\providecommand{\cmsLeft}{top}}{\providecommand{\cmsLeft}{left}}
\ifthenelse{\boolean{cms@external}}{\providecommand{\cmsRight}{bottom}}{\providecommand{\cmsRight}{right}}
\ifthenelse{\boolean{cms@external}}{\providecommand{\CL}{C.L.\xspace}}{\providecommand{\CL}{CL\xspace}}
\ifthenelse{\boolean{cms@external}}{\providecommand{\NA}{\ensuremath{\cdots}\xspace}}{\providecommand{\NA}{\text{---}\xspace}}
\newcolumntype{.}{D{.}{.}{-1}}
\newcolumntype{x}[1]{D{,}{\,}{#1}}

\cmsNoteHeader{EXO-12-041}
\title{Search for pair production of first and second generation leptoquarks in proton-proton collisions at \texorpdfstring{$\sqrt{s} = 8$\TeV}{sqrt(s)=8 TeV}}

\date{\today}
\abstract{
    A search for pair production of first and second generation leptoquarks is performed in final states containing either two charged leptons and two jets, or one charged lepton, one neutrino and two jets, using proton-proton collision data at $\sqrt{s}=8$~TeV.  The data, corresponding to an integrated luminosity of 19.7~$\fbinv$, were recorded with the CMS detector at the LHC.  First-generation scalar leptoquarks with masses less than 1010 (850)~GeV are excluded for $\beta = 1.0$~$(0.5)$, where $\beta$ is the branching fraction of a leptoquark decaying to a charged lepton and a quark. Similarly, second-generation scalar leptoquarks with masses less than 1080 (760)~GeV are excluded for $\beta = 1.0$~$(0.5)$.  Mass limits are also set for vector leptoquark production scenarios with anomalous vector couplings, and for R-parity violating supersymmetric scenarios of top squark pair production resulting in similar final-state signatures.  These are the most stringent limits placed on the masses of vector leptoquarks and RPV top squarks to date.
}

\hypersetup{
    pdfauthor={CMS Collaboration},
    pdftitle={Search for pair production of first and second generation leptoquarks in proton-proton collisions at sqrt(s) = 8 TeV},
    pdfsubject={CMS},
    pdfkeywords={CMS, physics, exotica, leptoquark}
}

\maketitle

\newpage

\section{Introduction}
\label{introduction}

The structure of the standard model (SM) of particle physics exhibits a symmetry between quarks and leptons. This paper reports on a search with the CMS detector at the CERN LHC for leptoquark (LQ) particles.  These particles, which manifest a fundamental connection between quarks and leptons, are hypothesized by a variety of extensions to the SM such as grand unified theories~\cite{gut0,gut1,gut2,gut3,gut4,gut5,gut6,gut7}, extended technicolor models~\cite{technicolor1,technicolor2,technicolor3}, superstring-inspired models~\cite{superstring_e6}, and composite models with lepton and quark substructure~\cite{composite}. Leptoquarks carry both baryon (B) and lepton (L) quantum numbers and thus couple to leptons and quarks. They carry fractional electric charge, are color triplets under SU(3)$_C$, and can be either scalar or vector particles. Other properties such as their weak isospin, the helicity of the quarks and leptons to which they couple, and their fermion number $\mathrm{F=(3B+L)}$ depend on the specific structure of each model. Interpretations of direct searches for LQs at particle colliders rely on effective theories, such as the one described in Ref.~\cite{mBRW}, which require LQs to have renormalizable interactions, to obey SM gauge group symmetries, and to couple only to SM fermions and gauge bosons. In order to ensure proton stability, in effective theories LQs are generally constrained to conserve lepton and baryon numbers separately. Moreover, existing experimental limits~\cite{lq_constraints1,FCNC} on lepton number violation, flavor changing neutral currents, and other rare processes favor three generations of LQs with no intergenerational mixing, which is the scenario considered here.

 A search for pair production of first and second generation leptoquarks is performed in final states containing either two charged leptons and two jets, or one charged lepton, one neutrino and two jets, using proton-proton collision data at $\sqrt{s}=8$~TeV.  The data, corresponding to an integrated luminosity of 19.7~$\fbinv$, were recorded with the CMS detector at the LHC.  At hadron colliders, LQs would be produced in pairs or singly; this paper concentrates on LQ pair production. Recent CMS results for single LQ production are documented in Ref.~\cite{singleLQcms}.

The production and decay of a scalar LQ are characterized by its mass ($M_{\mathrm{LQ}}$), its decay branching fraction $\beta$ into a charged lepton and a quark, and the Yukawa coupling $\lambda_{\ell q}$ characterizing the LQ-lepton-quark vertex. The interaction of scalar LQs with SM bosons is completely determined by these three parameters~\cite{mBRW}.  The interaction of vector LQs with the SM bosons additionally depends on two anomalous couplings $\lambda_G$ and $\kappa_G$, which relate to the anomalous magnetic and electric quadrupole moments of the LQ that can be present in the $\mathrm{gLQ\overline{LQ}}$ and $\mathrm{ggLQ\overline{LQ}}$ vertices~\cite{anomCouplings}, where g represents a gluon and $\mathrm{\overline{LQ}}$ represents the anti-LQ. Four scenarios for the values of the anomalous couplings are typically considered in this case:  minimal couplings (MC), $\lambda_G=0$, $\kappa_G=1$; Yang--Mills (YM) type couplings, $\lambda_G=\kappa_G=0$; minimal-minimal (MM) couplings, $\lambda_G=\kappa_G=-1$; and the case of absolute minimal (AM) cross section with respect to the $\lambda_G$, $\kappa_G$ parameters for each value of LQ mass.

\begin{figure*}[htbp]
    \centering
    {\includegraphics[width=.3\textwidth]{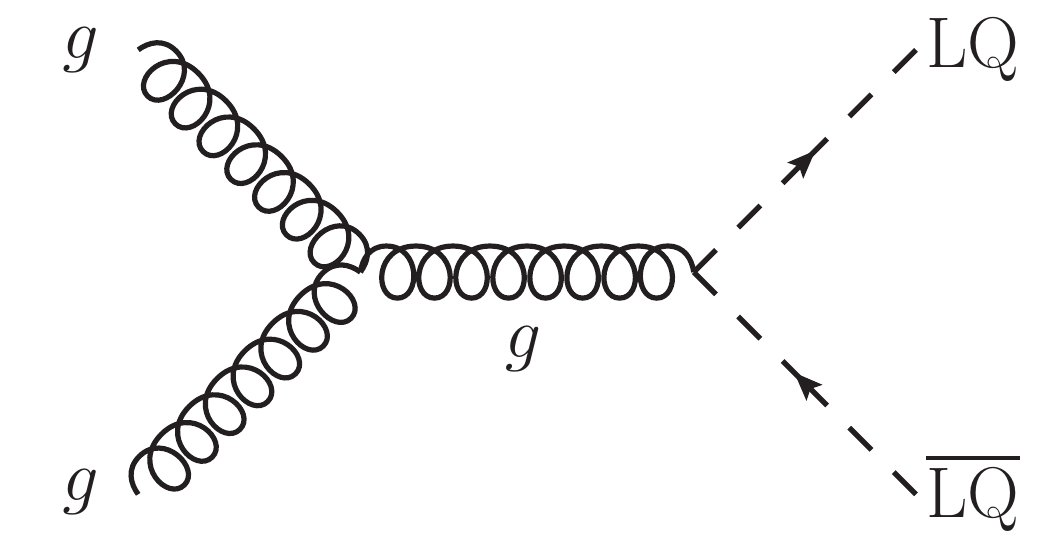}}
    {\includegraphics[width=.3\textwidth]{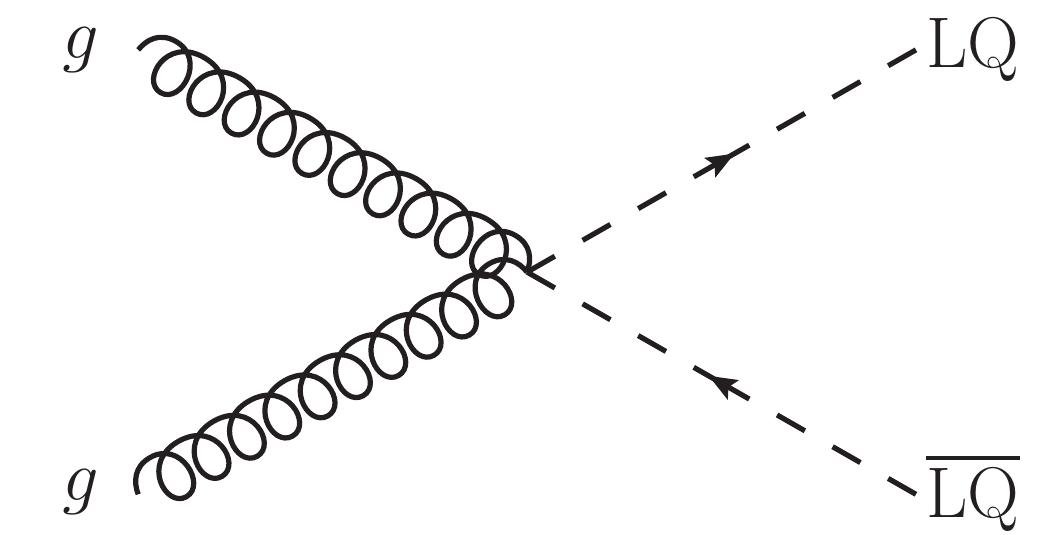}}\\
    {\includegraphics[width=.3\textwidth]{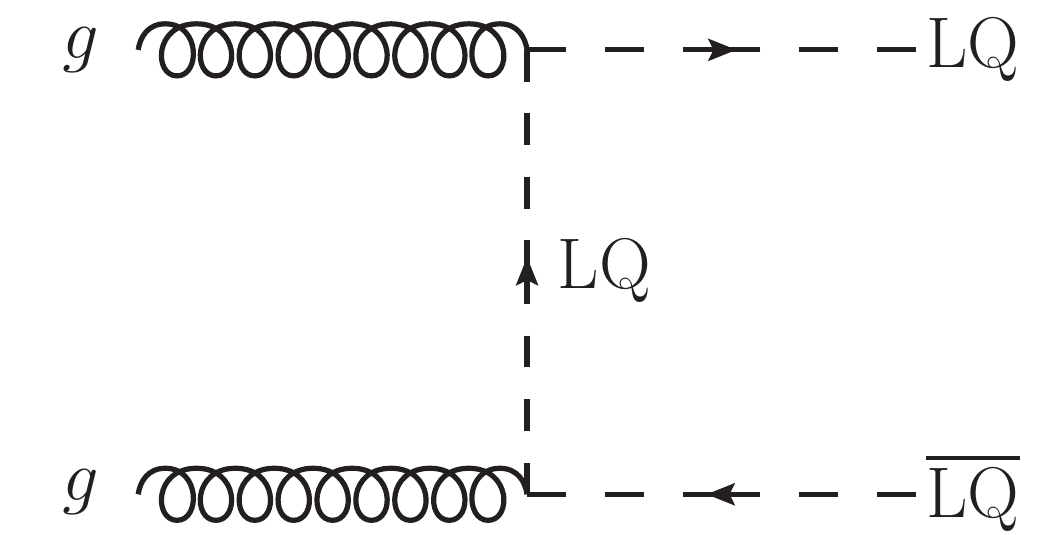}}
    {\includegraphics[width=.3\textwidth]{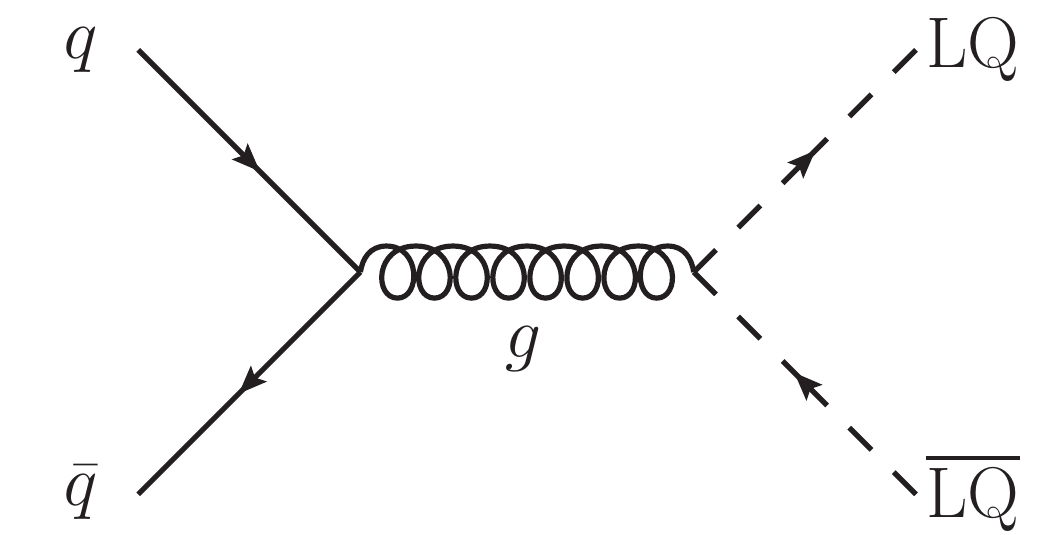}}
    \caption{Dominant leading order diagrams for the pair production of scalar leptoquarks.}
    \label{lqDiagrams}
\end{figure*}

 LQ pair production arises predominantly through gluon-gluon fusion and quark-antiquark annihilation, shown in Fig.~\ref{lqDiagrams}, which have been calculated using next-to-leading order (NLO) QCD corrections~\cite{kramer}. The dominant pair production mechanisms for scalar LQs do not depend on $\lambda_{\ell q}$ and the search sensitivity can be considered $\lambda_{\ell q}$-independent as long as $\lambda_{\ell q}$ is sufficiently large so that LQs decay within a few mm of the primary vertex.

\begin{figure}[htbp]
  \begin{center}
    \includegraphics[width=.4\textwidth]{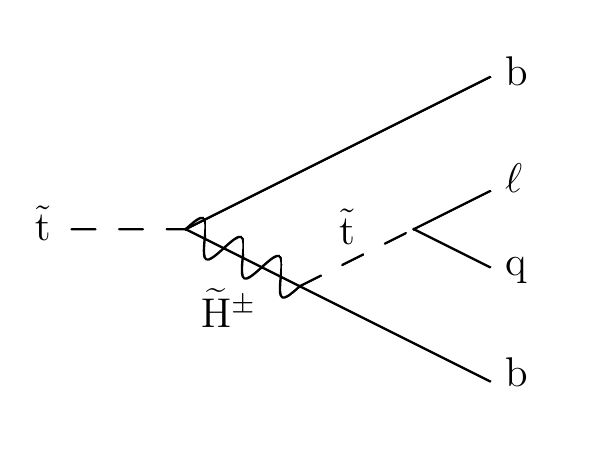}
    \caption{Diagram of the Higgsino-mediated top squark decay via the RPV $\lambda^{\prime}_{132}$ ($\ell$=\Pe) or $\lambda^{\prime}_{232}$ ($\ell$=\Pgm) coupling.}
    \label{rpvFig}
  \end{center}
\end{figure}

Other scenarios of physics beyond the SM could also lead to the prediction of particles with LQ-type couplings. One such theory is supersymmetry (SUSY), which postulates a symmetry between fermions and bosons, and predicts in some models the existence of quark superpartners (squarks), such as the top quark superpartner (top squark, $\PSQt$), decaying into LQ-like final states if R-parity is violated (RPV)~\cite{rpv}.  We consider one such model~\cite{evans}, where top squark decay is mediated by a Higgsino (\sHig) with a mass $M_{\sHig}=M_{\PSQt}-100\GeV$ with a 100$\%$ branching fraction.  The Higgsino in turn produces an off-shell top squark, which decays to a charged lepton and a quark, as shown in Fig.~\ref{rpvFig}.  The top squark decays via the RPV $\lambda^{\prime}_{ijk}$ vertex, where $\lambda^{\prime}_{ijk}$ represents the Yukawa coupling of the RPV term of the superpotential, and the $ijk$ indices represent the family numbers of the interaction superfields, which correspond to $\lambda^{\prime}_{132}$ for the electron final state and $\lambda^{\prime}_{232}$ for the muon final state.  Limits have not previously been set on this model.

The final-state event signatures of the decay of pair-produced LQs can be classified as: dilepton and jets (both LQ and $\mathrm{\overline{LQ}}$ decay into a charged lepton and a quark); single lepton, missing transverse momentum and jets (one LQ decays into a charged lepton and a quark, while the other decays into a neutrino and a quark); and missing transverse momentum and jets (both LQ and $\mathrm{\overline{LQ}}$ decay into neutrinos and quarks). The three signatures correspond to branching fractions of $\beta^2$, 2$\beta(1-\beta)$, and $(1-\beta)^2$, respectively. The charged leptons can be either electrons, muons, or tau leptons, corresponding to the three generations of LQs. Only final states containing electrons and muons are considered here, and two distinct signatures: one with two high transverse momentum ($\PT$) charged leptons and two high $\PT$ jets (denoted as \lljj), and the other with one high $\PT$ charged lepton, large missing transverse momentum, and two high $\PT$ jets (denoted as \lnujj).
These final states are analyzed in the context of scalar LQs, vector LQs~\cite{belyaev} and the RPV SUSY scenario previously mentioned.

The most stringent limits on the pair-production of scalar LQs come from the LHC experiments.  The ATLAS experiment excluded first (second) generation LQs with masses below {1050 (1000)\GeV} for $\beta = 1$, and {900 (850)\GeV} for $\beta = 0.5$, using {20\fbinv} of $\sqrt{s}=8\TeV$ data~\cite{atlas_lq_8tev}.  Using $\sqrt{s}=7\TeV$ proton-proton collisions data corresponding to an integrated luminosity of {5.0\fbinv}, the CMS experiment excluded first- (second-)generation pair-produced scalar LQs with masses below 830 (840)\GeV for $\beta = 1$  and 640 (650)\GeV for $\beta = 0.5$~\cite{cmslq-2011}.  CMS excluded third-generation pair-produced scalar LQs with masses below 740\GeV for $\beta = 1$, using {19.7\fbinv} of $\sqrt{s}=8\TeV$ data~\cite{cms_lq3}.  ATLAS excluded third-generation pair-produced scalar LQs with masses below 534\GeV for $\beta = 1$, using {4.7\fbinv} of $\sqrt{s}=7\TeV$ data~\cite{atlas_lq3}.  The HERA experiments H1~\cite{hera2} and ZEUS~\cite{hera1} produced $\lambda$-dependent results for LQ models, excluding scalar LQ masses up to roughly 500--650 (300)\GeV for $\lambda$ = 1.0 (0.3).  Searches for scalar LQs have also been performed by the Tevatron experiments \DZERO~\cite{dzero_lq1,dzero_lq2,dzero_enujj} and CDF~\cite{cdf_lq1,cdf_lq2,cdf_nunujj}.  The most stringent limits on vector LQs have been reported by \DZERO~\cite{dzero_vlq1,dzero_vlq2,dzero_vlq3} and CDF~\cite{cdf_vlq}.

\section{The CMS detector}
\label{cms}
The central feature of the CMS apparatus is a superconducting solenoid of 6\unit{m} internal diameter, providing a magnetic field of 3.8\unit{T}. Within the solenoid volume are a silicon pixel and strip tracker, a lead tungstate crystal electromagnetic calorimeter (ECAL), and a brass and scintillator hadron calorimeter (HCAL), each composed of a barrel and two endcap sections. Muons are measured in gas-ionization detectors embedded in the steel flux-return yoke outside the solenoid. Extensive forward calorimetry complements the coverage provided by the barrel and endcap detectors.  A more detailed description of the CMS detector, together with a definition of the coordinate system used and the relevant kinematic variables, can be found in Ref.~\cite{cms-jinst}.

The inner tracking system of CMS consists of a silicon pixel and strip tracker, providing the required
granularity and precision for the reconstruction of vertices of
charged particles in the range of the azimuthal angle $0 \leq \phi < 2\pi$ and
pseudorapidity $|\eta|<2.5$. The crystal ECAL and the brass and scintillator
sampling HCAL are used to measure the energies of
photons, electrons, and hadrons within $|\eta|<3.0$.  The electron momentum is estimated by combining the energy measurement in the ECAL with the momentum measurement in the tracker. The momentum resolution for electrons with $\pt {\approx} 45$\GeV from $\Z \rightarrow \Pe \Pe$ decays ranges from 1.7\% for nonshowering electrons in the barrel region to 4.5\% for showering electrons in the endcaps~\cite{Khachatryan:2015hwa}.

The CMS detector is nearly hermetic, which allows for a measurement of missing transverse momentum.
The three muon systems surrounding the solenoid cover the region $|\eta|<2.4$ and are composed of drift tubes in the barrel region $(|\eta|<1.2)$,
cathode strip chambers in the endcaps $(0.9<|\eta|<2.4)$, and
resistive-plate chambers in both the barrel region and the endcaps
$(|\eta|<1.6)$.
Events are recorded based on a trigger decision using
information from the CMS detector subsystems.
The first level (L1) of the CMS trigger system, composed of custom hardware processors, uses information from the calorimeters and muon detectors to select the most interesting events in a fixed time interval of less than 4\mus. The high-level trigger (HLT) processor reduces the event rate from 100\unit{kHz} at L1 to roughly 400\unit{Hz}.

\section{Data and simulation samples}
\label{samples}

The data used in this paper correspond to an integrated luminosity of
{($19.7 \pm 0.5$)~$\fbinv$}. The integrated luminosity is measured as described in Ref.~\cite{lumi_new}.

For the searches in the \eejj~and \enujj~channels, events are selected by triggers requiring at least
one electron with $\PT>30\GeV$, at least one jet with
$\PT>100\GeV$, and at least one additional jet with $\PT>25\GeV$.
For the determination of the hadronic multijet background in the
\eejj~and the \enujj~channels, events are selected using single-photon triggers, which require at least
one ECAL energy deposit.

Events in the \mumujj~and the \munujj~channels are selected if they pass a single-muon trigger selection that requires a muon with $\PT>$ 40\GeV,  $|\eta| < 2.1$.  There are no isolation requirements. This selection is also used to provide a sample of \emujj~events for the determination of the \ttbar~background in both \lljj~channels.

Simulated signal and background samples are produced and fully reconstructed using a simulation of the CMS detector based on {\sc GEANT4}~\cite{GEANT4}.  These simulations include additional collisions in a single bunch crossing (pileup) with a distribution matched to the number of pileup events observed during the various data-taking periods.

Signal samples for scalar LQ masses from 300 to 1200\GeV in 50\GeV steps were generated at the leading-order (LO) level with the \PYTHIA event generator~\cite{PYTHIA}
and CTEQ6L1~\cite{CTEQ} parton distribution function (PDF) set. These samples are used to study the acceptance, while NLO cross sections are used for comparison in the limit-setting procedure.  With the exception of the RPV SUSY sample described below, \PYTHIA 6.422 with the Z2 tune~\cite{tunez2} was used.  The search limits being $\lambda_{\ell q}$-independent, these samples were generated with a coupling strength $\lambda_{\ell q} = 0.3$. The vector LQ signal samples were generated with the \CALCHEP version 3.4 event generator~\cite{calcHEP} and CTEQ6L PDF set using the model with vector LQ implemented in Ref.~\cite{belyaev}. Vector LQ masses between 200 and 1800\GeV were generated in 100\GeV steps, for the four scenarios of the anomalous couplings $\lambda_G$ and $\kappa_G$ described in Section~\ref{introduction}. Samples of RPV SUSY events were produced with $\eejj$~and $\mumujj$~final state signatures.  These samples were produced for top squark masses from 300 to 1000\GeV in 50\GeV steps using \PYTHIA 8.175~\cite{PYTHIA8} and their decays were simulated withd \MADGRAPH 5.1.1~\cite{MG5}.  Top squark production cross sections were calculated at the NLO + next-to-leading-logarithm (NLL) level using \PROSPINO~\cite{prospino} and the {\sc nll-fast} program~\cite{NLLfast1,NLLfast2}, using the CTEQ6M PDF set.

The main sources of background for these searches are $\ttbar$, single top quark, $\cPZ/\gamma^*$+jets, $\PW$+jets, diboson ($\cPZ\cPZ/\PW\cPZ/\PW\PW$)+jets, and multijet production. Backgrounds in the \lljj~channels from multijet production and $\ttbar$ events are estimated from data control regions, while single top quark, $\cPZ/\gamma^*$+jets, $\PW$+jets, and diboson ($\cPZ\cPZ/\PW\cPZ/\PW\PW$)+jets backgrounds are estimated using simulated events. In the \lnujj~channel, the $\ttbar$ background is also estimated using simulated events.  The simulated samples of $\ttbar$, $\cPZ/\gamma^*$+jets, and $\PW$+jets are generated with \MADGRAPH;
single top quark samples ($s$-, $t$-, and t$\PW$- channels) are generated with \POWHEG version 1.0~\cite{Nason:2004rx,Frixione:2007vw,Alioli:2010xd,Alioli:2009je}; and samples of $\PVV$, where $\PV$ represents either a $\PW$ or $\cPZ$ boson, are generated with \PYTHIA.
The simulations with \MADGRAPH and \PYTHIA use the CTEQ6L1 PDF set. The simulations with \POWHEG use the CTEQ6M PDF set.

The $\PW$+jets and $\cPZ/\gamma^*$+jets samples are normalized to next-to-NLO (NNLO) inclusive
cross sections calculated with \FEWZ version 3.1~\cite{FEWZ}. Single top quark and
$\PVV$ samples are normalized to NLO inclusive cross sections calculated
with \MCFM version 6.6~\cite{ mcfm_t, mcfm_Wt, mcfm_tch,mcfm_diboson}.
Results from Refs.~\cite{Czakon:2013goa,Czakon:2013tha} are used to normalize the $\ttbar$ sample at the NNLO + next-to-NLL level.

\section{Event reconstruction and selection}
\label{eventselection}

Electron candidates are created by matching an electromagnetic cluster in the ECAL in $\eta$ and $\phi$ to a reconstructed track in the inner tracking system. The ECAL cluster must have a shower shape and longitudinal profile consistent with that of an electromagnetic shower. The matching reconstructed track can lack a hit in at most one pixel layer, and must be within 0.02 (0.05)\unit{cm} of the matched primary vertex in the barrel (endcap). The resulting electron candidates are required to pass a set of criteria optimized for electrons with energies of hundreds of \GeV\cite{Khachatryan:2015hwa}.   In particular, they must have transverse momenta \PT$>35\GeV$ and $|\eta| < 2.5$, excluding the transition region between the barrel and endcap detectors, $1.442 < |\eta| < 1.560$, where the electron reconstruction is suboptimal. The transverse momentum sum of tracks in a cone of {$\Delta R = \sqrt{(\Delta \phi )^2 + (\Delta \eta) ^2} = 0.5$} around the electron candidate's track must be less than 5\GeV, which reduces the chance of jets being misidentified as electrons. Tracks used in this momentum sum, known as tracker isolation, must be within 0.2\unit{cm} of the z coordinate of the electron candidate's matching primary vertex to eliminate tracks coming from other proton-proton collisions in the same bunch crossing. The transverse energy sum of the calorimeter energy deposits falling in the $\Delta R = 0.5$ cone is required to be less than about 3\% of the candidate's transverse energy. This energy sum, known as calorimeter isolation, has an extra contribution accounting for the average contribution of additional proton-proton collisions in the same bunch crossing.

Muons are reconstructed as tracks combining hit segments in the muon system and hits in the inner tracking system~\cite{MuId}. Muons are required to have $\PT > 45\GeV$ and to be contained in the fiducial volume used for the HLT muon selection, $|\eta| < 2.1$. In addition, muons are required to satisfy a set of identification criteria optimized for high~\PT.  They require at least one muon detector segment be included in the muon track fit, and segments in at least two muon stations be geometrically matched to a track in the inner tracking system. Isolated muons are selected by requiring that the sum of the transverse momenta of all tracks in the tracker in a cone of {$\Delta R = 0.3$} around the muon track (excluding the muon track itself), divided by the muon $\PT$, is less than 0.1. To have a precise $\PT$~measurement and to suppress muons from decays in flight, at least 8 tracker layers with associated hits are required, and at least one hit in the pixel detector. To reject muons from cosmic rays, the transverse impact parameter with respect to the primary vertex is required to be less than 2~mm and the longitudinal distance of the tracker-only track with respect to the primary vertex is required to be less than 5~mm, where the primary vertex is defined as the reconstructed vertex for which the $\PT^2$~sum of the assigned tracks is largest~\cite{primVert}.

Events are reconstructed using a particle-flow (PF) algorithm~\cite{CMS-PAS-PFT-09-001, CMS-PAS-PFT-10-001}, which identifies and measures stable particles by combining information from all CMS sub-detectors. The missing transverse momentum vector \ptvecmiss is defined as the projection on the plane transverse to the beams of the negative vector of the momenta of all particles reconstructed with the PF algorithm in the event, and the missing transverse energy (\MET) is defined as the magnitude of the \ptvecmiss vector. Jets are reconstructed using the anti-$k\rm _{\mathrm{T}}$~\cite{akt_jets, fastjetmanual} algorithm with a
distance parameter of $0.5$. The jet energy is calibrated using the $\PT$ balance of dijet and $\gamma+$jet events in both data and simulation~\cite{JetCorr}.  The PF jet energy resolution is 15\% at 10\GeV, 8\% at 100\GeV, and 4\% at 1\TeV, to be compared to about 40\%, 12\%, and 5\% obtained when the calorimeters alone are used for jet clustering.  The leading (sub-leading) jet is required to have $\PT >125$~$(45)\GeV$. All jets are required to have $|\eta| < 2.4$. Furthermore, only jets having a spatial separation from electron or muon candidates of $\Delta R > 0.3$ are considered.

\subsection{\texorpdfstring{The \lljj~channel}{The lljj channel event selection}}
\label{sec:lljjSelection}
An initial selection is made to obtain events containing at least two charged lepton candidates (either two electrons or two muons) and at least two jets for this channel. The two highest $\PT$ leptons and the two highest $\PT$ jets are considered as the decay products
from a pair of LQs. They must satisfy the identification criteria described above.  Further, the invariant mass of the two leptons, $M_{\ell\ell}$, is required to be larger than 50\GeV. Muons are required to be spatially separated from one another by $\Delta R > 0.3$.  The scalar sum of the transverse momenta of the selected final state leptons and jets in the event $S_{\mathrm{T}} = \PT(\ell_1) + \PT(\ell_2) + \PT(\mathrm{j_1}) + \PT(\mathrm{j_2})$ is required to be larger than 300\GeV.  No charge requirement is placed on the leptons.
After this initial selection, the signal-to-background separation is optimized by maximizing $S/\sqrt{S+B}$, where $S$ and $B$ represent numbers of signal and background events, respectively.  This is done by varying cuts on certain kinematic variables, and selecting the combination of cuts with the maximum $S/\sqrt{S+B}$.
Three variables are optimized for each LQ mass hypothesis in both \lljj~channels: $S_{\mathrm{T}}$; $M_{\ell\ell}$, used to remove most of the contribution from the $\cPZ/\gamma^*$+jets background; and $M^{\mathrm{min}}_{\ell, {\mathrm{jet}}}$, defined as the smaller of the two lepton-jet invariant masses, given the combination that minimizes the $\mathrm{LQ-\overline{LQ}}$ invariant mass difference.

The \eejj~and \mumujj~channels are optimized separately and the optimized thresholds are summarized in Tables~\ref{tab:optimization_ee} and~\ref{tab:optimization_mumu}, respectively. For the mass hypotheses beyond 1\TeV the same set of final selections as those for the 1\TeV mass hypothesis is used.

\begin{table*}[htbp]
\topcaption{Optimized thresholds for different LQ mass hypotheses of the \eejj~signal.}
\begin{center}
\scriptsize
\begin{scotch}{rrrrrrrrrrrrrrrrr}
      & \multicolumn{15}{c}{LQ mass [\GeVns{}]} \\
      & 300 & 350 & 400 & 450 & 500 & 550 & 600 & 650 & 700 & 750 & 800 & 850 & 900 & 950 & $\geq 1000$ \\
      \hline
      \multicolumn{1}{r}{\st~[\GeVns{}]}  & 435 & 485 & 535 & 595 & 650 & 715 & 780 & 850 & 920 & 1000 & 1075 & 1160 & 1245 & 1330 & 1425 \\
      \multicolumn{1}{r}{$M_{\mathrm{ee}}$~[\GeVns{}]}  & 110 & 110 & 115 & 125 & 130 & 140 & 145 & 155 & 160 & 170 & 175 & 180 & 190 & 195 & 205 \\
      \multicolumn{1}{r}{$M^{\mathrm{min}}_{\mathrm{ej}}$~[\GeVns{}]}  & 50 & 105 & 160 & 205 & 250 & 290 & 325 & 360 & 390 & 415 & 435 & 450 & 465 & 470 & 475 \\
    \end{scotch}
    \normalsize
\label{tab:optimization_ee}
\end{center}
\end{table*}

\begin{table*}[htbp]
\topcaption{Optimized thresholds for different LQ mass hypotheses of the \mumujj~signal.}
\begin{center}
\scriptsize
\begin{scotch}{rrrrrrrrrrrrrrrrr}
      & \multicolumn{15}{c}{LQ mass [\GeVns{}]} \\
      & 300 & 350 & 400 & 450 & 500 & 550 & 600 & 650 & 700 & 750 & 800 & 850 & 900 & 950 & $\geq 1000$ \\
      \hline
      \multicolumn{1}{r}{\st~[\GeVns{}]} & 380 & 460 & 540 & 615 & 685 & 755 & 820 & 880 & 935 & 990 & 1040 & 1090 & 1135 & 1175 & 1210 \\
      \multicolumn{1}{r}{$M_{\Pgm\Pgm}$~[\GeVns{}]}  & 100 & 115 & 125 & 140 & 150 & 165 & 175 & 185 & 195 & 205 & 215 & 220 & 230 & 235 & 245 \\
      \multicolumn{1}{r}{$M^{\mathrm{min}}_{\Pgm\cPj}$~[\GeVns{}]} & 115 & 115 & 120 & 135 & 155 & 180 & 210 & 250 & 295 & 345 & 400 & 465 & 535 & 610 & 690 \\
      \end{scotch}
      \normalsize
\label{tab:optimization_mumu}
\end{center}
\end{table*}

\subsection{\texorpdfstring{The \lnujj~channel}{The lnujj channel}}
Events in this channel are selected to contain exactly one charged lepton (electron or muon),
at least two jets, and $\MET > 55\GeV$.  Leptons and jets must meet the criteria described above.  Events containing a second lepton (electron or muon) are vetoed for the \lnujj~selections.
In addition, in order to reject events with mis-reconstructed $\MET$, the angle in the transverse plane between the direction of the leading $\PT$ jet and the \ptvecmiss vector,~$\Delta \phi(\ptvecmiss,\mathrm{j_1})$ is required to be larger than 0.5.  For the same reason, the electron or muon and the \ptvecmiss are required to be separated by $\Delta \phi(\ptvecmiss,\ell) > 0.8$. In the \enujj~channel, the angular separation $\Delta R$ between the electron and either of the jets is required to be larger than 0.7 in order to reduce the contamination from QCD multijet background in that channel.
Events are required to have $M_{\mathrm{T}} >50 \GeV$, where $M_{\mathrm{T}}$, the transverse mass of the charged lepton and undetected particles, is defined as $M_{\mathrm{T}}= \sqrt{\smash[b]{2 \pt^{\ell}   \MET \left( 1 - \cos{\Delta \phi} \right) }}$, where $\pt^{\ell}$ is the lepton $\PT$ and $\Delta \phi$ is the difference in azimuthal angle between the charged lepton momentum direction and the \ptvecmiss vector.
Lastly, events are selected to have $S_{\mathrm{T}}>300\GeV$, where the scalar transverse energy $S_{\mathrm{T}}$ is defined in this case to be $S_{\mathrm{T}}= \PT(\ell) + \MET + \PT(\mathrm{j_1}) + \PT(\mathrm{j_2})$.

After this initial selection, the following variables are used to optimize a final selection for each LQ mass hypothesis using the method described above: $M_{\mathrm{T}}$; $S_{\mathrm{T}}$; and $M_{\ell\cPj}$, defined as the invariant mass of the lepton-jet pair that minimizes the difference in the $M_{\mathrm{T}}$ of the lepton-jet and $\MET$-jet pairs. The \enujj~channel uses $\MET$ as an additional optimization variable.

The optimized thresholds for the \enujj~and the \munujj~channels are summarized in Tables~\ref{tab:optimization_enu} and~\ref{tab:optimization_munu}, respectively. Mass hypotheses beyond 950 (1000)\GeV for the \enujj~(\munujj)~channel use the same set of final selections as those for the 950 (1000)\GeV mass hypothesis.

\begin{table*}[htbp]
  \scriptsize
  \topcaption{Optimized thresholds for different LQ mass hypotheses of the \enujj~signal.}
  \begin{scotch}{rrrrrrrrrrrrrrrrr}
   & \multicolumn{14}{c}{LQ Mass [\GeVns{}]} \\
    & 300 & 350 & 400 & 450 & 500 & 550 & 600 & 650 & 700 & 750 & 800 & 850 & 900 & ${\ge}950$ \\
    \hline
    \multicolumn{1}{r}{$S_{\mathrm{T}}$~[\GeVns{}]}  & 495 & 570 & 645 & 720 & 800 & 880 & 960 & 1040 & 1120 & 1205 & 1290 & 1375 & 1460 & 1545 \\
    \multicolumn{1}{r}{$M_{\mathrm{ej}}$~[\GeVns{}]}  & 195 & 250 & 300 & 355 & 405 & 455 & 505 & 555 & 600 & 645 & 695 & 740 & 780 & 825 \\
    \multicolumn{1}{r}{$M_{\mathrm{T}}$~[\GeVns{}]}  & 125 & 150 & 175 & 200 & 220 & 240 & 255 & 270 & 280 & 290 & 295 & 300 & 300 & 300 \\
    \multicolumn{1}{r}{\MET~[\GeVns{}]}  & 90 & 95 & 100 & 110 & 115 & 125 & 135 & 145 & 155 & 170 & 180 & 195 & 210 & 220 \\
    \end{scotch}
\label{tab:optimization_enu}
\normalsize
\end{table*}

\begin{table*}[htbp]
\scriptsize
\topcaption{Optimized thresholds for different LQ mass hypotheses of the \munujj~signal.}
\begin{scotch}{rrrrrrrrrrrrrrrrr}
   & \multicolumn{15}{c}{LQ Mass [\GeVns{}]} \\
    & 300 & 350 & 400 & 450 & 500 & 550 & 600 & 650 & 700 & 750 & 800 & 850 & 900 & 950& ${\ge}1000$ \\
    \hline
    \multicolumn{1}{r}{$S_{\mathrm{T}}$~[\GeVns{}]}  & 455 & 540 & 625 & 715 & 800 & 890 & 980 & 1070 & 1160& 1250 & 1345 & 1435 & 1530 & 1625 & 1720 \\
    \multicolumn{1}{r}{$M_{\mu\cPj}$~[\GeVns{}]}  & 125 & 150 & 175 & 200 & 225 & 250 & 280 & 305 & 330 & 355 & 380 & 410 & 435 & 465 & 490\\
    \multicolumn{1}{r}{$M_{\mathrm{T}}$~[\GeVns{}]}  & 155 & 180 & 205 & 225 & 245 & 260 & 275 & 290 & 300 &310 & 315 & 320 & 320 & 325 & 320 \\
\end{scotch}
\label{tab:optimization_munu}
\normalsize
\end{table*}

\section{Background estimation}
\label{backgrounds}

\def\WTM{M^{\mathrm{T}}_{\ell\nu}}
\def\ST{S_{\mathrm{T}}}

The main SM processes that can mimic the LQ signal in the \lljj~channels are: processes that lead to the production of genuine dilepton events such as $\cPZ/\gamma^*$+jets, \ttbar, and $\PVV$+jets; and processes which produce either 0 or 1 genuine leptons and at least one hadronic jet which leads to a mis-identified lepton such as multijet events, single t production, and $\PW$+jets. The contributions from single top quarks, $\PVV$+jets, and $\PW$+jets are estimated from simulation and are small once the full event selection is applied. The contribution from the principal background, $\cPZ/\gamma^*$+jets, is estimated with using simulated events normalized to the data in a control region, where the non-$\cPZ/\gamma^*$+jets backgrounds have been removed from the data control region using the identical selection in simulation. The $\cPZ/\gamma^*$+jets simulation is rescaled to agree with this modified data sample at the \lljj~initial selection level within a \cPZ~boson enriched region of 70 (80) $< \mathrm{M_{\ell\ell}} < 110$ (100)\GeV for the electron (muon) channel. The resulting correction factor is $R_{\cPZ} = 0.97 \pm 0.01$~(stat) for \eejj, and $R_{\cPZ} = 0.92 \pm 0.01$~(stat) for \mumujj. The contribution from $\ttbar$ events with two leptons of the same flavor is estimated from a data sample containing one electron and one muon. This data sample is dominated by $\ttbar$ processes, which are expected to yield \emujj~events with the same probability as (\eejj~+~\mumujj) events. The data sample is therefore reweighted to account for: the different branching fraction of the \emujj~final state, which is twice that of the \eejj~or \mumujj~final states; the differences in electron and muon identification and isolation efficiencies; and the differences in trigger efficiencies. This sample can then be used to estimate the contribution from the \ttbar~process in the \lljj~channels for both the initial and final selections in all kinematic distributions.

The multijet background in the \eejj~channel is determined from a data control region containing exactly two electron candidates that pass loosened identification criteria on the cluster shape and no isolation requirements, and at least two jets. Each electron candidate in this sample is weighted by the probability that an electron candidate passing such loosened requirements additionally passes all final electron requirements.
This probability is measured as a function of $\PT$ in three $\eta$ regions ($|\eta| < 1.442$, $1.56 < |\eta| < 2.00$, and $2.00 < |\eta| < 2.50$), using a data sample dominated by multijet events, collected with a single-photon trigger and containing one and only one electron candidate and two or more jets.
The contribution from multijet processes in the \mumujj~channel is determined using a multijet-enriched data sample of same-sign dimuon events with no muon isolation criteria imposed. The same-sign nonisolated data sample is reweighted according to a same-sign/opposite-sign ratio and an isolation acceptance factor calculated using simulation. After reweighting, the same-sign nonisolated data sample is used to predict the multijet contribution to the final \mumujj~selection, which is shown to be negligible.

\begin{figure*}[!htb]
       \centering
       {\includegraphics[width=.45\textwidth]{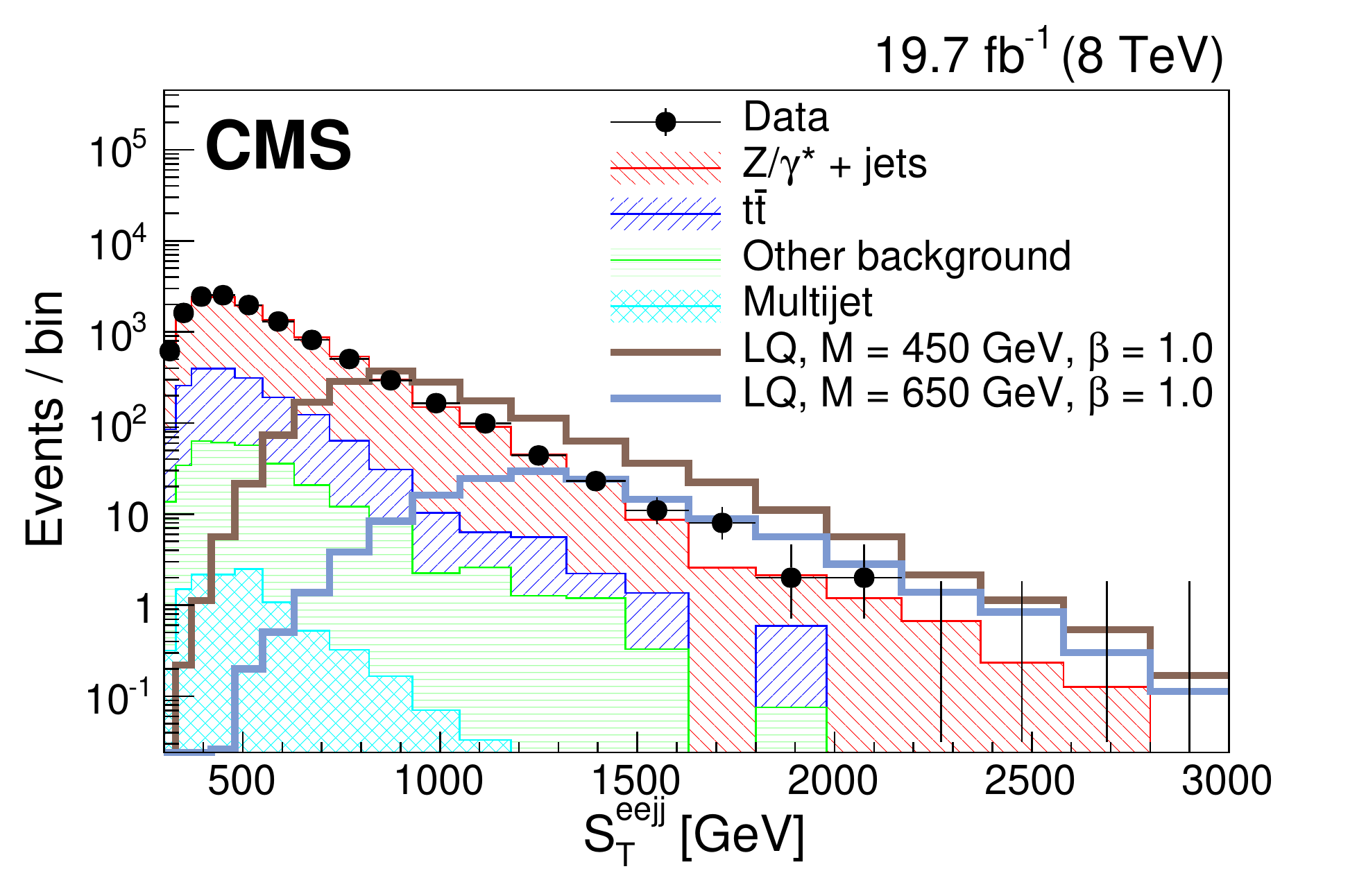}}
       {\includegraphics[width=.45\textwidth]{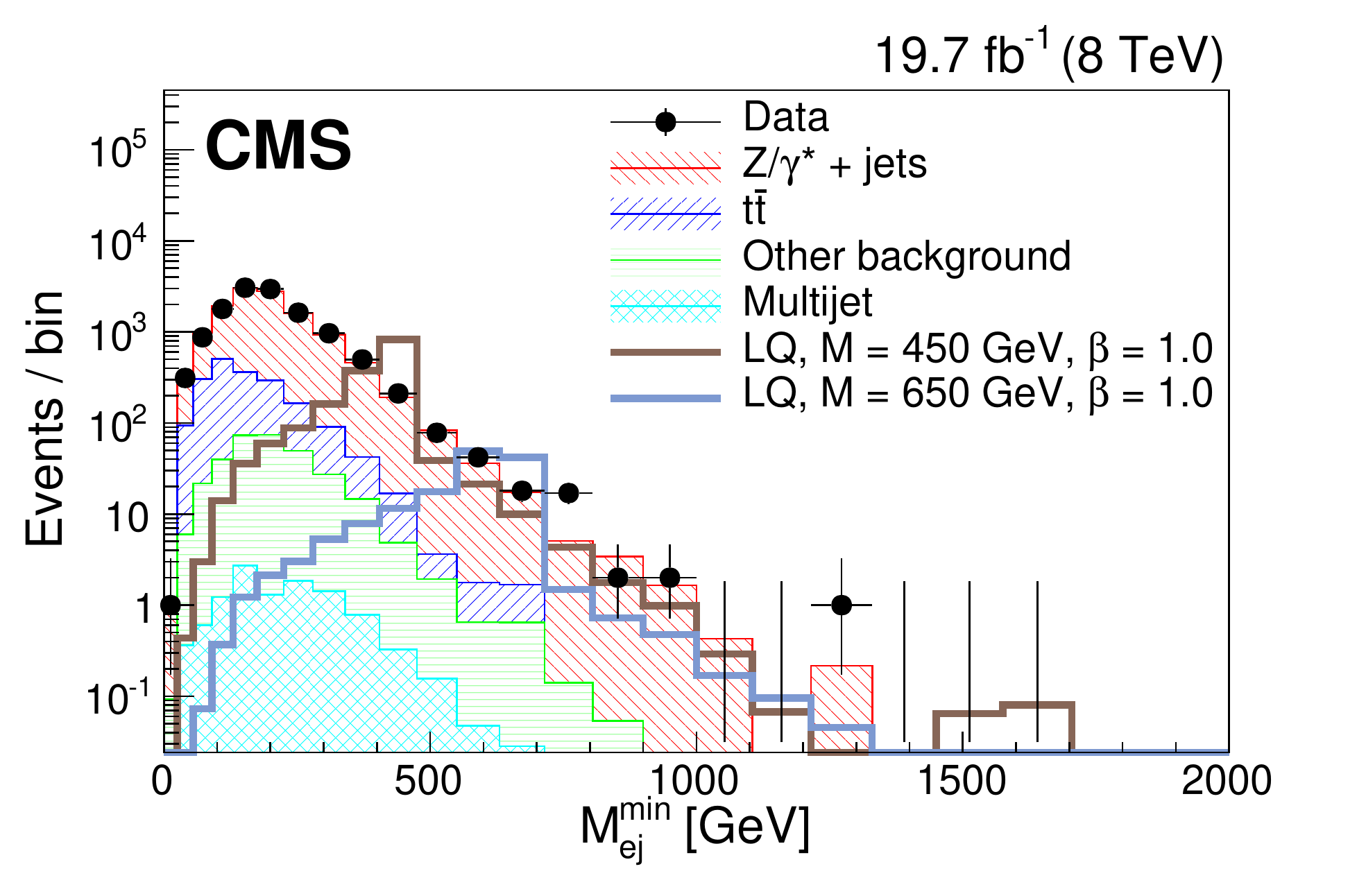}}
       {\includegraphics[width=.45\textwidth]{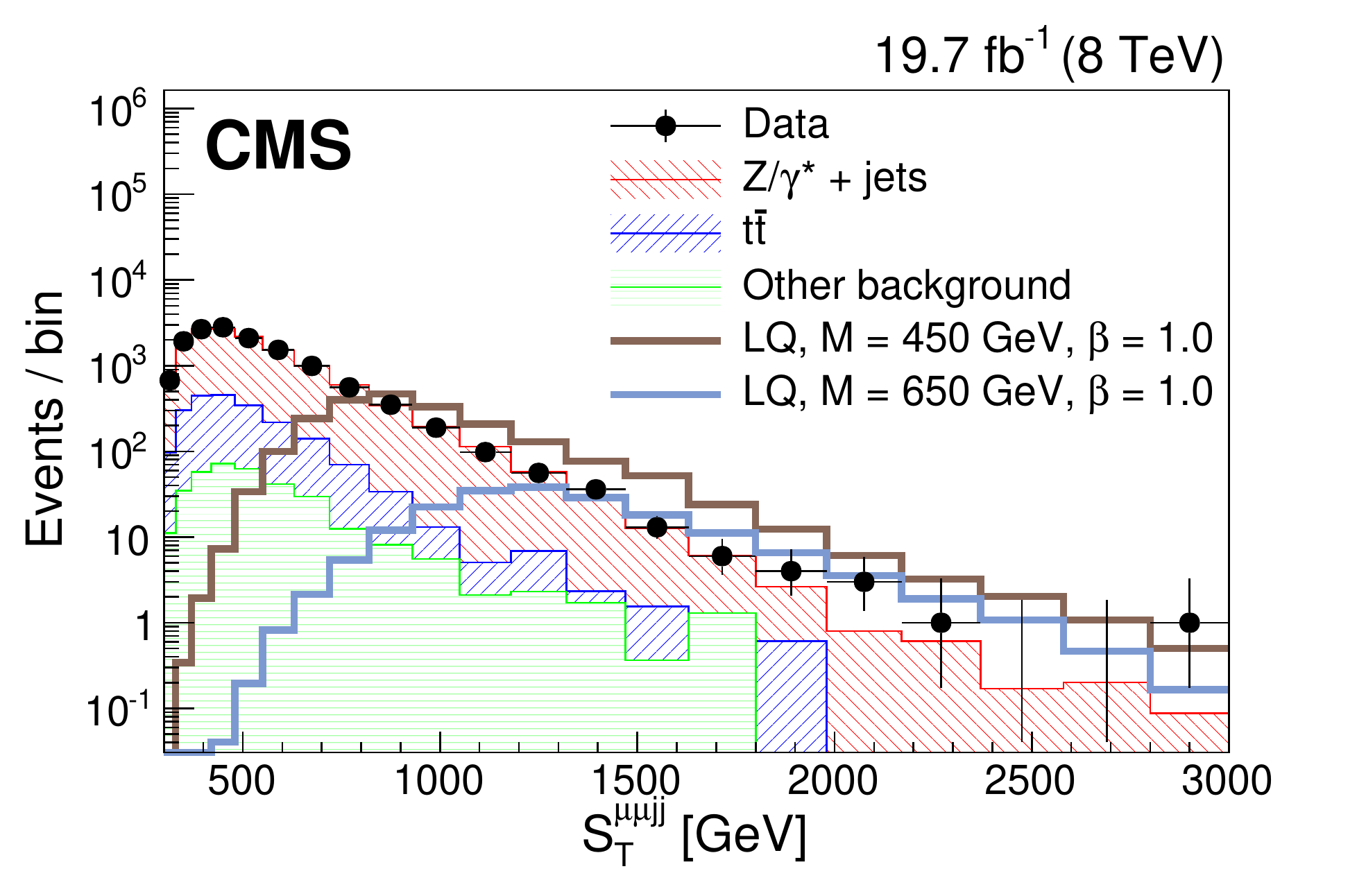}}
       {\includegraphics[width=.45\textwidth]{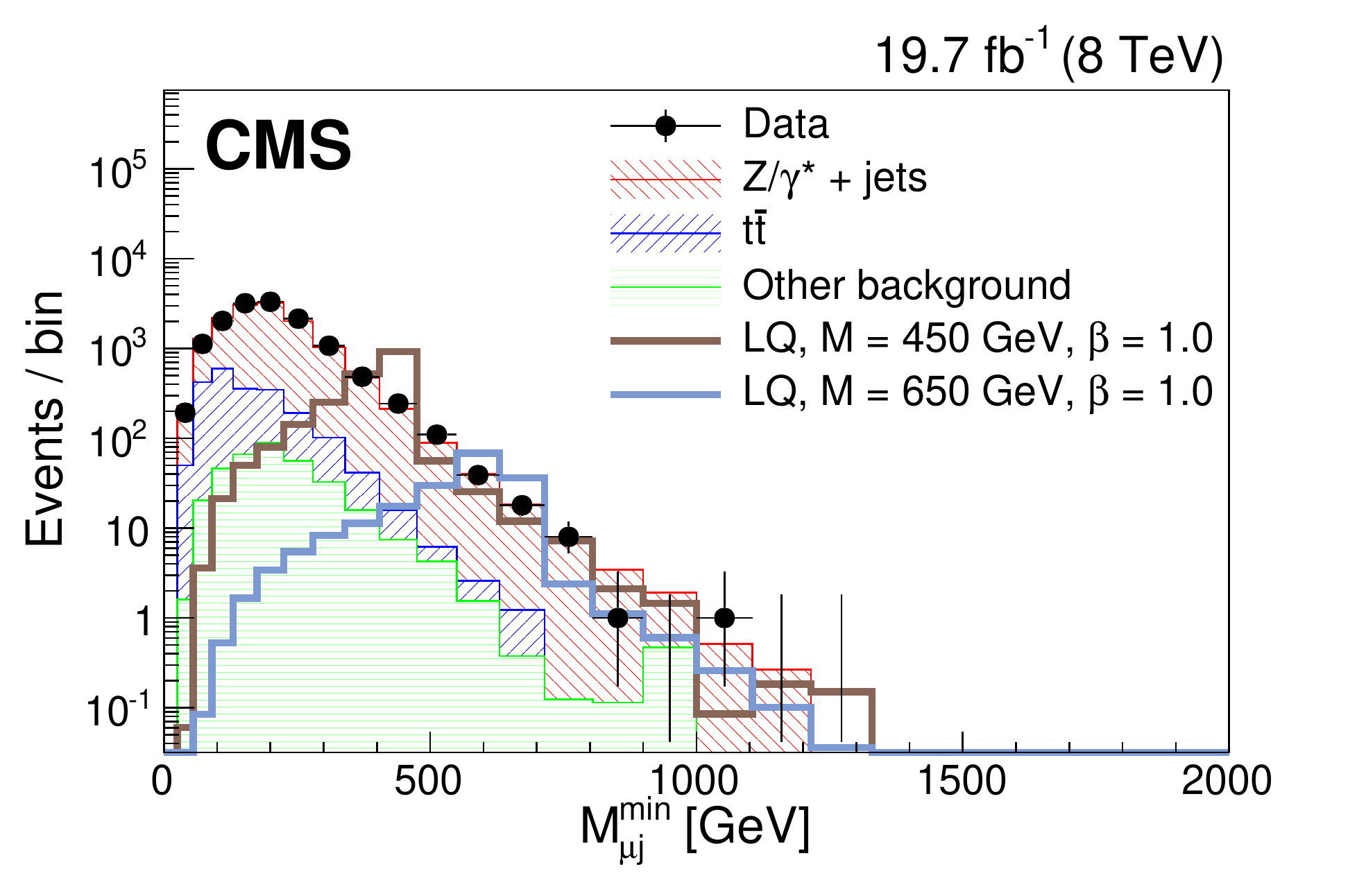}}
\caption{Distributions of $S_{\mathrm{T}}$ (left) and $M^{\mathrm{min}}_{\ell\cPj}$ (right) at the initial selection level in the \eejj~(top) and \mumujj~(bottom) channels. ``Other background" includes: diboson, $\PW$+jets, $\gamma$+jets, and single top quark contributions.  The horizontal lines on the data points show the variable bin width.
	  \label{figapp:presel_ee_and_mumu}}
\end{figure*}

All final state distributions in the \eejj~and \mumujj~channels of the background prediction and of data, at the initial selection level, have been studied and show agreement within uncertainties.  The specific distributions of $S_{\mathrm{T}}$ and $M^{\mathrm{min}}_{\ell\cPj}$ are shown in Fig.~\ref{figapp:presel_ee_and_mumu}.  Systematic uncertainties, discussed in the next section, are not included in these plots.

The primary backgrounds that can mimic the LQ signal in the \lnujj~channels fall into three categories: events with genuine $\PW$ bosons such as those from $\PW$+jets, \ttbar, single top quark production, and $\PW\PW$ and  $\PW\cPZ$ processes; events with misidentified leptons and misreconstructed \MET in the final state caused mostly by the misidentification of jets as leptons in multijet processes; and events with $\cPZ$ bosons such as those from $\cPZ/\gamma^*$+jets and $\cPZ\cPZ$ processes, where only one lepton passes the identification and selection requirements.

The contributions from the leading backgrounds ($\PW$+jets and $\ttbar$) are determined using simulated events normalized to the data in control regions. The signal-depleted region $70< M_{\mathrm{T}} <110\GeV$ is used to determine both the $\PW$+jets and the $\ttbar$~normalization factors using
two mutually exclusive selections.  Selecting events with fewer than four jets produces a sample enhanced with $\PW$+jets, and selecting events with at least four jets produces a sample enhanced with $\ttbar$~events. The results of these two selections are used to derive normalization factors from the following set of equations:
\begin{linenomath}
\begin{equation}
\begin{aligned}
& N_1 = R_{\ttbar} N_{1,\ttbar} + R_{\PW} N_{1,{\PW}} +  N_{1,O} \\
& N_2 = R_{\ttbar} N_{2,\ttbar} + R_{\PW} N_{2,{\PW}} +  N_{2,O}
\end{aligned}
\end{equation}
\end{linenomath}
where $N_{i}$, $N_{i,\ttbar}$, $N_{i,{\PW}}$, and $N_{i,O}$
are the number of events in data, $\PW$+jets, $\ttbar$, and other backgrounds passing selection $i$. The solution of the system yields the following normalization factors for the \munujj~channel: $R_{\ttbar} = 0.99 \pm 0.02$~(stat) and $R_{\PW} = 0.95 \pm 0.01$~(stat). Similar factors are obtained for the \enujj~channel: $R_{\ttbar} = 0.97 \pm 0.02$~(stat)~$\pm$~$0.01$~(syst) and $R_{\PW} = 0.85 \pm 0.01$~(stat)~$\pm$~$0.01$~(syst), where the systematic uncertainties are associated with the estimate of the multijet background in this particular channel. The value of $R_{\PW}$ in the \enujj~channel is affected by the lower efficiency of the trigger used in selecting $\PW$+jets events.

The multijet background in the \enujj~channel is determined from data, using the previously described probability that an electron candidate satisfying loosened requirements also passes the final electron requirements. The probability is used to weight a sample of events containing: exactly one electron candidate passing the loosened identification criteria, at least two jets, and large \MET.
The contribution from multijet processes is determined in the \munujj~channel using a sample of muon-enriched multijet simulated events with no muon isolation condition imposed. In the multijet-enriched region with $\MET < 10\GeV$, the muon-enriched multijet simulated events are reweighted to agree with data, and a muon isolation acceptance rate is calculated using the data as the number of events passing the isolation condition divided by the total number of events. After reweighting and an adjustment by the muon isolation acceptance factor, the nonisolated muon-enriched multijet simulated events are used to estimate the multijet contribution passing the final selection, which is determined in the \munujj~channel to be negligible.

The contribution from the remaining backgrounds (diboson, single top quark, and $\cPZ/\gamma^*$+jets) is small and is determined entirely from simulation.

\begin{figure*}[!htb]
       \centering
       {\includegraphics[width=.45\textwidth]{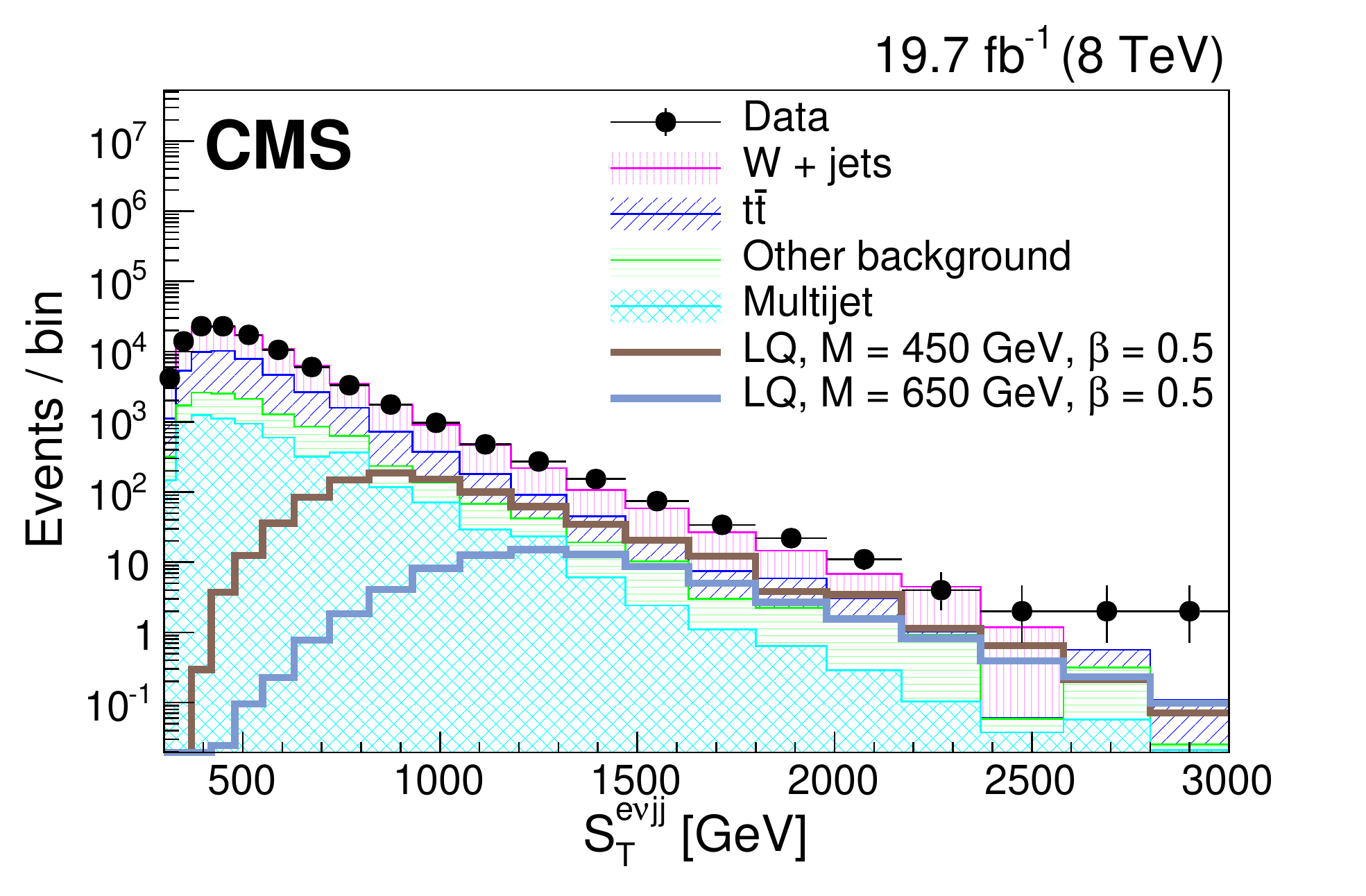}}
       {\includegraphics[width=.45\textwidth]{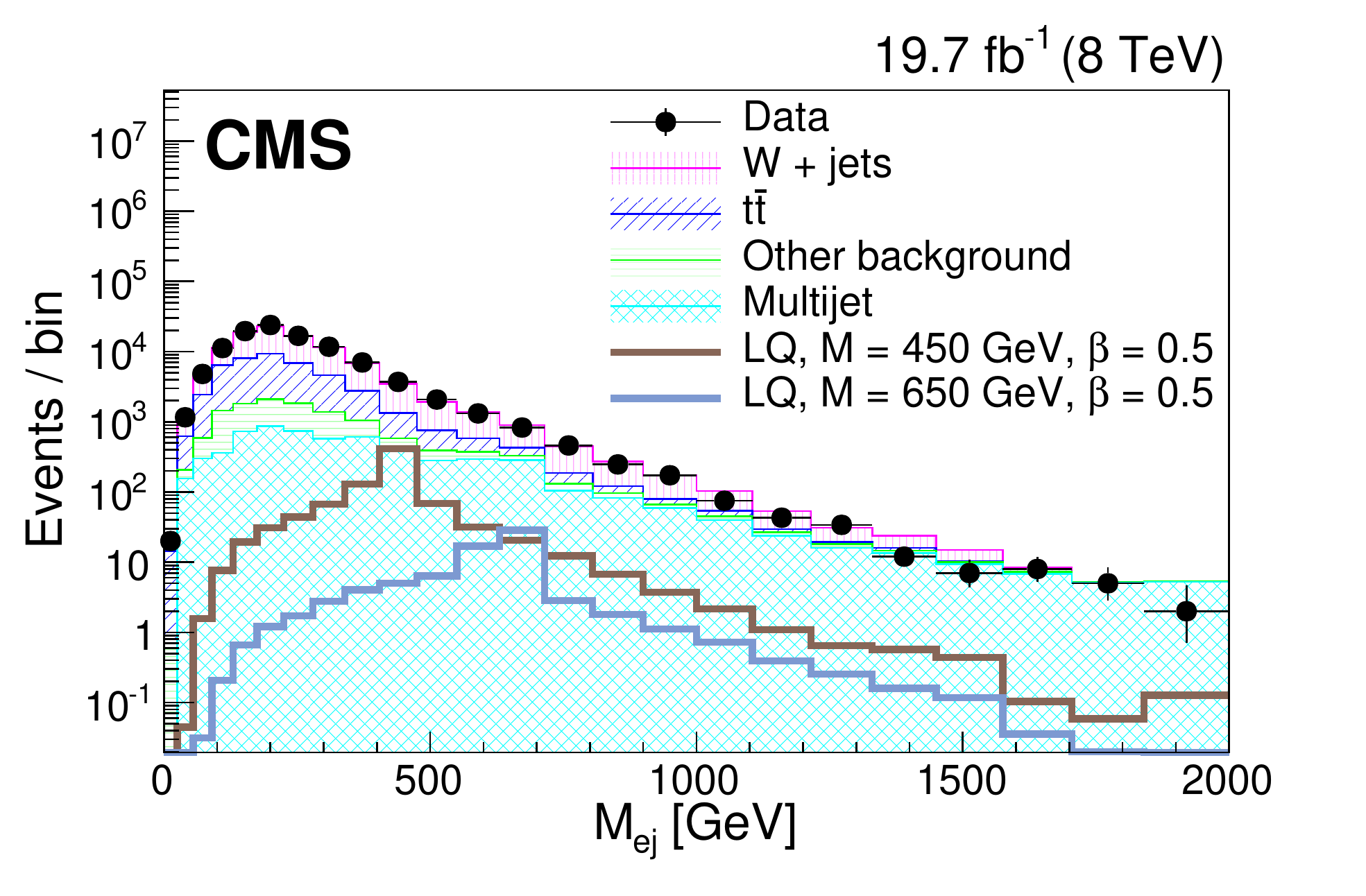}}
       {\includegraphics[width=.45\textwidth]{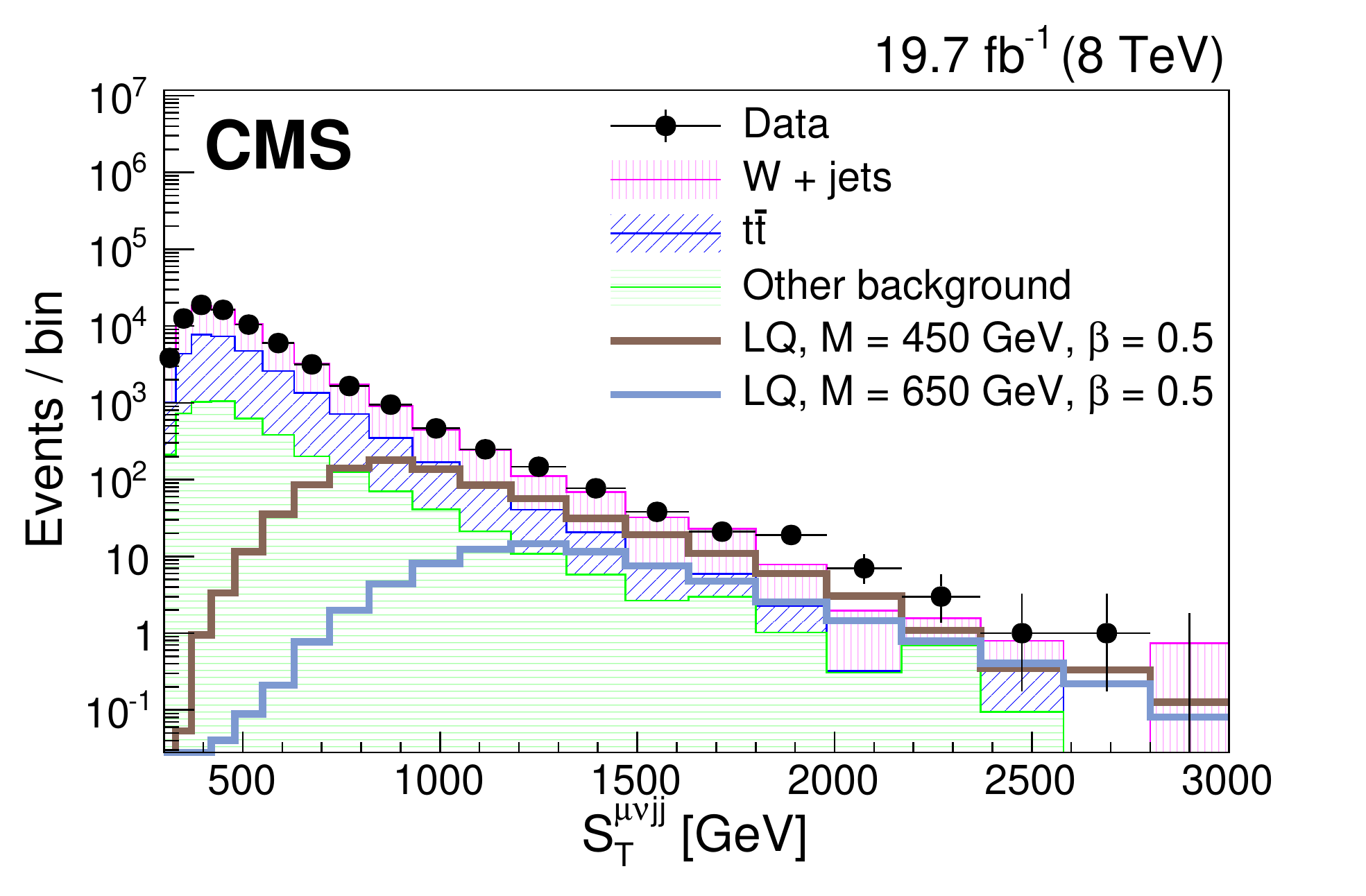}}
       {\includegraphics[width=.45\textwidth]{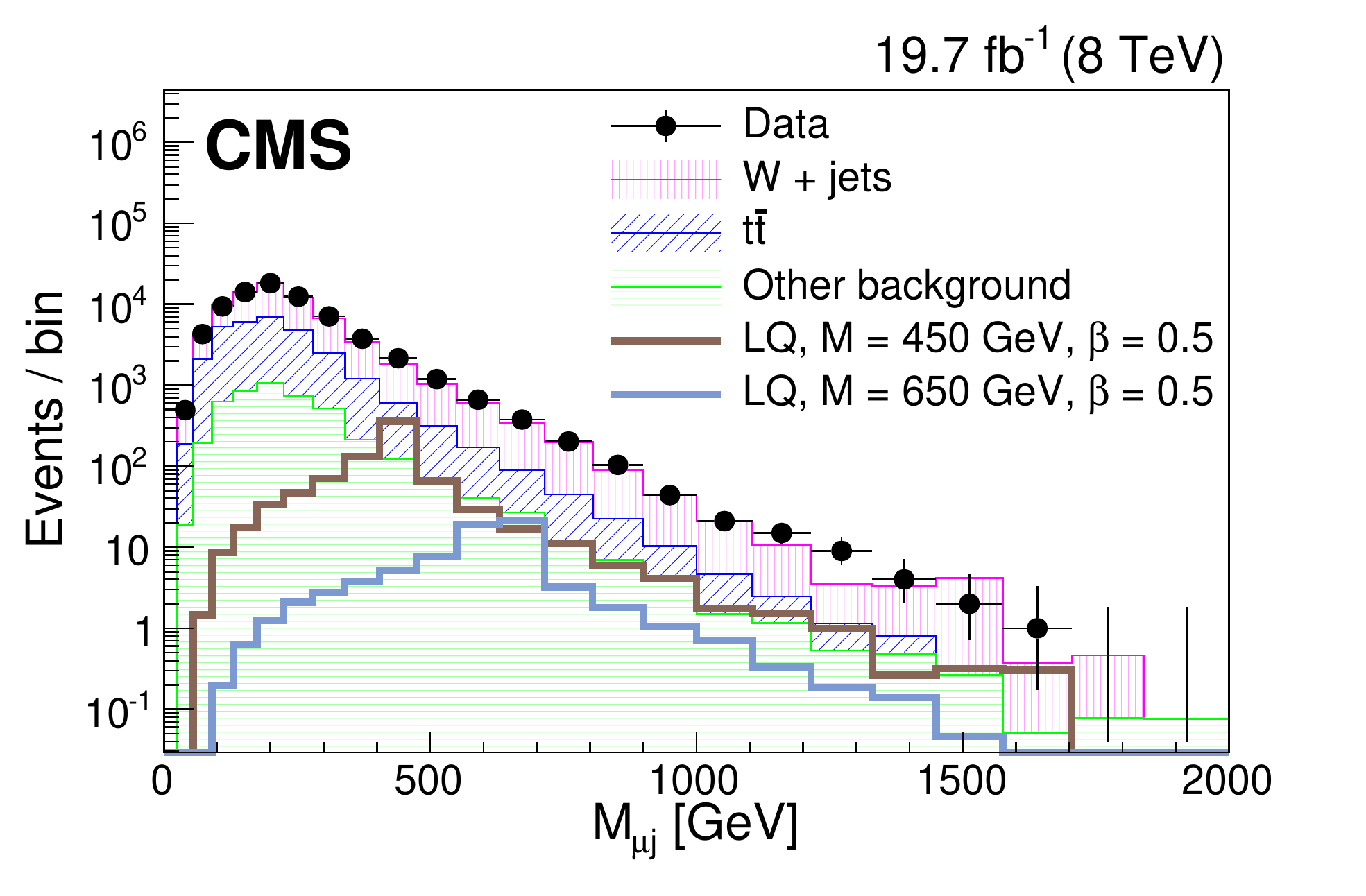}}
        \caption{Distributions of $S_{\mathrm{T}}$ (left), and $M_{\mathrm{\ell\cPj}}$ (right) at the initial selection level in the \enujj~(top) and \munujj~(bottom) channels. ``Other background" includes: diboson, $\cPZ/\gamma^*$+jets, and single top quark contributions.  The horizontal lines on the data points show the variable bin width.
	  \label{figapp:lq_enu_and_munu}}
\end{figure*}

As with the \eejj~and \mumujj~channels, all final state distributions in the \enujj~and \munujj~channels of the background prediction and of data, at the initial selection level, have been studied and also show agreement within uncertainties.  The specific distributions of $S_{\mathrm{T}}$ and $M^{\text{min}}_{\ell\cPj}$ for these channels are shown in Fig.~\ref{figapp:lq_enu_and_munu}.  Systematic uncertainties, discussed in the next section, are not included in these plots.

\section{Systematic uncertainties}
\label{systematics}

The sources of systematic uncertainty considered in this analysis are described below. To determine the final uncertainty in signal and background predictions, each quantity is individually varied within its uncertainty, and the entire analysis is repeated to find the change in the predicted number of background and signal events.

The dominant sources of systematic uncertainty are the jet energy scale and resolution uncertainties, which are estimated by assigning a $\PT$- and $\eta$-dependent uncertainty in jet energy corrections, as described in Ref.~\cite{JetCorr}, and by varying the jet $\PT$ by the uncertainty. The uncertainties in jet energy resolution are assessed by modifying the $\PT$ difference between the reconstructed and particle-level jets by an $\eta$-dependent value~\cite{JetCorr} ranging between 5\% and 30\% for most jets. Charged-lepton momentum scale and resolution uncertainties also introduce uncertainties in the overall event acceptance. A $\PT$-dependent muon momentum uncertainty of 5\%$\times$($\PT$), where $\PT$ is expressed in $\TeV$, and a $\PT$-dependent muon momentum resolution uncertainty ranging between 1 and 4\% are used~\cite{MuId}.
For electrons in the ECAL barrel and endcap region, an energy scale uncertainty of 2\%~\cite{EXO-12-061} and an electron energy resolution uncertainty of 10\%~\cite{Khachatryan:2015hwa} are used.
The effects of these uncertainties are assessed by modifying the electron momentum scale and resolution in the simulation according to these uncertainties.  A 2$\%$ per muon uncertainty in the muon reconstruction, identification, and isolation requirements, and a 1$\%$ per muon uncertainty in the muon HLT efficiency are assigned in the \mumujj~and \munujj~channels. An additional uncertainty is assigned for the \mumujj~and \munujj~channels because of the effect on the muon momentum determination of the uncertainty on the alignment of the muon system.  In simulation, a $\phi$ modulation can be seen in the difference between the inverse of the muon momentum as determined by the tracker with that determined by the tracker plus the muon system.  Corrections were derived, but produced minimal differences, so instead a small uncertainty is added to account for possible alignment effects.  In the \lnujj~analyses, the uncertainty in the charged lepton and jet energy and momentum scales and resolutions are propagated to the measurement of $\MET$.

Other important sources of systematic uncertainty are related to the modeling of the backgrounds in the simulation. The uncertainties in the $\cPZ/\gamma^*$+jets, $\PW$+jets, and $\ttbar$ background shapes are determined using simulated {\sc MadGraph} samples for which the renormalization and factorization scales and matrix element to parton shower matching thresholds have been varied up and down by a factor of two. The uncertainty of the scale factors for the normalization of the $\cPZ/\gamma^*$+jets background is determined to be $1\%$ in both \lljj~channels. A similar uncertainty for the normalization of the $\PW$+jets background is determined to be $2\%(1\%)$ in the \enujj~(\munujj)~channel. The scale factor for the normalization of the \ttbar background is determined to have an uncertainty of $2\%$ in the \enujj~and \munujj~channels. The scale factor for the normalization of the \emujj~sample used for the \ttbar~background estimate in the \eejj~channel is determined to have an $8\%$ uncertainty.

The estimate of the multijet background from data in the \eejj~(\enujj) channel has an uncertainty of $60\%$~$(30\%)$.  This uncertainty is assessed by probing the precision of the method used to measure this type of background on an independent data control sample.

An uncertainty in the modeling of pileup is determined by re-weighting the MC events to match with a number of pileup events 6$\%$~larger or smaller than what is observed in data, and an uncertainty of 2.6$\%$ is assigned to the value of the integrated luminosity~\cite{lumi_new}.

Lastly, the uncertainty in the signal acceptance, background acceptance, and cross section due to the PDF choice is estimated for signal (background) to be: 2\% (3\%) in the \eejj~channel; 3\% (3-25\%) in the \enujj~channel;  2\% (2-12\%) in the \mumujj~channel; and 2\%~(1-21\%) in the \munujj~channel, following the {\sc PDF4LHC} procedure~\cite{pdf4lhc1,pdf4lhc2}.

The systematic uncertainties for both signal and background are summarized in Table~\ref{tab:SysUncertainties_all} for all channels, corresponding to the final selection optimized for $M_{\mathrm{LQ}}=650\GeV$, which is representative of other high mass LQ values.

\begin{table*}[htbp]
\begin{center}
\topcaption{Systematic uncertainties (in \%) for signal ($S$) and background ($B$) in all channels for the $M_{\mathrm{LQ}}=650\GeV$ final selection.}
\footnotesize
\begin{scotch}{l........}
     Systematic & \multicolumn{2}{c}{\eejj} & \multicolumn{2}{c}{\mumujj} & \multicolumn{2}{c}{\enujj} & \multicolumn{2}{c}{\munujj}\\
     Uncertainties                     &\multicolumn{1}{c}{$S\,[\%]$} &\multicolumn{1}{c}{$B\,[\%]$} &\multicolumn{1}{c}{$S\,[\%]$}  &\multicolumn{1}{c}{$B\,[\%]$} &\multicolumn{1}{c}{$S~[\%]$}  &\multicolumn{1}{c}{$B\,[\%]$}   &\multicolumn{1}{c}{$S\,[\%]$}   &\multicolumn{1}{c}{$B\,[\%]$} \\ \hline
     Jet energy scale                  & 0.30   & 0.52   & 0.42    & 0.14   & 1.6     & 2.2      & 0.02     & 1.9    \\
     Electron energy scale             & 0.97   & 6.4    & \multicolumn{1}{c}{\NA}       & \multicolumn{1}{c}{\NA}      & 2.8     & 3.3      & \multicolumn{1}{c}{\NA}        & \multicolumn{1}{c}{\NA}     \\
     Electron Reco/ID/Iso              & 4.0    & {<}0.01  & \multicolumn{1}{c}{\NA}       & \multicolumn{1}{c}{\NA}      & 2.0     & {<}0.01    & \multicolumn{1}{c}{\NA}        & \multicolumn{1}{c}{\NA}     \\
     Muon momentum scale               & \multicolumn{1}{c}{\NA}      & \multicolumn{1}{c}{\NA}      & 0.63    & 1.7    & \multicolumn{1}{c}{\NA}       & \multicolumn{1}{c}{\NA}        & 0.19     & 13 \\
     Muon Reco/ID/Iso                  & \multicolumn{1}{c}{\NA}      & \multicolumn{1}{c}{\NA}      & 4       & 0.48   & \multicolumn{1}{c}{\NA}       & \multicolumn{1}{c}{\NA}        & 2.0       & 0.19     \\
     Jet resolution                    & 0.01   & 0.23   & 0.23    & 0.86   & 0.09    & 0.46     & 0.78     & 2.2    \\
     Electron resolution               & 0.46   & 0.22   & \multicolumn{1}{c}{\NA}       & \multicolumn{1}{c}{\NA}      & 0.61    & 0.53     & \multicolumn{1}{c}{\NA}        & \multicolumn{1}{c}{\NA}     \\
     Muon resolution                   & \multicolumn{1}{c}{\NA}      & \multicolumn{1}{c}{\NA}      & 0.14    & 0.39   & \multicolumn{1}{c}{\NA}       & \multicolumn{1}{c}{\NA}        & 0.15     & 7.1     \\
     Muon alignment                    & \multicolumn{1}{c}{\NA}      & \multicolumn{1}{c}{\NA}      & 0.1     & 0.54   & \multicolumn{1}{c}{\NA}       & \multicolumn{1}{c}{\NA}        & 1.0      & 2.8    \\
     Trigger                           & {<}0.01  & {<}0.01  & {<}0.01   & {<}0.01  & {<}0.01   & {<}0.01    & 1.0      & 0.10    \\
     \ttbar normalization              & \multicolumn{1}{c}{\NA}      & 2.1   & \multicolumn{1}{c}{\NA}       & 0.35   & \multicolumn{1}{c}{\NA}       & 1.5      & \multicolumn{1}{c}{\NA}        & 0.60     \\
     \ttbar shape                      & \multicolumn{1}{c}{\NA}      & \multicolumn{1}{c}{\NA}      & \multicolumn{1}{c}{\NA}       & \multicolumn{1}{c}{\NA}      & \multicolumn{1}{c}{\NA}       & 3.0      & \multicolumn{1}{c}{\NA}        & 1.4    \\
     $\PW$+jets normalization          & \multicolumn{1}{c}{\NA}      & {<}0.01  & \multicolumn{1}{c}{\NA}       & 0.01   & \multicolumn{1}{c}{\NA}       & 0.12     & \multicolumn{1}{c}{\NA}        & 0.63    \\
     $\PW$+jets shape                  & \multicolumn{1}{c}{\NA}      & {<}0.01  & \multicolumn{1}{c}{\NA}       & 0.23   & \multicolumn{1}{c}{\NA}       & 0.87     & \multicolumn{1}{c}{\NA}        & 13    \\
     $\cPZ/\gamma^*$+jets normalization & \multicolumn{1}{c}{\NA}     & 0.75   & \multicolumn{1}{c}{\NA}       & 0.59   & \multicolumn{1}{c}{\NA}       & {<}0.01      & \multicolumn{1}{c}{\NA}        & 0.07    \\
     $\cPZ/\gamma^*$+jets shape         & \multicolumn{1}{c}{\NA}     & 12     & \multicolumn{1}{c}{\NA}       & 12     & \multicolumn{1}{c}{\NA}       & {<}0.01      & \multicolumn{1}{c}{\NA}        & 1.5     \\
     Multijet modeling                 & \multicolumn{1}{c}{\NA}      & 0.10   & \multicolumn{1}{c}{\NA}       & \multicolumn{1}{c}{\NA}      & \multicolumn{1}{c}{\NA}       & 5        & \multicolumn{1}{c}{\NA}        & \multicolumn{1}{c}{\NA}      \\
     PDF                               & 2.0    & 2.1    & 2.0     & 4.8    & 3.0     & 13       & 3.0      & 5.1    \\
     Pileup                            & 0.04   & 0.38   & 0.16    & 0.22   & 0.14    & 1.2      & 0.06     & 1.3    \\
     Integrated luminosity             & 2.6    & 0.10   & 2.6     & 0.31   & 2.6     & 0.47     & 2.6      & 0.25     \\ \hline
     Total                             & 5.3    & 14     & 5.2     & 13     & 5.5     & 15       & 4.7      & 21 \\
\end{scotch}
\label{tab:SysUncertainties_all}
\end{center}
\end{table*}

\section{Results}
\label{results}

Data and background predictions are compared for every channel and each mass optimization point, after the optimized final selection criteria are applied to both signal and background. The first part of this section details such comparisons. There are no significant deviations from SM background predictions. Limits are set on the cross section times branching fraction for the hypothesis of scalar LQ pair production as a function of $M_{\mathrm{LQ}}$ and $\beta$. The expected and observed limits for scalar LQ pair production are detailed in the second part of this section. Additional interpretations of the results in the context of vector LQ pair production and of RPV SUSY production with $\lljj$ and $\lnujj$ signatures are described in the last part of this section.

\subsection{Data and background comparison}

\begin{figure*}[!htbp]
       \centering
       {\includegraphics[width=.45\textwidth]{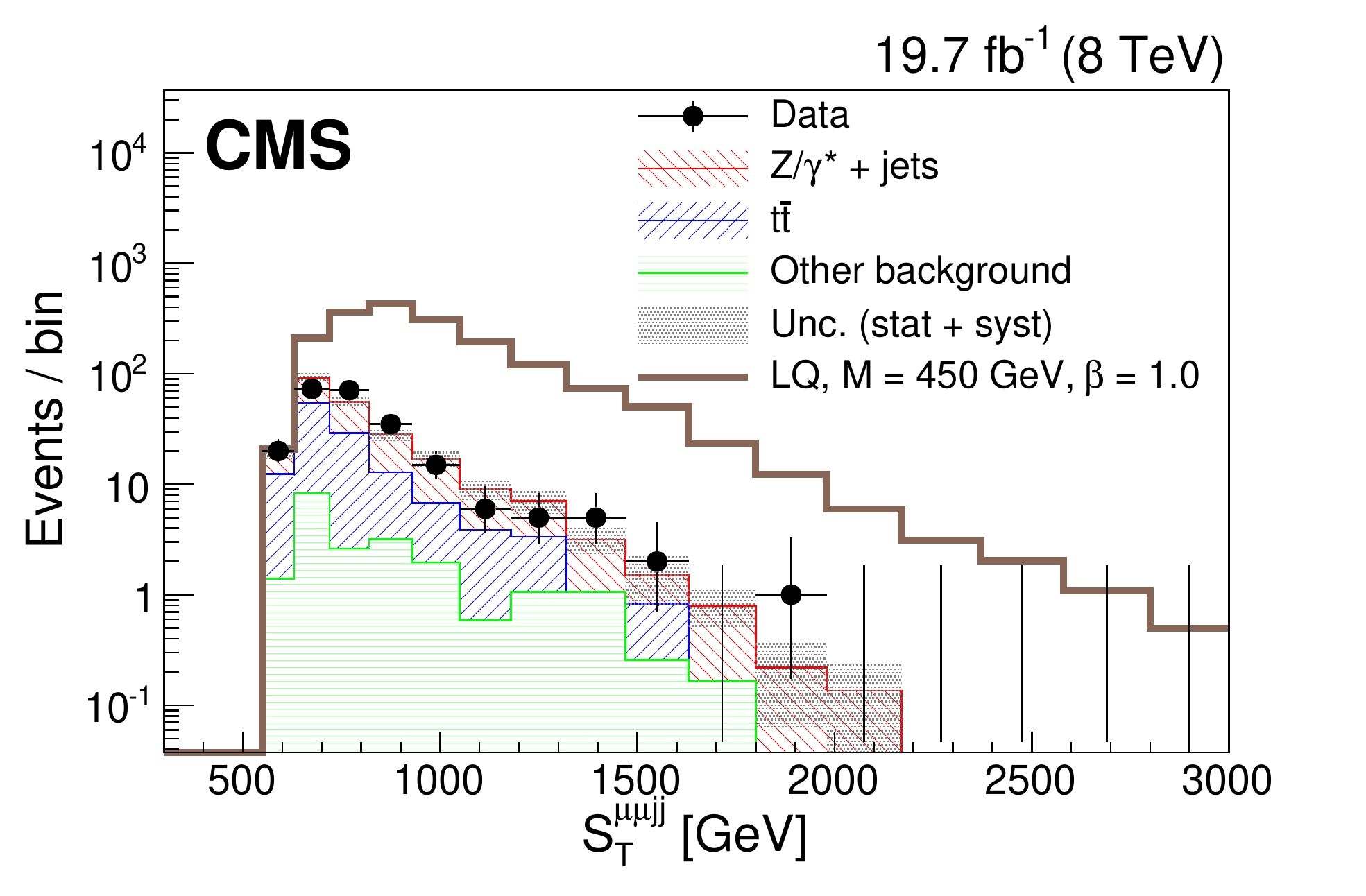}}
       {\includegraphics[width=.45\textwidth]{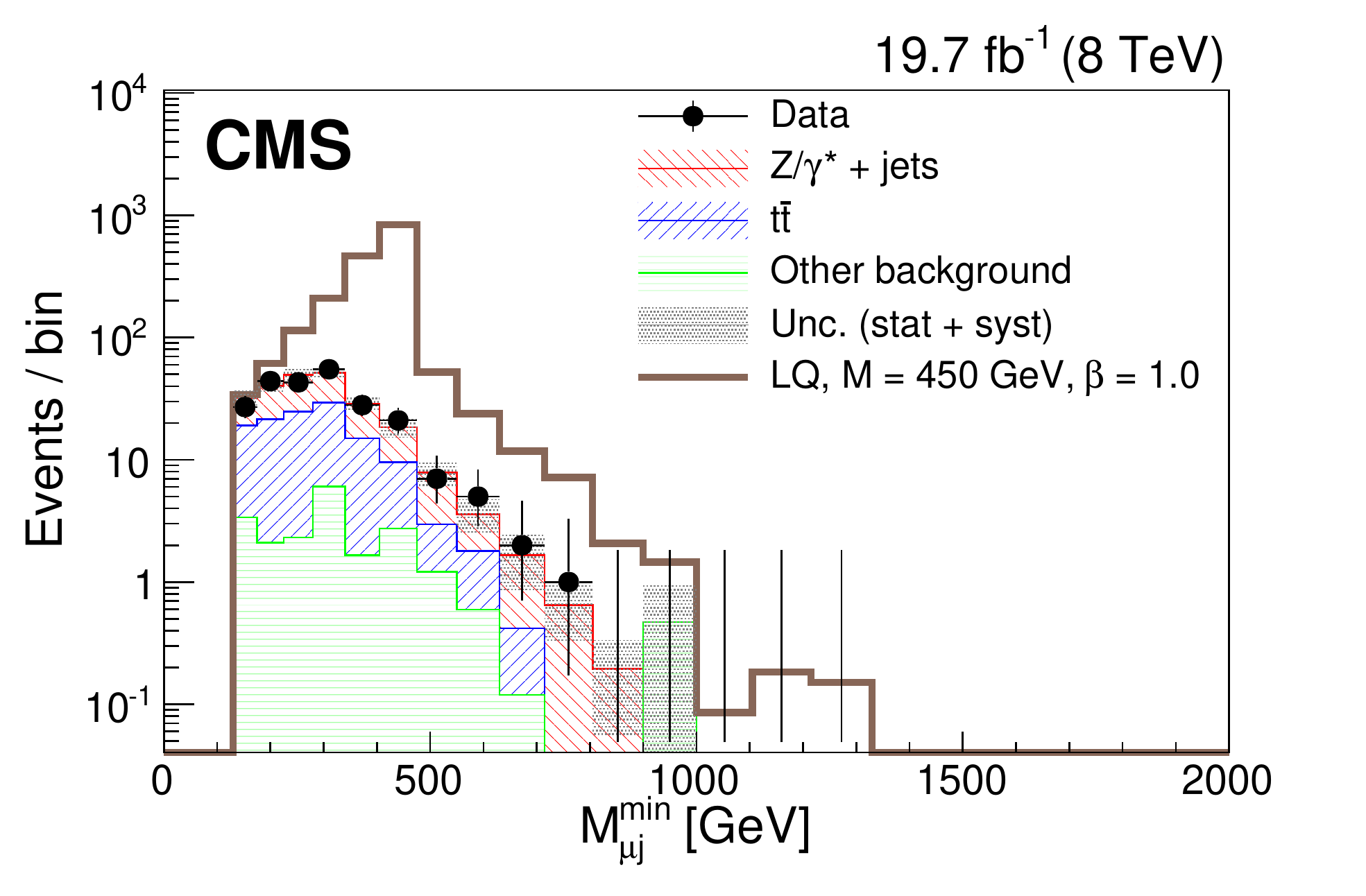}}
       {\includegraphics[width=.45\textwidth]{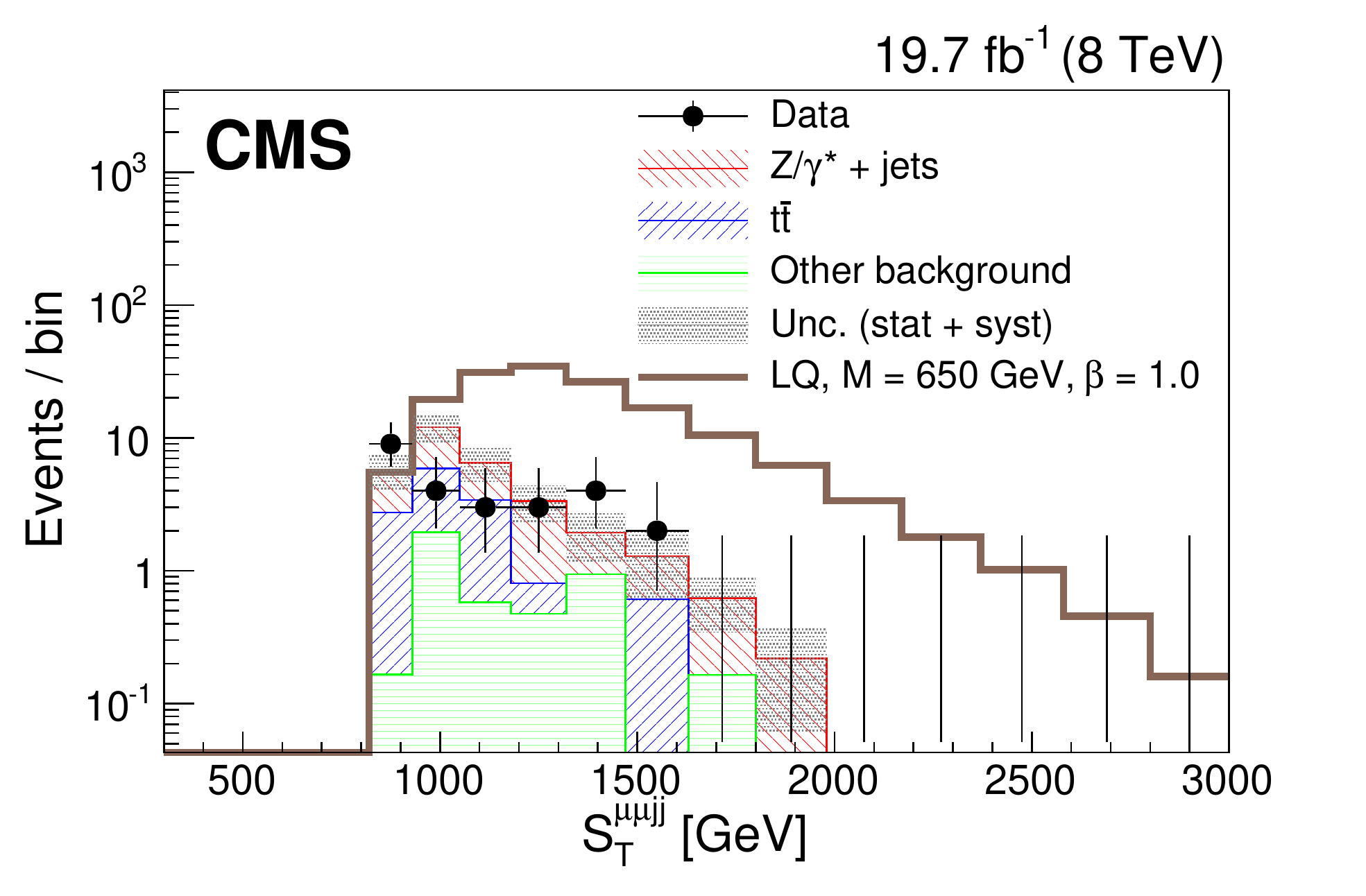}}
       {\includegraphics[width=.45\textwidth]{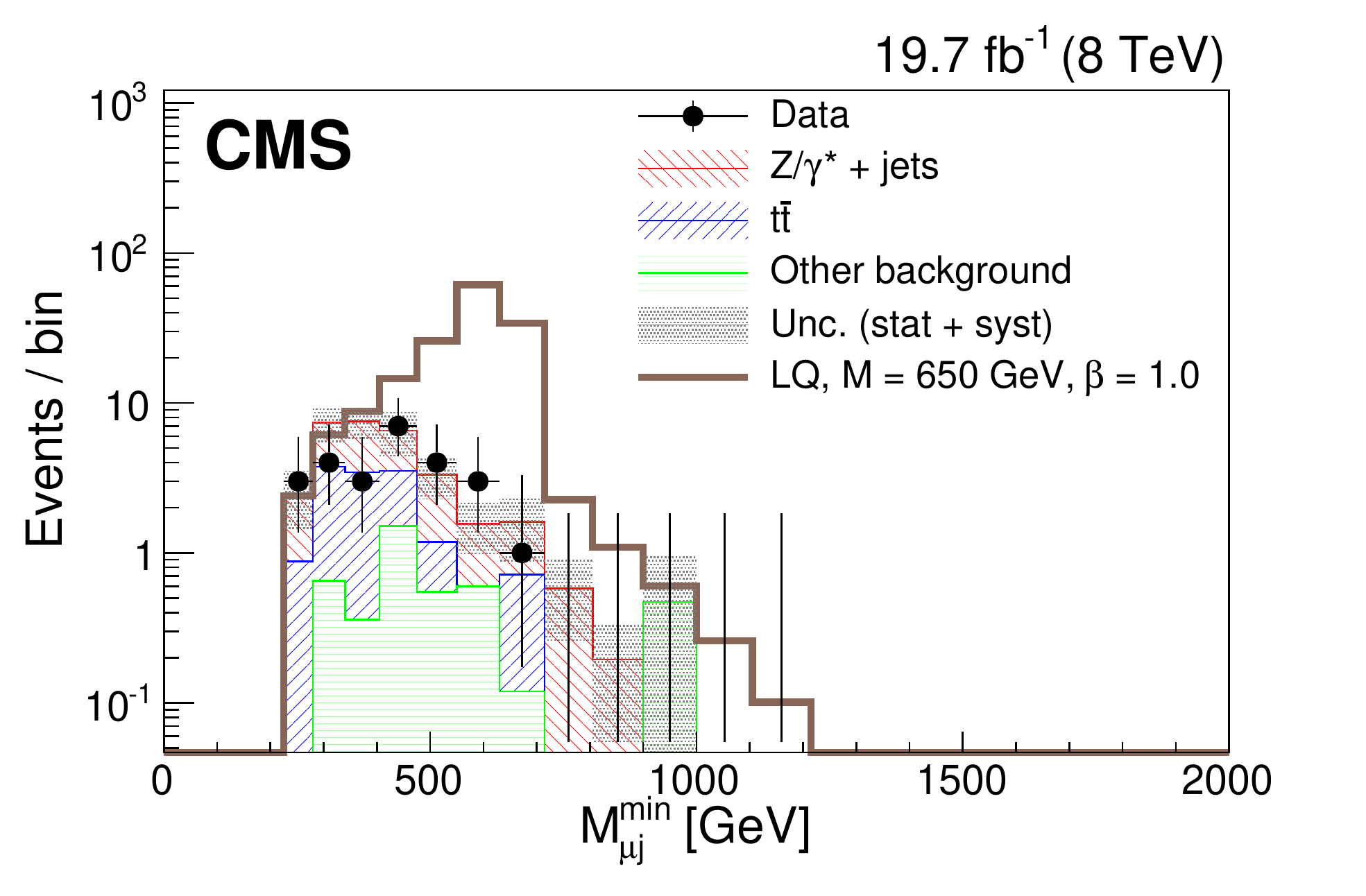}}
       \caption{Distributions of $S_{\mathrm{T}}$ (left) and  $M^{\mathrm{min}}_{\Pgm\cPj}$ (right) for the final selection for a LQ mass of 450\GeV (top) and 650\GeV (bottom) in the \mumujj~channel. The dark shaded region indicates the statistical and systematic uncertainty in the total background prediction. ``Other background" includes diboson, $\PW$+jets, and single top quark contributions.  The horizontal lines on the data points show the variable bin width.
	  \label{figapp:stmlq450_and_650_mumu}}
\end{figure*}

\begin{figure*}[!htbp]
       \centering
       {\includegraphics[width=.45\textwidth]{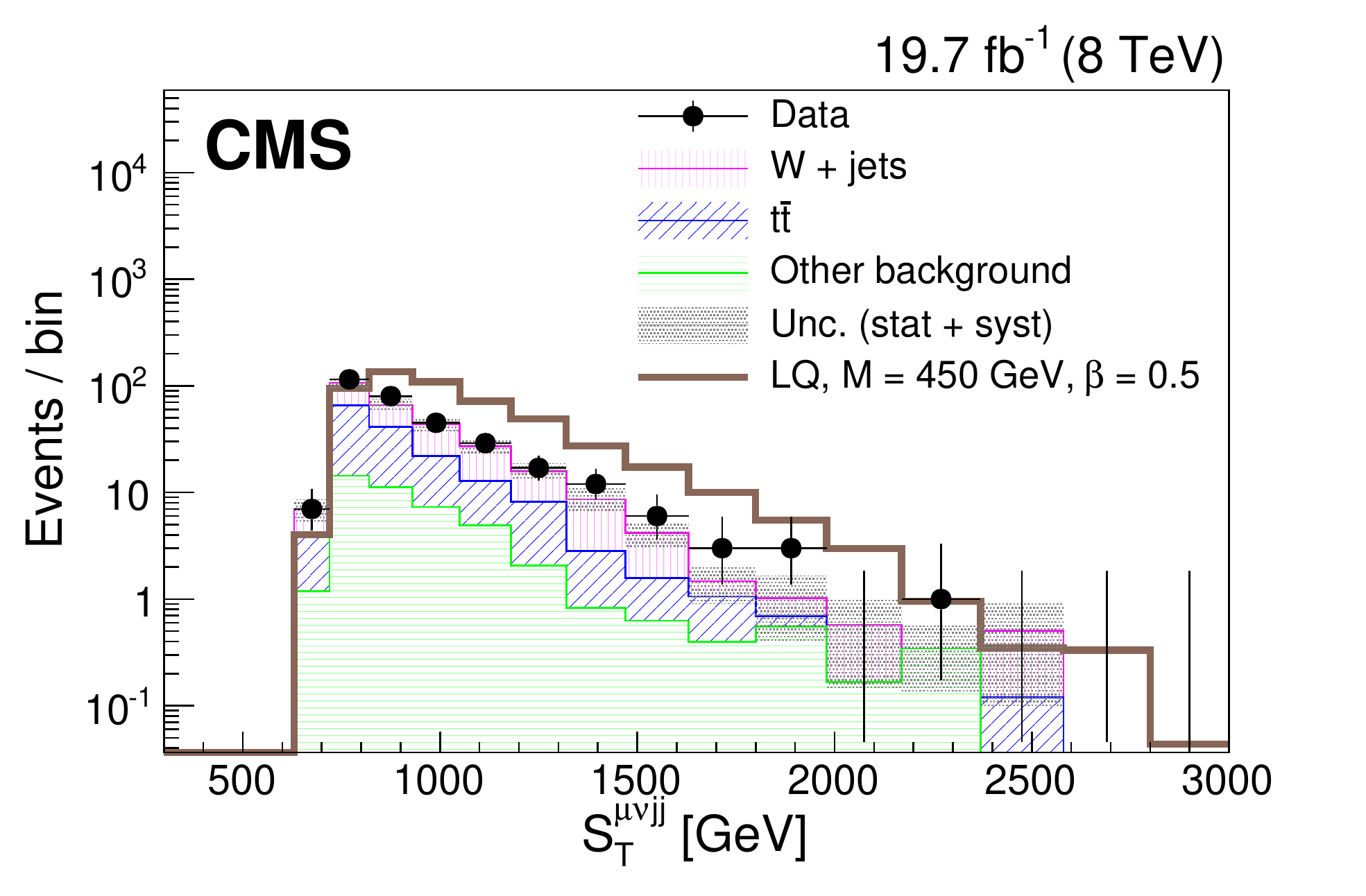}}
       {\includegraphics[width=.45\textwidth]{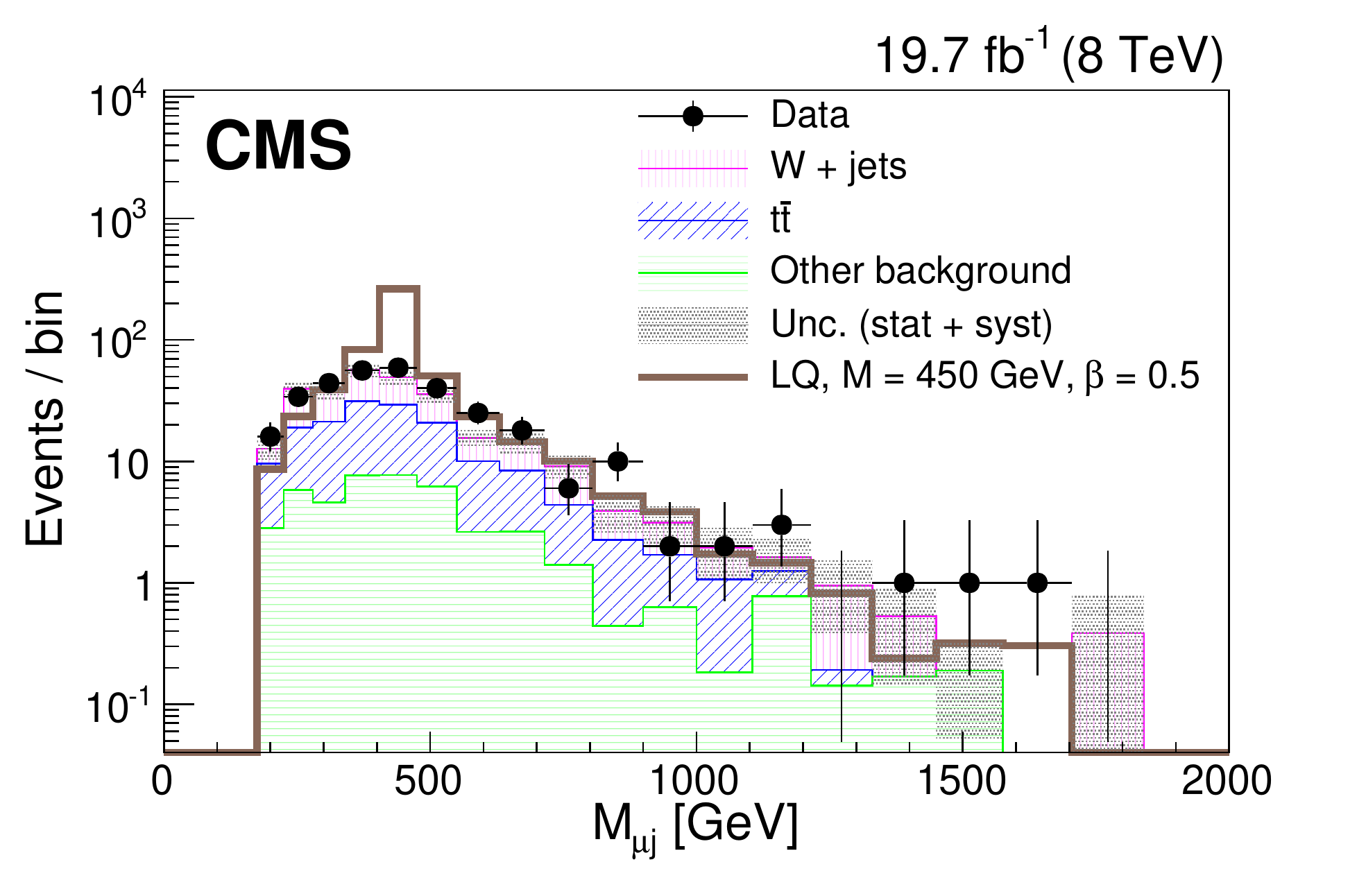}}
       {\includegraphics[width=.45\textwidth]{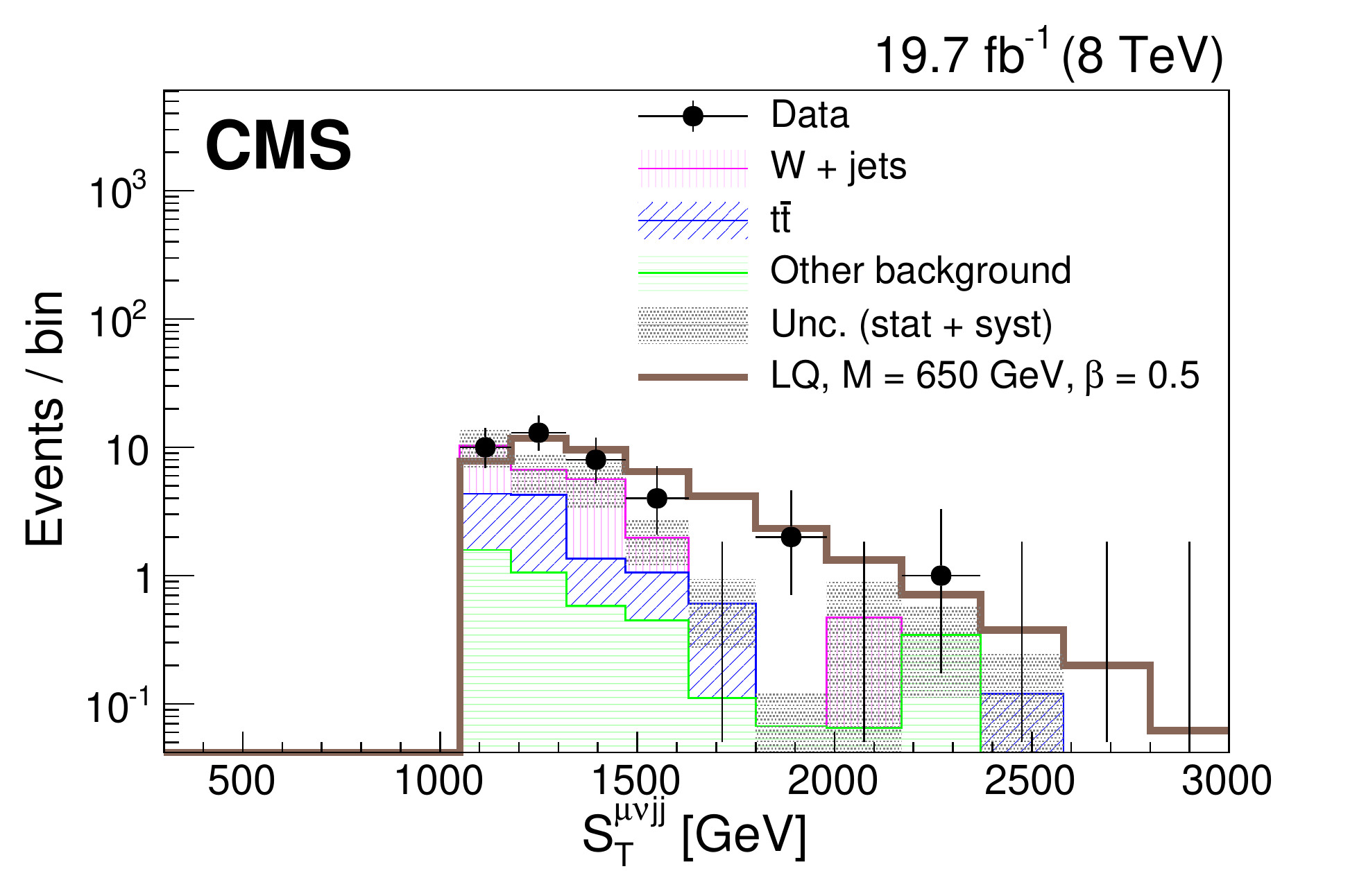}}
       {\includegraphics[width=.45\textwidth]{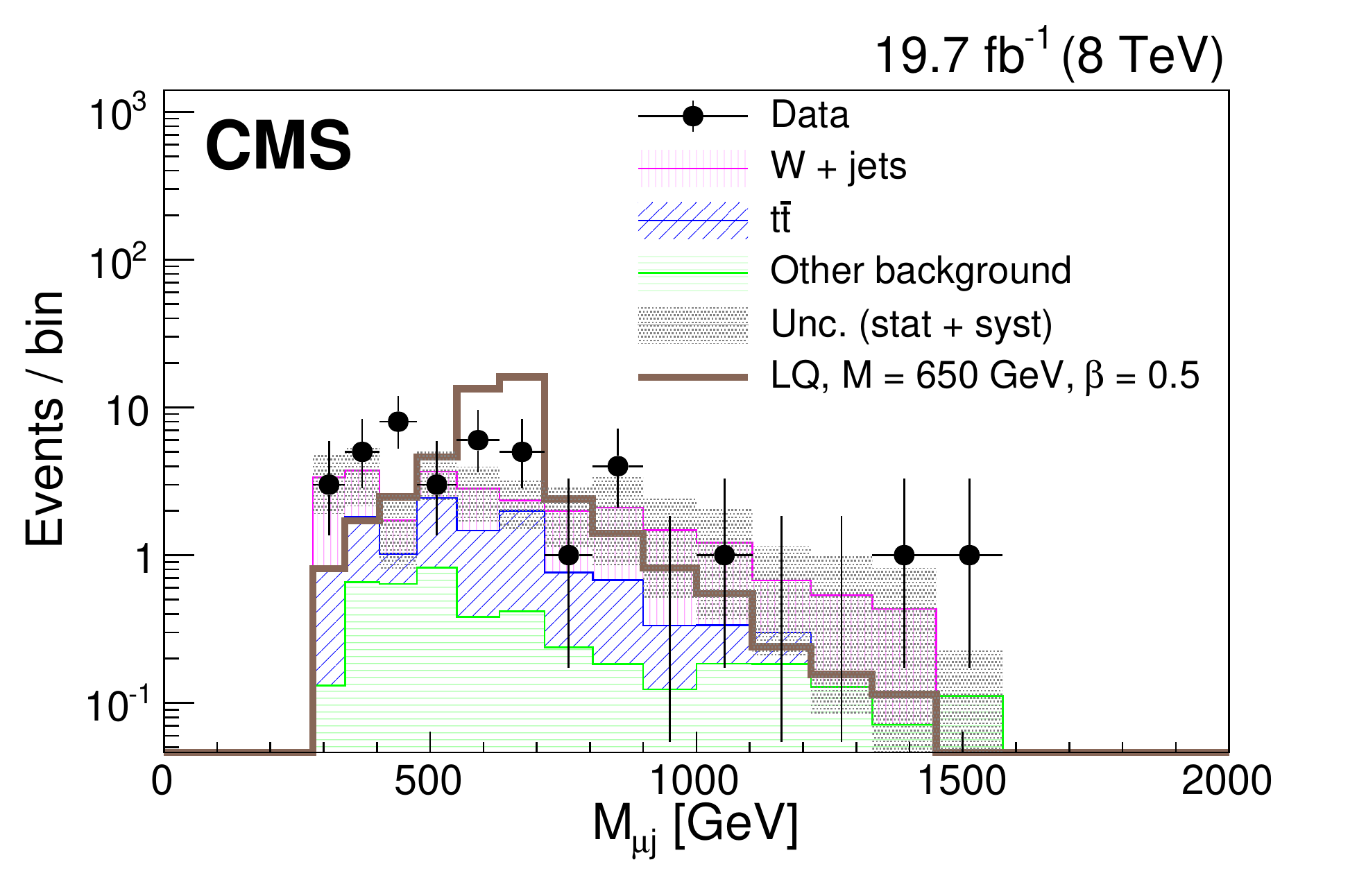}}
       \caption{Distributions of $S_{\mathrm{T}}$ (left) and $ M_{\Pgm\cPj}$ (right) for the final selection for a LQ mass of 450\GeV (top) and 650\GeV (bottom) in the \munujj~channel. The dark shaded region indicates the statistical and systematic uncertainty in the total background prediction. ``Other background" includes diboson, $\cPZ/\gamma^*$+jets, and single top quark contributions. The horizontal lines on the data points show the variable bin width.
	  \label{figapp:stmlq450_and_650_munu}}
\end{figure*}

Agreement is found between data and background predictions in both the \mumujj~and \munujj~channels, as shown in Fig.~\ref{figapp:stmlq450_and_650_mumu} for the \mumujj~channel, which displays $S_{\mathrm{T}}$ and $ M^{\mathrm{min}}_{\Pgm\cPj}$ for signal LQ masses of 450 and 650\GeV, and in Fig.~\ref{figapp:stmlq450_and_650_munu} for the \munujj~channel, which displays $S_{\mathrm{T}}$ and $M_{\Pgm\cPj}$ for the same signal LQ mass points.

The numbers of events selected in data, and the various backgrounds at final selection as a function of $M_{\mathrm{LQ}}$ for the \mumujj~and \munujj~channels are summarized in Tables~\ref{tab:finalselection_mumu} and~\ref{tab:finalselection_munu}, respectively.  Since mass hypotheses at 1\TeV and beyond share the same final selections, they also share the same background yields.

\begin{table*}[htb]
\topcaption{Event yields for the \mumujj~analysis for $\beta=1.0$ for all values of $M_{\mathrm{LQ}}$. Uncertainties are Poisson uncertainties in the simulated background, except for the second uncertainty for ``All background", which gives the total systematic uncertainty as detailed in Section~\ref{systematics}.  Systematic uncertainties are dominated by jet energy scale and simulation shape uncertainties.}
\begin{center}
    \resizebox{\textwidth}{!}{
\begin{scotch}{cx{10}x{8}x{8}x{8}x{12}c}
$M_{\mathrm{LQ}}\unit{[\GeVns{}]}$ &   \multicolumn{1}{c}{Signal} &                \multicolumn{1}{c}{$\cPZ/\gamma^*$+jets} &              \multicolumn{1}{c}{$\ttbar$} &                 \multicolumn{1}{c}{\PVV, \PW, single t} &                         \multicolumn{1}{c}{All background} &                                           Data \\
\hline
300 &        16240,\pm\,110  &      819.0,\pm\,9.2  &      666,\pm\,19  &           88,\pm\,5.4  &                     1573,\pm\, 22 \,\pm\,56  &                1659 \\
350 &        7570,\pm\,48  &        351.7,\pm\,6.0  &    405,\pm\,15  &         58.3,\pm\,4.5  &                   815,\pm\, 17 \,\pm\,21  &                 797 \\
400 &        3658,\pm\,22  &        200.4,\pm\,4.5  &    202,\pm\,11  &         31.5,\pm\,3.3  &                   434,\pm\, 12 \,\pm\,17  &                 439 \\
450 &        1816,\pm\,11  &        110.2,\pm\,3.3  &    103.7,\pm\,27.5  &          20.6,\pm\,2.7  &                   234.6,\pm\, 8.6 \,\pm\,11.2  &            233 \\
500 &        938.1,\pm\,5.5  &      69.9,\pm\,2.6  &     61.0,\pm\,5.7  &             13.2,\pm\,2.2  &                   144.2,\pm\, 6.7 \,\pm\,8.5  &             135 \\
550 &        498.8,\pm\,2.9  &      39.6,\pm\,1.9  &     29.4,\pm\,3.9  &          8.0,\pm\,1.8  &                    77,\pm\, 4.7 \,\pm\,5.2  &                84 \\
600 &        274.7,\pm\,1.6  &      25.8,\pm\,1.5  &     14.8,\pm\,2.8  &         6.5,\pm\,1.6  &                  47.1,\pm\, 3.6 \,\pm\,4.5  &              47 \\
650 &        157.1,\pm\,0.9  &     17.1,\pm\,1.2  &     10.3,\pm\,2.3  &           4.3,\,_{-1.3}^{+1.4}  &        31.7,\,_{-2.9}^{+2.9}\,\pm\,4.2  &     25 \\
700 &        89.49,\pm\,0.52  &     10.71,\pm\,0.98  &   7.0,\pm\,2.0  &          2.9,\,_{-1.0}^{+1.2}   &           20.6,\,_{-2.4}^{+2.5}\,\pm\,4.3  &     15 \\
750 &        52.39,\pm\,0.30  &      6.95,\pm\,0.79  &    2.20,\pm\,0.98  &           1.1,\,_{-0.56}^{+0.8}   &         10.3,\,_{-1.4}^{+1.5}\,\pm\,2.7  &     11 \\
800 &        31.3,\pm\,0.18  &      3.90,\pm\,0.59  &     1.08,\pm\,0.62  &        0.77,\,_{-0.46}^{+0.73}   &        5.8,\,_{-1.0}^{+1.1}\,\pm\,1.55  &  9 \\
850 &        18.99,\pm\,0.11  &     1.96,\pm\,0.39  &    0.0,\,_{-0.0}^{+0.65}   &  0.75,\,_{-0.46}^{+0.73}   &        2.71,\,_{-0.65}^{+0.97}\,\pm\,1.07  &  5 \\
900 &        11.290,\pm\,0.067  &    1.10,\pm\,0.29  &     0.0,\,_{-0.0}^{+0.65}   &  0.30,\,_{-0.20}^{+0.60}   &          1.38,\,_{-0.40}^{+0.82}\,\pm\,0.44  &  3 \\
950 &        6.907,\pm\,0.041  &    0.76,\pm\,0.25  &    0.0,\,_{-0.0}^{+0.65}   &  0.12,\,_{-0.12}^{+0.78}   &        0.87,\,_{-0.32}^{+0.85}\,\pm\,0.49  &  1 \\
1000 &       4.175,\pm\,0.026  &    0.41,\pm\,0.18  &    0.0,\,_{-0.0}^{+0.65}   &  0.0,\,_{-0.0}^{+0.77}   &          0.41,\,_{-0.18}^{+0.91}\,\pm\,0.27  &  0 \\
1050 &       2.778,\pm\,0.017  &    0.41,\pm\,0.18  &    0.0,\,_{-0.0}^{+0.65}   &  0.0,\,_{-0.0}^{+0.77}   &          0.41,\,_{-0.18}^{+0.91}\,\pm\,0.27  &  0 \\
1100 &       1.860,\pm\,0.011  &     0.41,\pm\,0.18  &    0.0,\,_{-0.0}^{+0.65}   &  0.0,\,_{-0.0}^{+0.77}   &          0.41,\,_{-0.18}^{+0.91}\,\pm\,0.27  &  0 \\
1150 &       1.2471,\pm\,0.0072  &  0.41,\pm\,0.18  &    0.0,\,_{-0.0}^{+0.65}   &  0.0,\,_{-0.0}^{+0.77}   &          0.41,\,_{-0.18}^{+0.91}\,\pm\,0.27  &  0 \\
1200 &       0.8202,\pm\,0.0047  &  0.41,\pm\,0.18  &    0.0,\,_{-0.0}^{+0.65}   &  0.0,\,_{-0.0}^{+0.77}   &          0.41,\,_{-0.18}^{+0.91}\,\pm\,0.27  &  0 \\
\end{scotch}
\label{tab:finalselection_mumu}
}
\end{center}
\end{table*}

\begin{table*}[htb]
\topcaption{Event yields for the \munujj~analysis for $\beta=0.5$ for all values of $M_{\mathrm{LQ}}$. Uncertainties are Poisson uncertainties in the simulated background, except for the second uncertainty for ``All background", which gives the total systematic uncertainty as detailed in Section~\ref{systematics}. Systematic uncertainties are dominated by the jet energy scale and simulation shape uncertainties.}
\begin{center}
    \resizebox{\textwidth}{!}{
\begin{scotch}{cx{10}x{8}x{8}x{8}x{12}c}
$M_{\mathrm{LQ}}\unit{[\GeVns{}]}$ &   \multicolumn{1}{c}{Signal} &                \multicolumn{1}{c}{$\PW$+jets} &              \multicolumn{1}{c}{$\ttbar$} &                  \multicolumn{1}{c}{\PVV, \cPZ, single t} &                         \multicolumn{1}{c}{All background} &                                           Data \\
\hline
300 &        5089,\pm\,58  &        1102,\pm\,22  &      1853,\pm\,15  &            331.3,\pm\,8.3  &                  3286,\pm\,28 \,\pm\,185  &               3549 \\
350 &        2352,\pm\,25  &        472,\pm\,14  &       640.0,\pm\,8.5  &            159.8,\pm\,5.7  &                  1272,\pm\,18 \,\pm\,70  &                1451 \\
400 &        1064,\pm\,11  &        213.9,\pm\,9.6  &    259.5,\pm\,5.4  &          84.6,\pm\,4.3  &                   558,\pm\,12 \,\pm\,38  &                 668 \\
450 &        526.7,\pm\,5.5  &      115.7,\pm\,7.1  &    116.3,\pm\,3.6  &          44.8,\pm\,2.9  &                   276.7,\pm\,8.5 \,\pm\,22  &              313 \\
500 &        263.6,\pm\,2.8  &      66.4,\pm\,5.3  &     56.1,\pm\,2.5  &           25.1,\pm\,2.1  &                   147.6,\pm\,6.2 \,\pm\,12.6  &            173 \\
550 &        142.7,\pm\,1.5  &      43.8,\pm\,4.4  &     26.1,\pm\,1.7  &           14.3,\pm\,1.6  &                   84.3,\pm\,5.0 \,\pm\,9.4  &              93 \\
600 &        78.1,\pm\,0.8  &      20.3,\pm\,2.7  &     13.7,\pm\,1.2  &           8.0,\pm\,1.1  &                    42,\pm\,3.2 \,\pm\,5.8  &                57 \\
650 &        44.62,\pm\,0.46  &     14.0,\pm\,2.3  &       7.97,\pm\,0.95  &          4.34,\pm\,0.72  &                  26.3,\pm\,2.6 \,\pm\,5.2  &              36 \\
700 &        25.27,\pm\,0.26  &     9.1,\pm\,1.8  &      5.20,\pm\,0.76  &           2.73,\,_{-0.46}^{+0.64}   &        17,\pm\,2.0 \,\pm\,4.7  &                25 \\
750 &        15.04,\pm\,0.15  &     7.0,\pm\,1.6  &      2.82,\pm\,0.56  &          1.93,\,_{-0.40}^{+0.60}   &          11.7,\pm\,1.8 \,\pm\,5.1  &              15 \\
800 &        9.080,\pm\,0.093  &     4.5,\pm\,1.4  &      1.47,\pm\,0.41  &          1.61,\,_{-0.37}^{+0.58}   &        7.6,\pm\,1.5  \,\pm\,3.5  &      11 \\
850 &        5.493,\pm\,0.056  &    1.08,\pm\,0.54  &    1.04,\pm\,0.35  &          1.16,\,_{-0.32}^{+0.55}   &        3.28,\,_{-0.74}^{+0.81}\,\pm\,1.04  &  7 \\
900 &        3.370,\pm\,0.035  &     0.62,\pm\,0.44  &    0.92,\pm\,0.32  &          0.9,\,_{-0.29}^{+0.53}   &         2.44,\,_{-0.64}^{+0.72}\,\pm\,0.89  &  3 \\
950 &        2.111,\pm\,0.022  &    0.4,\pm\,0.4  &      0.44,\pm\,0.22  &          0.51,\,_{-0.21}^{+0.49}   &        1.35,\,_{-0.52}^{+0.62}\,\pm\,0.6  &    3 \\
1000 &       1.322,\pm\,0.014  &    0.4,\pm\,0.4  &      0.26,\pm\,0.18  &         0.51,\,_{-0.21}^{+0.49}   &        1.17,\,_{-0.51}^{+0.61}\,\pm\,0.56  &  3 \\
1050 &       0.9338,\pm\,0.0092  &  0.4,\pm\,0.4  &      0.26,\pm\,0.18  &         0.51,\,_{-0.21}^{+0.49}   &        1.17,\,_{-0.51}^{+0.61}\,\pm\,0.56  &  3 \\
1100 &       0.6507,\pm\,0.0062  &  0.4,\pm\,0.4  &      0.26,\pm\,0.18  &         0.51,\,_{-0.21}^{+0.49}   &        1.17,\,_{-0.51}^{+0.61}\,\pm\,0.56  &  3 \\
1150 &       0.4457,\pm\,0.0041  &  0.4,\pm\,0.4  &      0.26,\pm\,0.18  &         0.51,\,_{-0.21}^{+0.49}   &        1.17,\,_{-0.51}^{+0.61}\,\pm\,0.56  &  3 \\
1200 &       0.3097,\pm\,0.0028  &  0.4,\pm\,0.4  &      0.26,\pm\,0.18  &         0.51,\,_{-0.21}^{+0.49}   &        1.17,\,_{-0.51}^{+0.61}\,\pm\,0.56  &  3 \\
\end{scotch}
\label{tab:finalselection_munu}
}
\end{center}
\end{table*}

\begin{figure*}[!htbp]
  \begin{center}
    {\includegraphics[width=.45\textwidth]{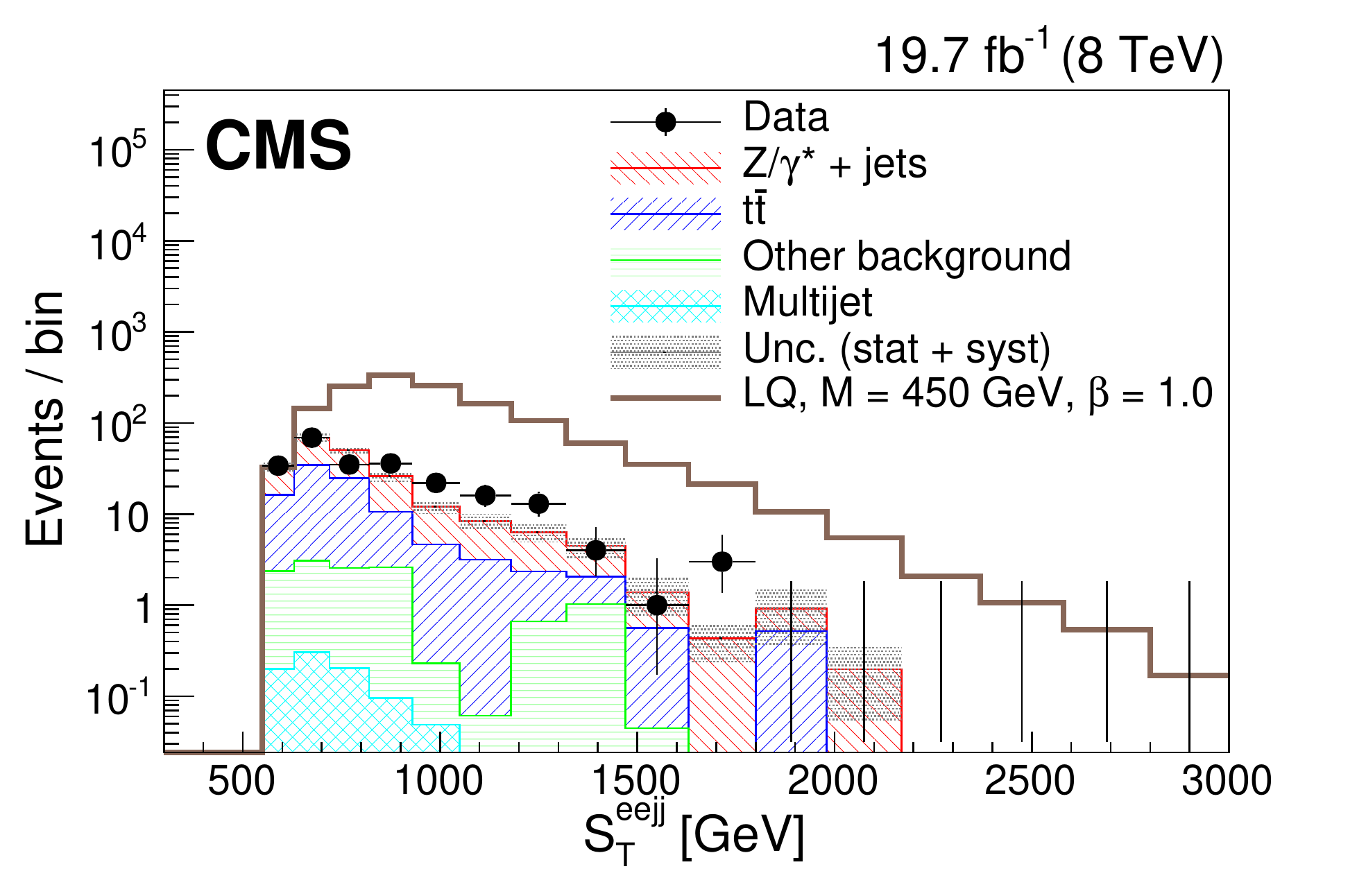}}
    {\includegraphics[width=.45\textwidth]{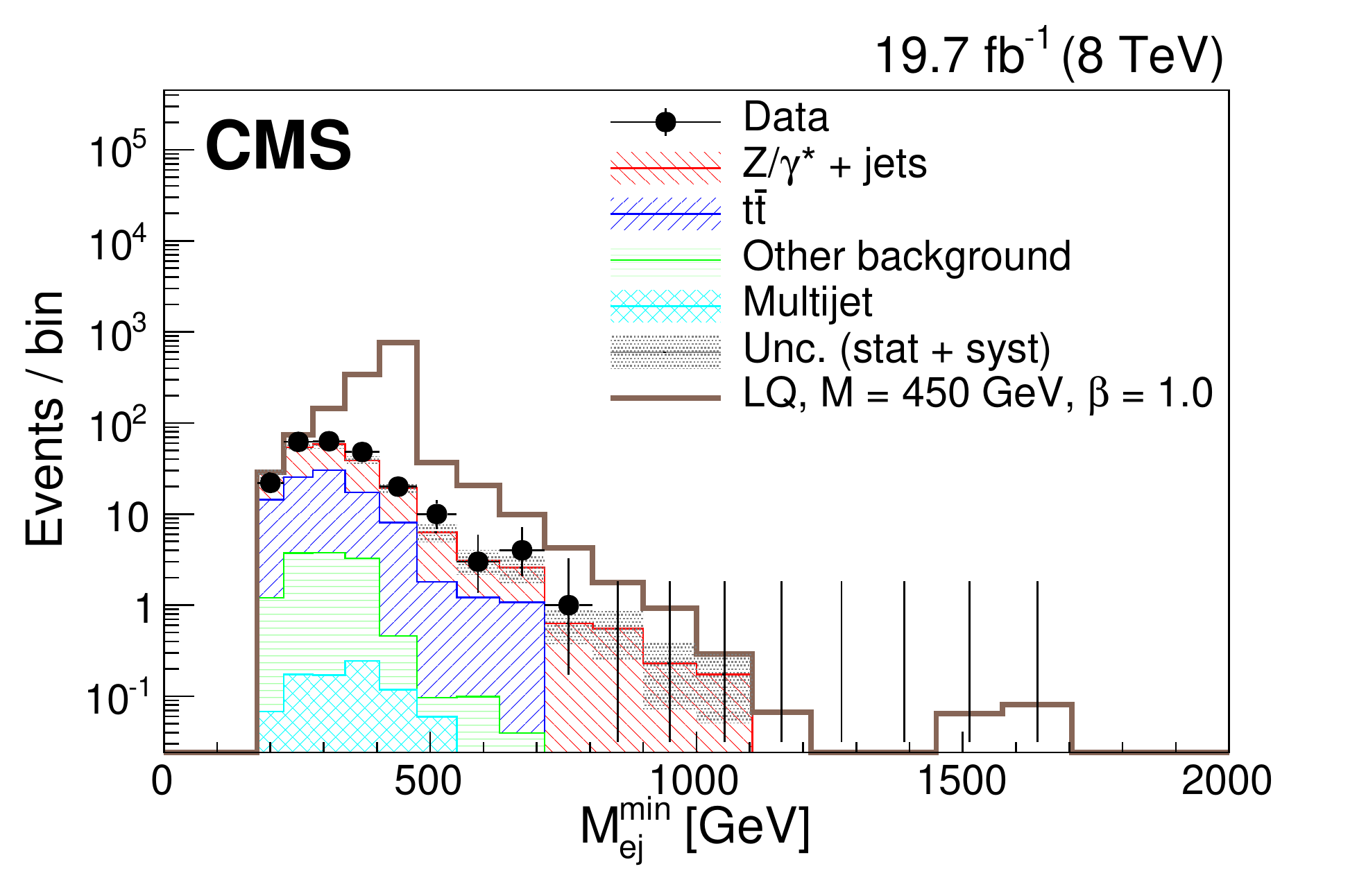}}
    {\includegraphics[width=.45\textwidth]{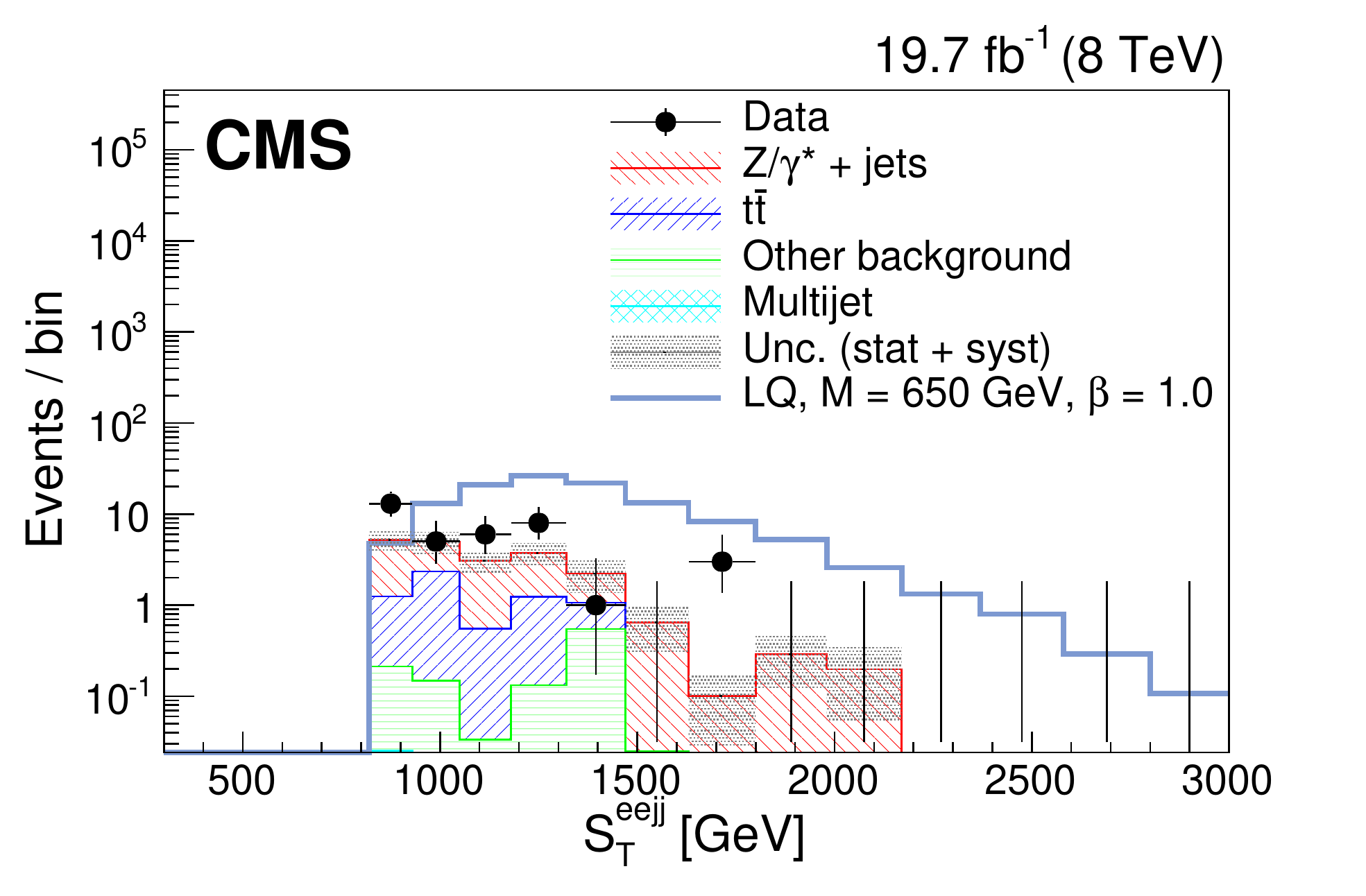}}
    {\includegraphics[width=.45\textwidth]{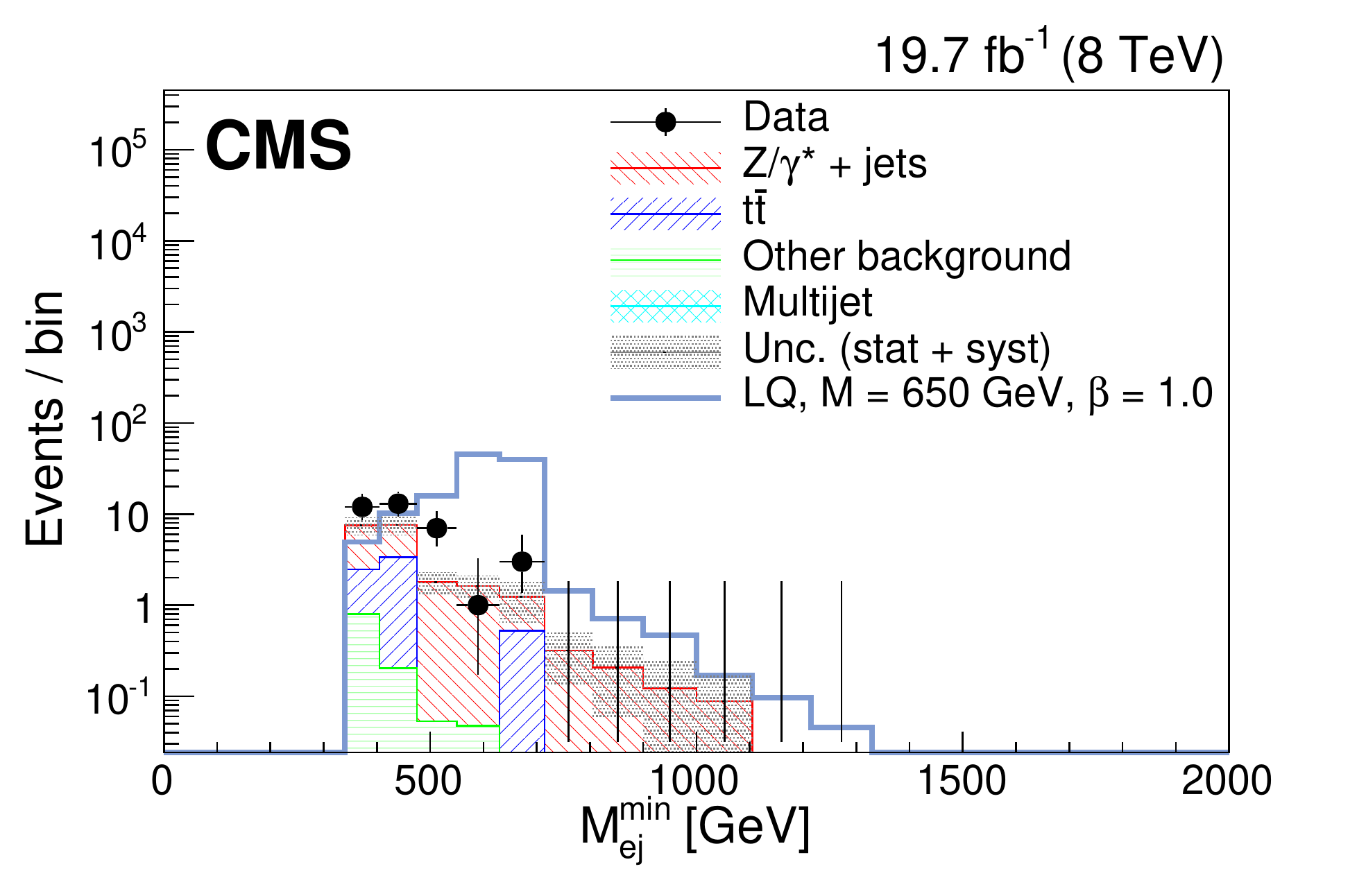}}
    \caption{
      The $\ST$ (left) and $M^{\mathrm{min}}_{\Pe\cPj}$ (right) distributions
      for events passing the \eejj~selection optimized for $M_{\text{LQ}} = 450\GeV$ (top) and $M_{\text{LQ}} = 650\GeV$ (bottom).
      The dark shaded region indicates the statistical and systematic uncertainty in the background total prediction. ``Other background" includes diboson, $\PW$+jets, and single top quark contributions.  The horizontal lines on the data points show the variable bin width.
    }
    \label{fig:eejj_finalSelection_LQM450_and_650}
  \end{center}
\end{figure*}

\begin{figure*}[!htbp]
  \begin{center}
    {\includegraphics[width=.45\textwidth]{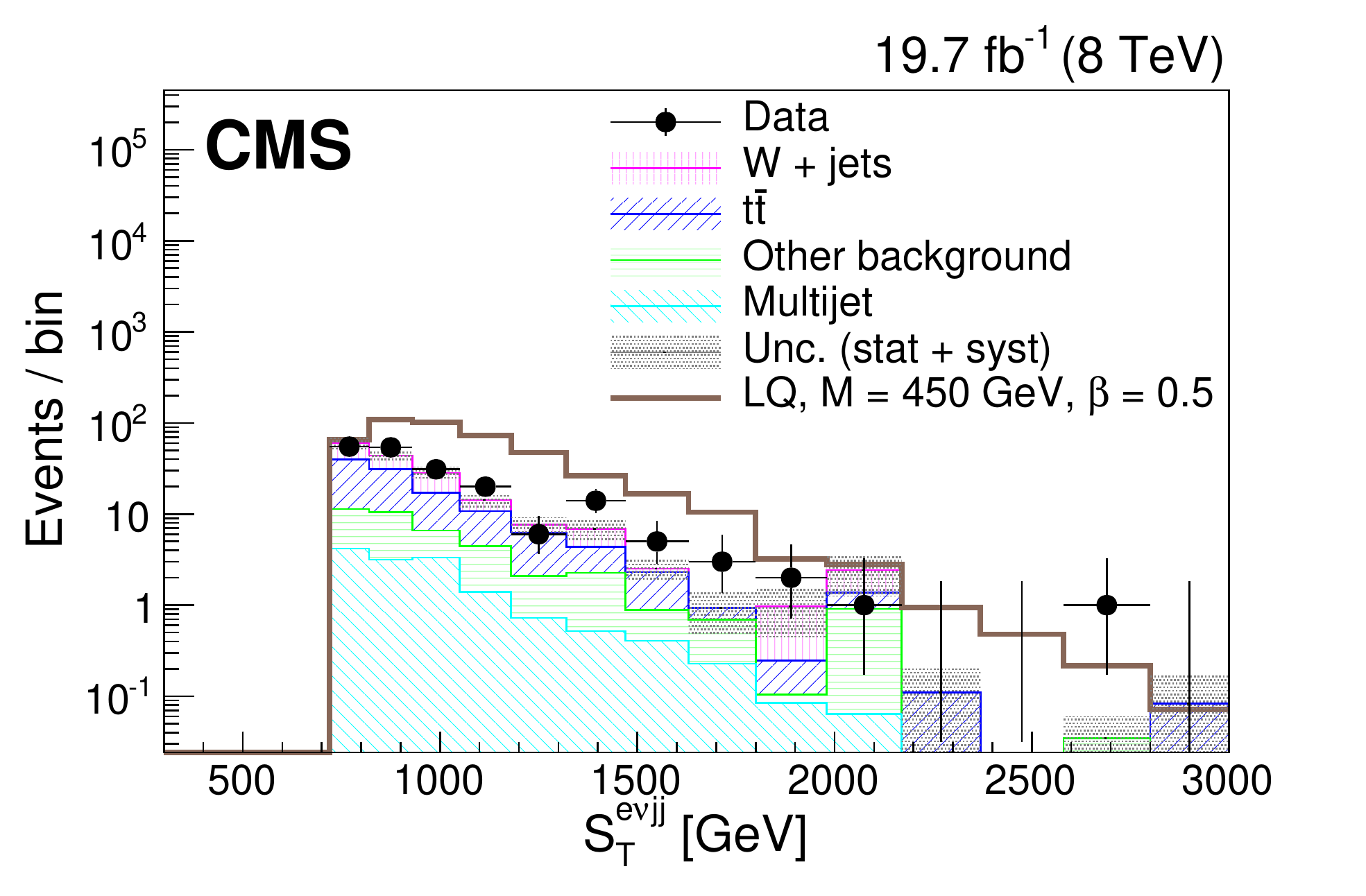}}
    {\includegraphics[width=.45\textwidth]{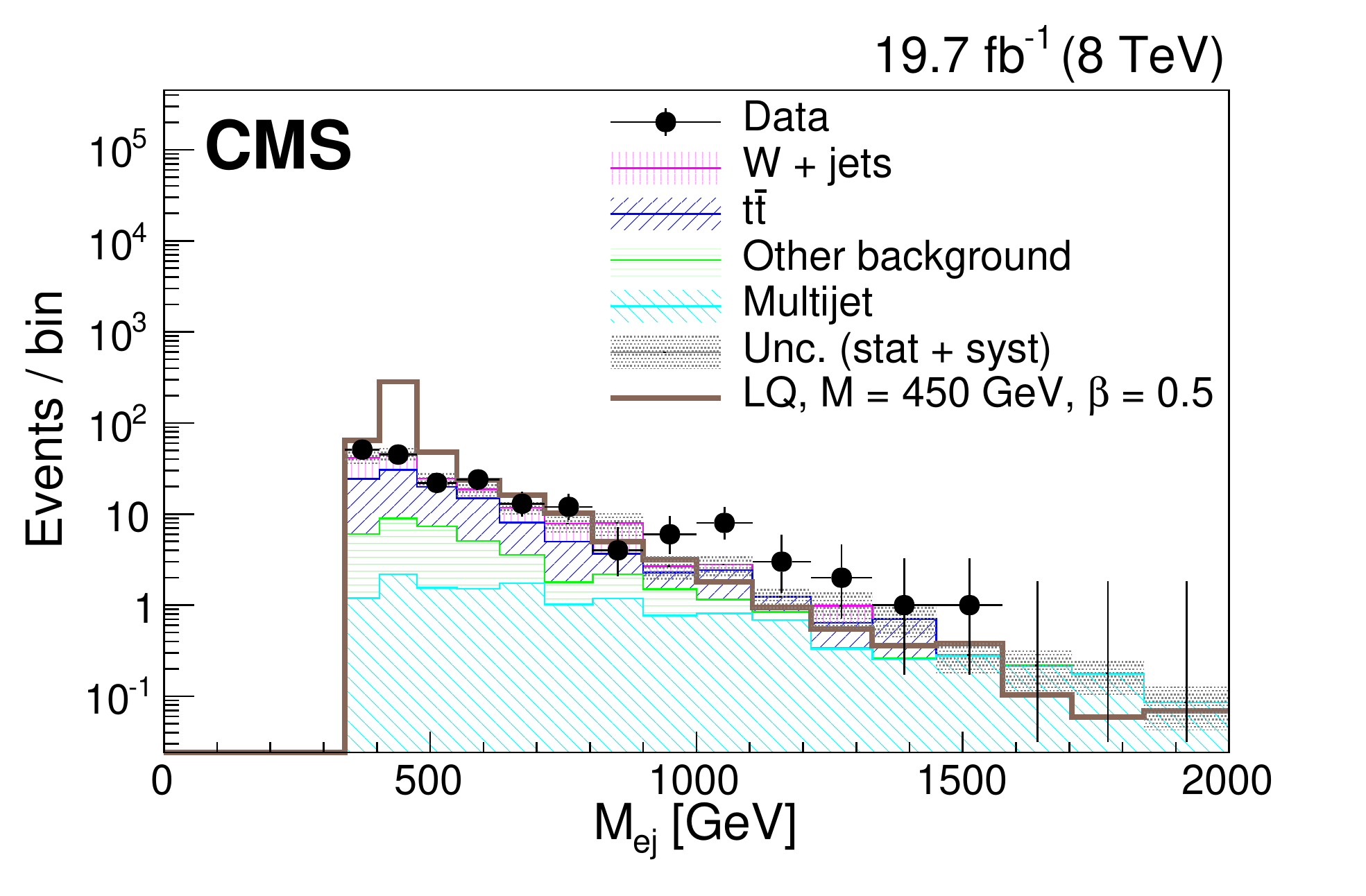}}
    {\includegraphics[width=.45\textwidth]{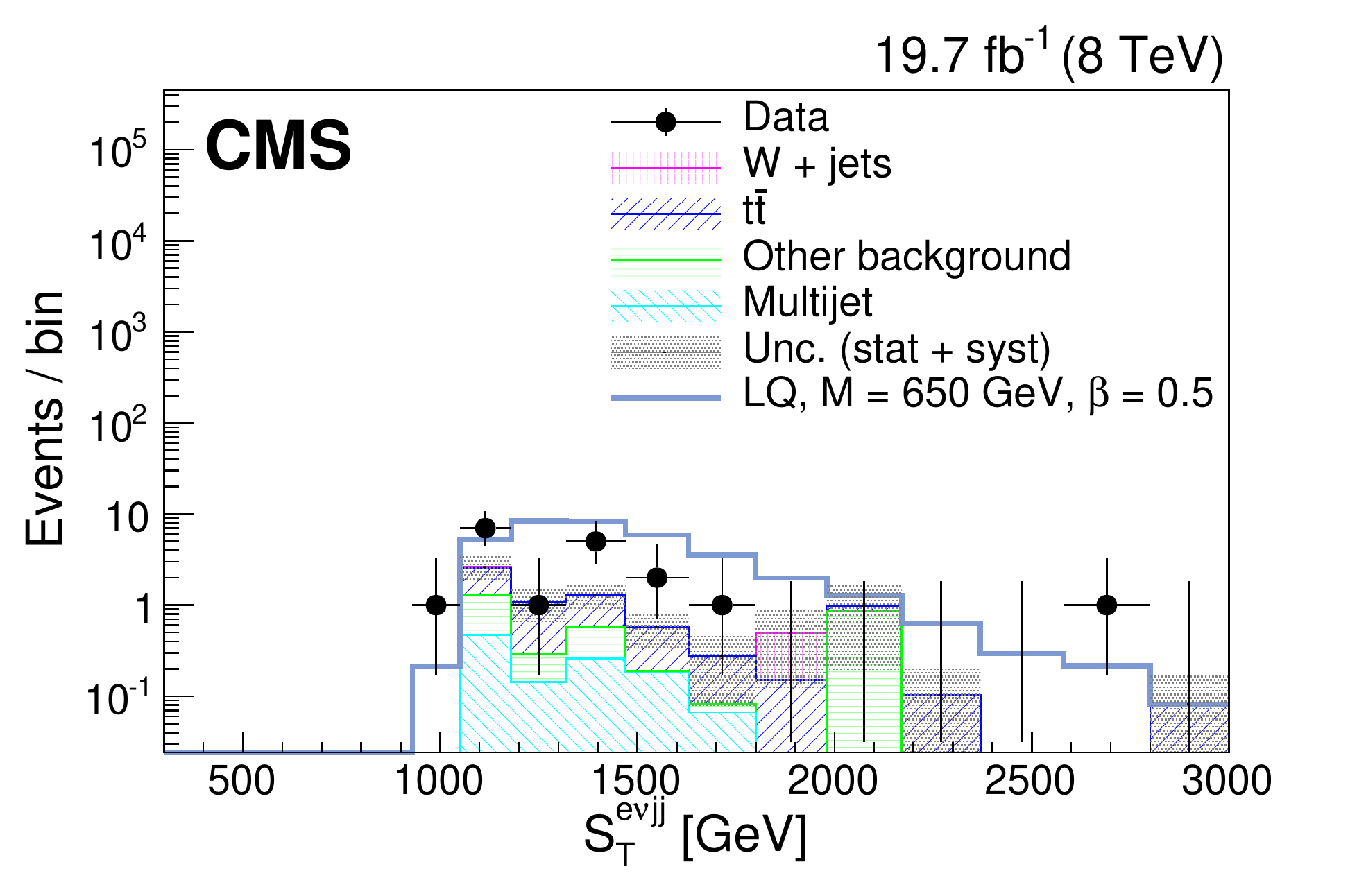}}
    {\includegraphics[width=.45\textwidth]{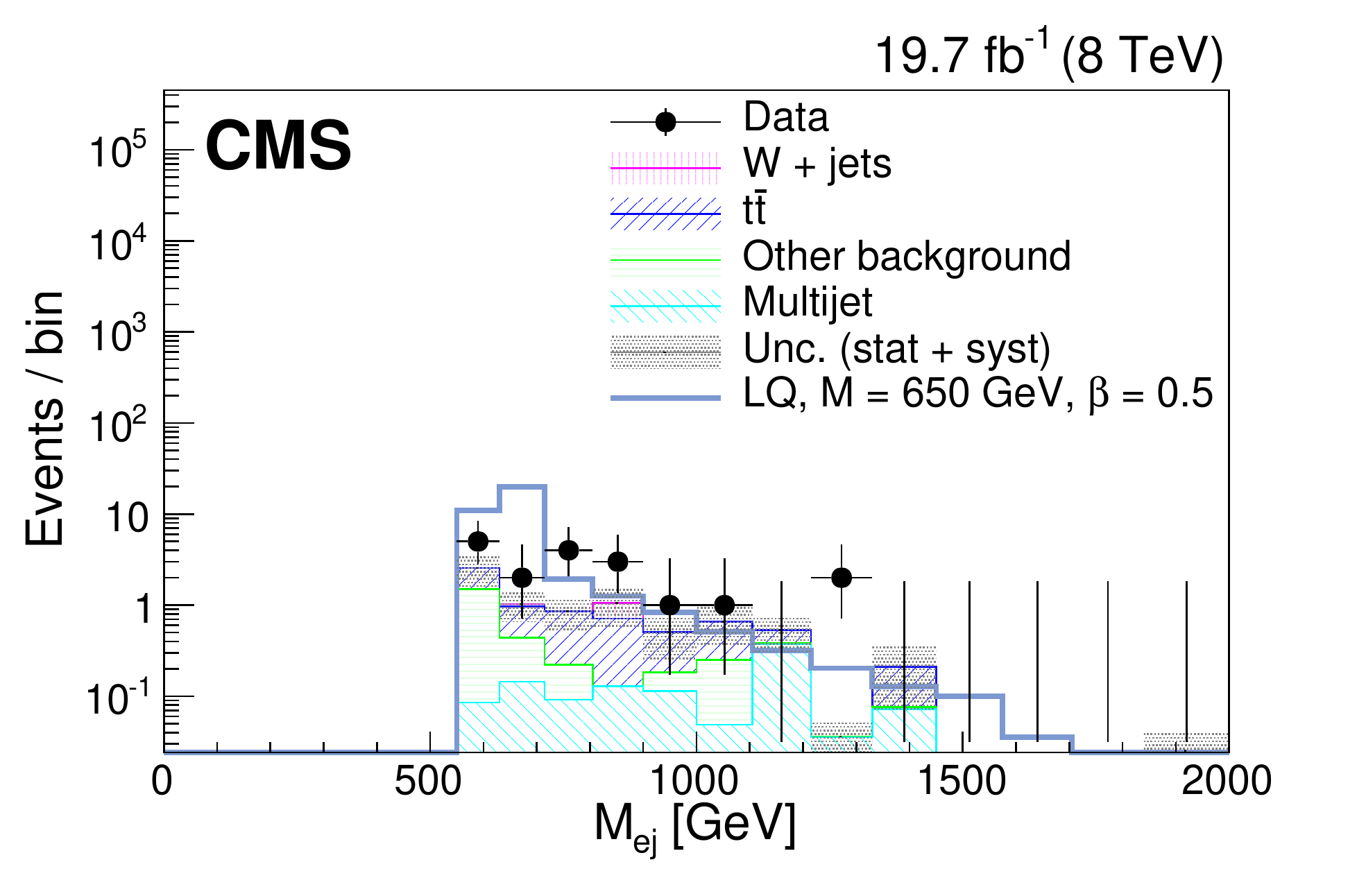}}
    \caption{
      The $\ST$ (left) and $M_{\Pe\cPj}$ (right) distributions
      for events passing the full \enujj~selection optimized for $M_{\text{LQ}} = 450\GeV$ (top) and $M_{\text{LQ}} = 650\GeV$ (bottom).
      The dark shaded region indicates the statistical and systematic uncertainty in the total background prediction. ``Other background" includes diboson, $\cPZ/\gamma^*$+jets, and single top quark contributions.  The horizontal lines on the data points show the variable bin width.
    }
    \label{fig:enujj_finalSelection_LQM450_and_650}
  \end{center}
\end{figure*}

In both the \eejj~and \enujj~channels, a broad data excess is observed for the selections optimized for a LQ mass greater than about 400\GeV,
as shown in Figs.~\ref{fig:eejj_finalSelection_LQM450_and_650} and~\ref{fig:enujj_finalSelection_LQM450_and_650} for two chosen selections,
and in Tables~\ref{tab:eejj_cuttable} and~\ref{tab:enujj_cuttable}.
This excess is most significant in the selection optimized for a LQ mass of 650\GeV, where  for the \eejj~(\enujj)~channel $20.5 \pm 2.1 \text{ (stat) } \pm 2.8 \text{ (syst) }$~($7.5 \pm 1.2 \text { (stat) } \pm 1.1\syst$) events are expected and
36 (18) events are observed, with a significance of 2.3 (2.6) standard deviations.

\begin{figure}[!htb]
  \begin{center}
    {\includegraphics[width=.45\textwidth]{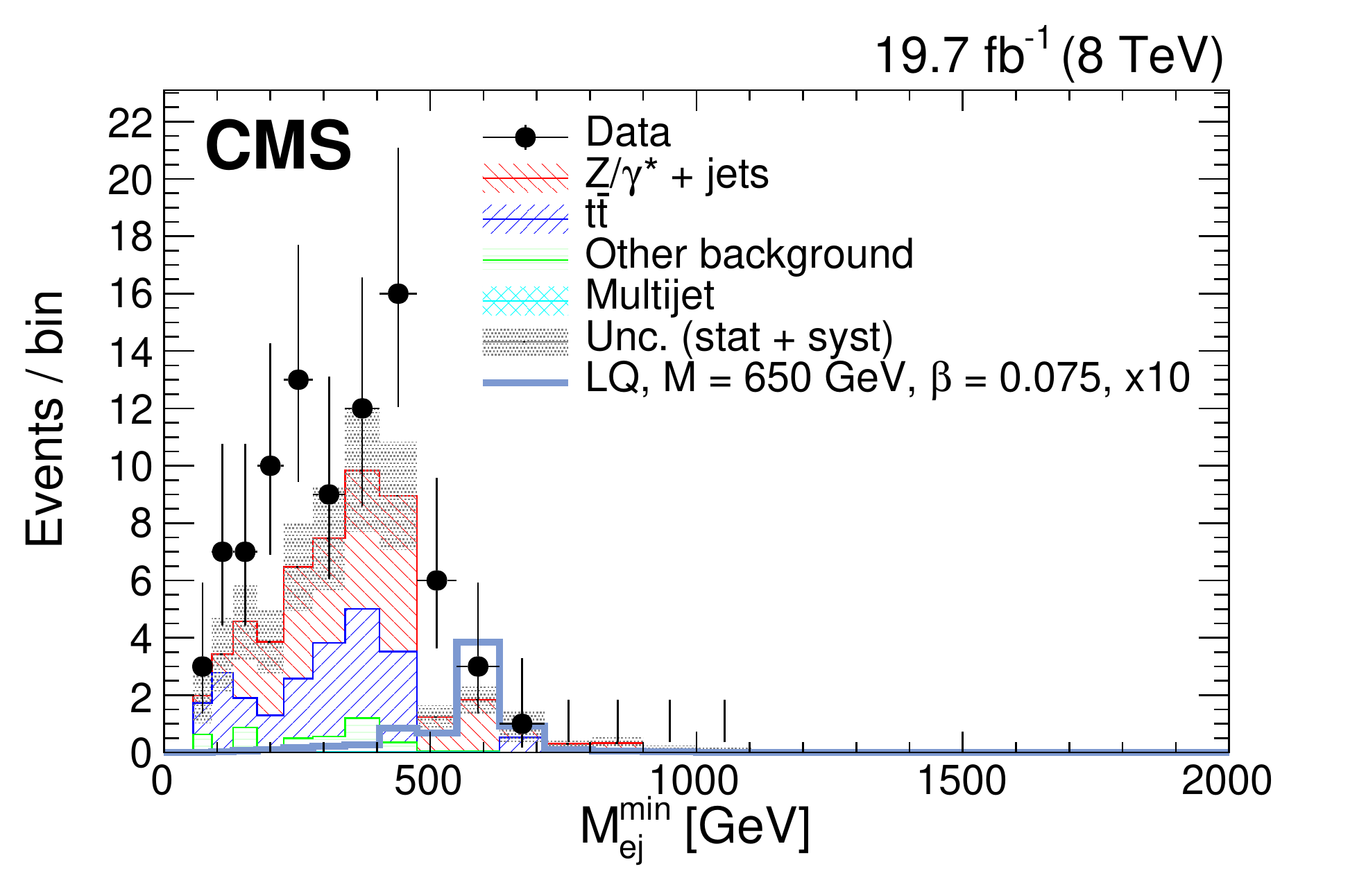}}
    {\includegraphics[width=.45\textwidth]{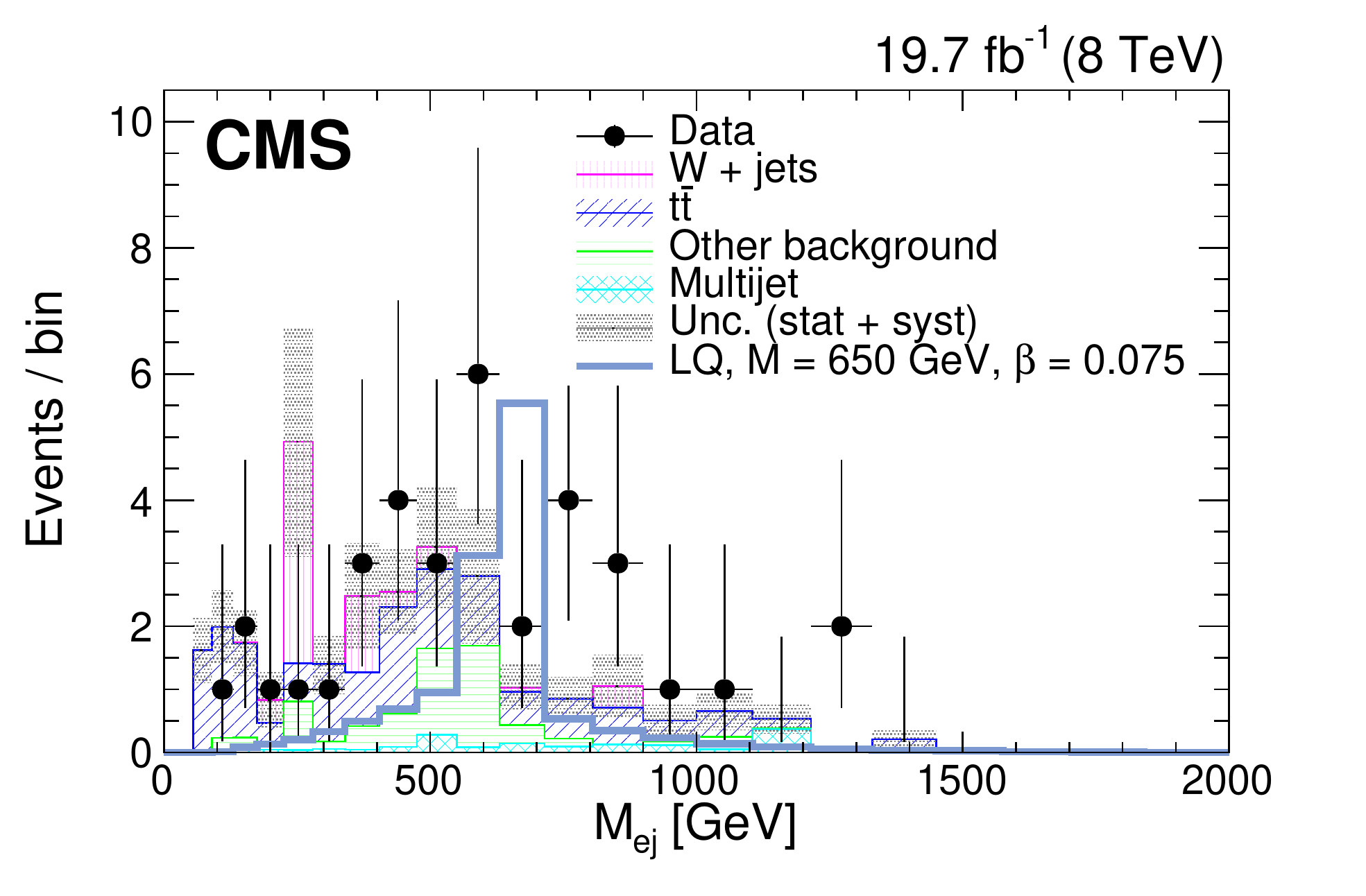}}
    \caption{
      The $M^{\text{min}}_{\Pe\cPj}$~distribution for the \eejj~channel (\cmsLeft) and
      the $M_{\Pe\cPj}$~distribution for the \enujj~channel (\cmsRight) after the selection criteria optimized for a LQ
      mass of 650\GeV have been applied. The dark shaded region indicates the statistical and systematic uncertainty in the total background prediction.
      The signal corresponds to a LQ mass of 650\GeV and $\beta = 0.075$.  The signal is multiplied by a factor of ten in the left plot. In the case of the \eejj~analysis, less than one signal event is
      expected to pass the selection.  The horizontal lines on the data points show the variable bin width.
   }
    \label{fig:nminus1}
  \end{center}
\end{figure}

An investigation of the kinematic distributions in both channels shows that the excesses are background-like.
In particular, unlike a LQ hypothesis, the excesses do not peak sharply in the $M^{\mathrm{min}}_{\mathrm{\Pe\cPj}}$ and the $M_{\mathrm{\Pe\cPj}}$ distributions, as shown in Fig.~\ref{fig:nminus1}. For comparison, the distributions that would result from a LQ mass hypothesis of 650\GeV and $\beta = 0.075$~are also shown (this is the value of $\beta$ that, for a LQ mass of 650\GeV, would produce 10 events in the \enujj~selection optimized for such a LQ mass, which is about the size of the excess).
The intrinsic width of scalar LQs is $\frac{\lambda_{\ell q}^{2}}{16\pi}\times M_{\text{LQ}}$.
The LQ signal events were generated with $\lambda_{\ell q} = 0.3$. This corresponds to an intrinsic width of about 1.2\GeV for a LQ with mass close to 650\GeV, which is negligible compared to the experimental resolution. Significantly higher values of $\lambda_{\ell q}$ (and consequently broader LQs) are strongly limited in this mass range by results from the HERA experiments~\cite{hera1,hera2}.

Further investigations of the characteristics of the data that survives the selections optimized for a LQ mass of 650\GeV show that
there are two events containing same-sign electrons out of the 36 events, and we expect the SM background to contribute about two events with same-sign electrons out of the about 20 predicted events, because of charge misidentification.
We have also verified that the excess is not enhanced if we require that the jets are identified as b-quark jets using the combined secondary vertex b-tagging algorithm~\cite{btag}.

A recently published search for heavy neutrinos and $\PW$ bosons with right-handed couplings~\cite{heavyNeutrinos} also observed an excess in the number of selected \eejj~events compared to the expectation from SM backgrounds. However, the excess in Ref.~\cite{heavyNeutrinos} is mostly localized in the region $1.8<M_{\eejj}<$ 2.2\TeV, where $M_{\eejj}$ is the invariant mass of the 2 leading electrons and 2 leading jets, while the excess observed in this analysis with the selection optimized for LQ mass of 650\GeV is broadly distributed between $M_{\eejj}$ values of 1 and 2\TeV.  Furthermore, only 30\% of the events populating the excess region in Ref.~\cite{heavyNeutrinos} survive the $M_{\mathrm{LQ}}=650\GeV$ selection.

In summary, the kinematic properties of the data in the excess regions for the \eejj~and the \enujj~channels are not found to be consistent with a LQ signal, and the size of the data excess is significantly less than that expected for a LQ with a mass of 650\GeV and $\beta \geq 0.5$. In the following section, limits are set on LQ production for both first and second generation.

\begin{table*}[htbp]
    \topcaption{Event yields for the \eejj~analysis for $\beta=1.0$ for all values of $M_{\mathrm{LQ}}$. Only statistical uncertainties are reported, except in the ``All background'' column, where systematic uncertainties are also reported.}
    \begin{center}
    \label{tab:eejj_cuttable}
    \resizebox{\textwidth}{!}{
	  \begin{scotch}{cx{8}x{6}x{6}x{8}x{8}x{12}c}
        $M_{\mathrm{LQ}}\unit{[\GeVns{}]}$ & \multicolumn{1}{c}{Signal} & \multicolumn{1}{c}{$\cPZ/\gamma^*$+jets} & \multicolumn{1}{c}{$\ttbar$} & \multicolumn{1}{c}{Multijet} & \multicolumn{1}{c}{VV, \PW, single t} & \multicolumn{1}{c}{All background} &  Data \\
        \hline
        300  & 13560,\pm\,80     & 462.2,\pm\,7.4&724,\pm\,20   &5.280,\pm\,0.052    &62.1,\pm\,4.6   &  1254,\pm\,22\,\pm\,76 & 1244 \\
        350  & 6474 ,\pm\,33     & 332.1,\pm\,6.2&352,\pm\,14   &3.220,\pm\,0.036    &37.7,\pm\,3.6   &  725,\pm\,16\,\pm\,48  & 736 \\
        400  & 3089 ,\pm\,15     & 203.2,\pm\,4.8&153.7,\pm\,9.1    &1.700,\pm\,0.023    &23.8,\pm\,2.9   &  382,\pm\,11\,\pm\,27  & 389 \\
        450  & 1508 ,\pm\,7.2    & 112.9,\pm\,3.5&86.9,\pm\,6.9     &0.890,\pm\,0.016    &11.8,\pm\,2.0   &  212,\pm\,8.0\,\pm\,18   & 233 \\
        500  & 767.4,\pm\,3.6    & 66.5,\pm\,2.7 &47.2,\pm\,5.1     &0.490,\pm\,0.011    &7.4,\pm\,1.6    &  122,\pm\,6.0\,\pm\,9.3    & 148 \\
        550  & 410.5,\pm\,1.9    & 37.4,\pm\,2.1 &25.8,\pm\,3.7     &0.2800,\pm\,0.0084   &3.7,\pm\,1.1    &  67.2,\pm\,4.4\,\pm\,5.2     & 81 \\
        600  & 225.7,\pm\,1.0    & 22.2,\pm\,1.6 &14.2,\pm\,2.8     &0.1500,\pm\,0.0065   &3.12,\pm\,1.00  &  39.7,\pm\,3.4\,\pm\,3.3     & 57 \\
        650  & 125.90,\pm\,0.58   & 14.0,\pm\,1.2 &5.4,\pm\,1.7      &0.0760,\pm\,0.0040  &1.05,\pm\,0.47  &  20.5,\pm\,2.1\,\pm\,2.8     & 36 \\
        700  & 72.88,\pm\,0.33   & 8.16,\pm\,0.93&4.3,\pm\,1.5      &0.0450,\pm\,0.0029  &0.21,\pm\,0.12  &  12.7,\pm\,1.8\,\pm\,2.3     & 17 \\
        750  & 43.10,\pm\,0.20   & 4.88,\pm\,0.69&1.55,\pm\,0.90    &0.0260,\pm\,0.0023  &0.078,\pm\,0.038&  6.5,\pm\,1.1\,\pm\,1.2      & 12 \\
        800  & 26.17,\pm\,0.12   & 2.93,\pm\,0.52&1.04,\pm\,0.73    &0.0190,\pm\,0.0022  &0.078,\pm\,0.038&  4.06,\pm\,0.90\,\pm\,0.93      & 7 \\
        850  & 15.980,\pm\,0.072  & 2.34,\pm\,0.48&0.52,\pm\,0.52    &0.0110,\pm\,0.0015  &0.042,\pm\,0.028&  2.91,\pm\,0.71\,\pm\,0.74      & 5 \\
        900  & 9.813,\pm\,0.044  & 1.23,\pm\,0.36&0.52,\pm\,0.52    &0.0069,\pm\,0.0012 &0.022,\pm\,0.020&  1.77,\pm\,0.63\,\pm\,0.39      & 3 \\
        950  & 6.086,\pm\,0.028  & 0.89,\pm\,0.29&0.00,\,_{-0}^{+1.14}   &0.00450,\pm\,0.00085&0.022,\pm\,0.020& 0.91,\,_{-0.30}^{+1.18}\,\pm\,0.28& 1 \\
        1000 & 3.860,\pm\,0.018  & 0.56,\pm\,0.22&0.00,\,_{-0}^{+1.14}   &0.00370,\pm\,0.00082&0.0025,\pm\,0.0025& 0.57,\,_{-0.22}^{+1.16}\,\pm\,0.18& 1 \\
        1050 & 2.576,\pm\,0.011  & 0.56,\pm\,0.22&0.00,\,_{-0}^{+1.14}   &0.00370,\pm\,0.00082&0.0025,\pm\,0.0025& 0.57,\,_{-0.22}^{+1.16}\,\pm\,0.18& 1 \\
        1100 & 1.6940,\pm\,0.0072 & 0.56,\pm\,0.22&0.00,\,_{-0}^{+1.14}   &0.00370,\pm\,0.00082&0.0025,\pm\,0.0025& 0.57,\,_{-0.22}^{+1.16}\,\pm\,0.18& 1 \\
        1150 & 1.1270,\pm\,0.0047 & 0.56,\pm\,0.22&0.00,\,_{-0}^{+1.14}   &0.00370,\pm\,0.00082&0.0025,\pm\,0.0025& 0.57,\,_{-0.22}^{+1.16}\,\pm\,0.18& 1 \\
        1200 & 0.7500,\pm\,0.0030& 0.56,\pm\,0.22&0.00,\,_{-0}^{+1.14}   &0.00370,\pm\,0.00082&0.0025,\pm\,0.0025& 0.57,\,_{-0.22}^{+1.16}\,\pm\,0.18& 1 \\
      \end{scotch}
    }
  \end{center}
\end{table*}

\begin{table*}[htbp]
    \topcaption{Event yields for the \enujj~analysis for $\beta=0.5$ for all values of $M_{\mathrm{LQ}}$. Only statistical uncertainties are reported, except in the ``All background'' column, where systematic uncertainties are also reported.}
    \begin{center}
    \label{tab:enujj_cuttable}
    \resizebox{\textwidth}{!}{
 	  \begin{scotch}{cx{8}x{6}x{6}x{8}x{8}x{12}c}

        $M_{\mathrm{LQ}}\unit{[\GeVns{}]}$ & \multicolumn{1}{c}{Signal} & \multicolumn{1}{c}{$\PW$+jets} & \multicolumn{1}{c}{$\ttbar$} & \multicolumn{1}{c}{Multijet} & \multicolumn{1}{c}{VV, \cPZ, single t} & \multicolumn{1}{c}{All background} & Data \\
        \hline
        300  & 4642,\pm\,50  & 822,\pm\,22  & 1191,\pm\,12 & 117.9,\pm\,1.5  & 210.5,\pm\,7.7   &  2342,\pm\,27\,\pm\,343  &   2455 \\
        350  & 2112,\pm\,21  & 276,\pm\,15  & 441.4,\pm\,7.2   & 59.11,\pm\,0.97  & 102.1,\pm\,5.4   &  879,\pm\,17\,\pm\,127   &   908 \\
        400  & 945.8,\pm\,9.3   & 110.4,\pm\,7.8   & 184.2,\pm\,4.7   & 32.88,\pm\,0.69  & 51.5,\pm\,3.8    &  379.0,\pm\,9.9\,\pm\,53.2     &   413 \\
        450  & 457.5,\pm\,4.5   & 53.1,\pm\,5.8    & 74.7,\pm\,3.0    & 14.13,\pm\,0.42  & 25.7,\pm\,2.7    &  167.6,\pm\,7.1\,\pm\,22.2     &   192 \\
        500  & 226.7,\pm\,2.2   & 20.5,\pm\,3.3    & 34.4,\pm\,2.0    & 7.76,\pm\,0.30   & 15.3,\pm\,2.1    &  78.0,\pm\,4.4\,\pm\,10.1       &   83 \\
        550  & 118.2,\pm\,1.2   & 8.6,\pm\,1.8     & 14.9,\pm\,1.4    & 3.89,\pm\,0.21   & 7.8,\pm\,1.6     &  35.4,\pm\,2.8\,\pm\,4.5       &   44 \\
        600  & 64.65,\pm\,0.64  & 2.3,\pm\,1.0     & 7.08,\pm\,0.93   & 2.29,\pm\,0.17   & 4.6,\pm\,1.2     &  16.3,\pm\,1.8\,\pm\,2.1       &   28 \\
        650  & 36.25,\pm\,0.36  & 0.41,\pm\,0.29   & 3.82,\pm\,0.70   & 1.18,\pm\,0.12   & 2.13,\pm\,0.92   &  7.5,\pm\,1.2\,\pm\,1.1        &   18 \\
        700  & 21.18,\pm\,0.21  & 0.41,\pm\,0.29   & 2.61,\pm\,0.60   & 0.85,\pm\,0.10   & 0.58,\pm\,0.24   &  4.45,\pm\,0.71\,\pm\,0.76        &   6 \\
        750  & 12.56,\pm\,0.12  & 0.00 ,\,_{-0}^{+0.94} & 1.75,\pm\,0.47   & 0.510,\pm\,0.091 & 0.27,\pm\,0.15   & 2.54 ,\,_{-0.50}^{+1.07}\,\pm\,0.50 &   4 \\
        800  & 7.412,\pm\,0.073  & 0.00 ,\,_{-0}^{+0.94} & 1.10,\pm\,0.37   & 0.317,\pm\,0.067 & 0.27,\pm\,0.15   & 1.70 ,\,_{-0.41}^{+1.02}\,\pm\,0.31 &   3 \\
        850  & 4.591,\pm\,0.045  & 0.00 ,\,_{-0}^{+0.94} & 0.90,\pm\,0.34   & 0.117,\pm\,0.029 & 0.140,\pm\,0.087 & 1.15 ,\,_{-0.35}^{+1.00}\,\pm\,0.24 &   2 \\
        900  & 2.853,\pm\,0.028  & 0.00 ,\,_{-0}^{+0.94} & 0.37,\pm\,0.21   & 0.076,\pm\,0.024 & 0.084,\pm\,0.069 & 0.53 ,\,_{-0.22}^{+0.97}\,\pm\,0.10 &   1 \\
        950  & 1.791,\pm\,0.017  & 0.00 ,\,_{-0}^{+0.94} & 0.37,\pm\,0.21   & 0.069,\pm\,0.023 & 0.084,\pm\,0.069 & 0.52 ,\,_{-0.22}^{+0.97}\,\pm\,0.10 &   1 \\
        1000 & 1.272,\pm\,0.011  & 0.00 ,\,_{-0}^{+0.94} & 0.37,\pm\,0.21   & 0.069,\pm\,0.023 & 0.084,\pm\,0.069 & 0.52 ,\,_{-0.22}^{+0.97}\,\pm\,0.10 &   1 \\
        1050 & 0.8788,\pm\,0.0074 & 0.00 ,\,_{-0}^{+0.94} & 0.37,\pm\,0.21   & 0.069,\pm\,0.023 & 0.084,\pm\,0.069 & 0.52 ,\,_{-0.22}^{+0.97}\,\pm\,0.10 &   1 \\
        1100 & 0.6063,\pm\,0.0049 & 0.00 ,\,_{-0}^{+0.94} & 0.37,\pm\,0.21   & 0.069,\pm\,0.023 & 0.084,\pm\,0.069 & 0.52 ,\,_{-0.22}^{+0.97}\,\pm\,0.10 &   1 \\
        1150 & 0.4196,\pm\,0.0032 & 0.00 ,\,_{-0}^{+0.94} & 0.37,\pm\,0.21   & 0.069,\pm\,0.023 & 0.084,\pm\,0.069 & 0.52 ,\,_{-0.22}^{+0.97}\,\pm\,0.10 &   1 \\
        1200 & 0.2894,\pm\,0.0021 & 0.00 ,\,_{-0}^{+0.94} & 0.37,\pm\,0.21   & 0.069,\pm\,0.023 & 0.084,\pm\,0.069 & 0.52 ,\,_{-0.22}^{+0.97}\,\pm\,0.10 &   1 \\
      \end{scotch}
    }
  \end{center}
\end{table*}

\subsection{Exclusion limits on scalar LQ pair-production}
\label{sec:exclusionlimits_scalar}

Upper limits are set on the scalar LQ production cross sections $\sigma$ using
the asymptotic CL$_\mathrm{S}$ modified-frequentist approach~\cite{cls1,cls2}. A log-normal probability density function is used to integrate over the systematic uncertainties described in Section~\ref{systematics}. Uncertainties of statistical nature are described with gamma distributions with
widths determined by the number of events in signal and background simulated samples or observed in data control regions.

\begin{figure}[!htb]
  \begin{center}
    {\includegraphics[width=.40\textwidth]{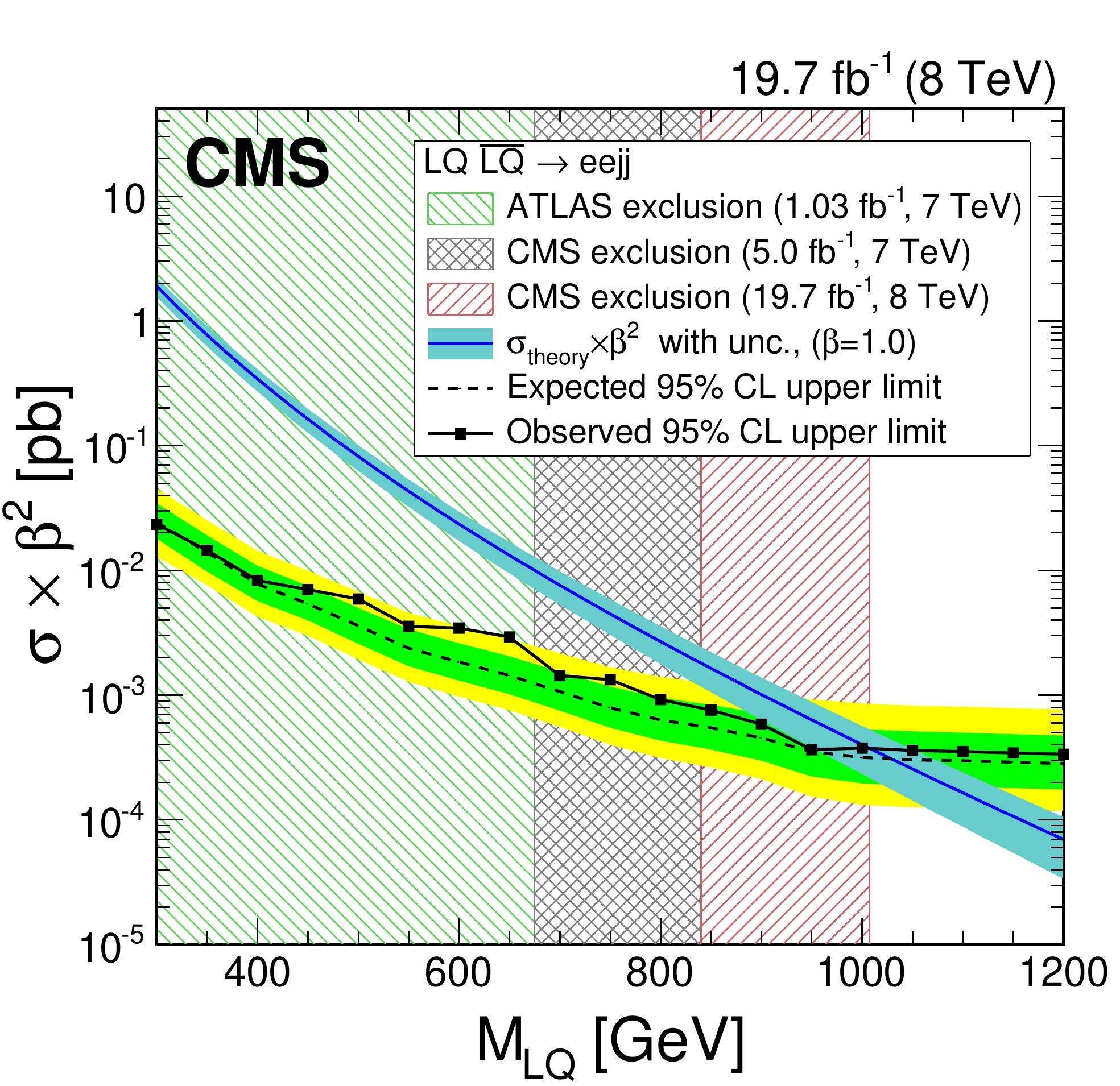}}
    {\includegraphics[width=.40\textwidth]{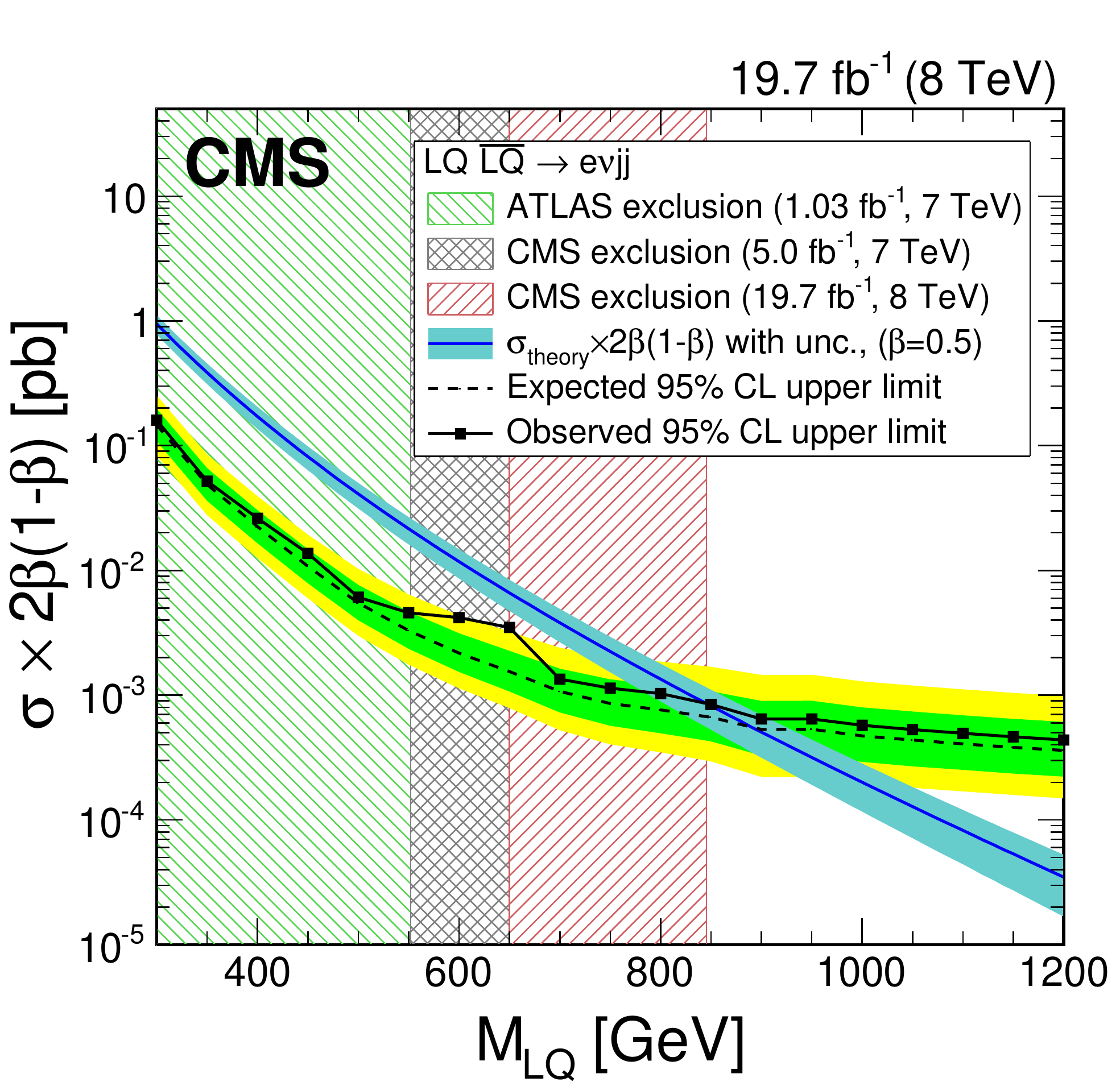}}
    \caption{
    Frame on \cmsLeft\ (\cmsRight): the expected and observed upper limits at $95\%$~\CL on the LQ pair production
    cross section times $\beta^2$ ($2\beta(1-\beta)$) as a function of the first generation LQ mass obtained with the \eejj~(\enujj) analysis. The expected limits and uncertainty bands represent the median expected limits and the 68\% and 95\% confidence intervals.
    The left shaded regions are excluded by Ref.~\cite{atlaslq1-2011} and the middle shaded regions are excluded by Ref.~\cite{cmslq-2011}. The right shaded region is excluded by the analysis presented in this paper.
    The $\sigma_{\rm theory}$ curves and their bands represent, respectively, the theoretical scalar LQ pair production cross section
    and the uncertainties due to the choice of PDF and renormalization/factorization scales.
    }
    \label{fig:limits_1d}
  \end{center}
\end{figure}

\begin{figure}[!htb]
  \centering
  \includegraphics[width=0.45\textwidth]{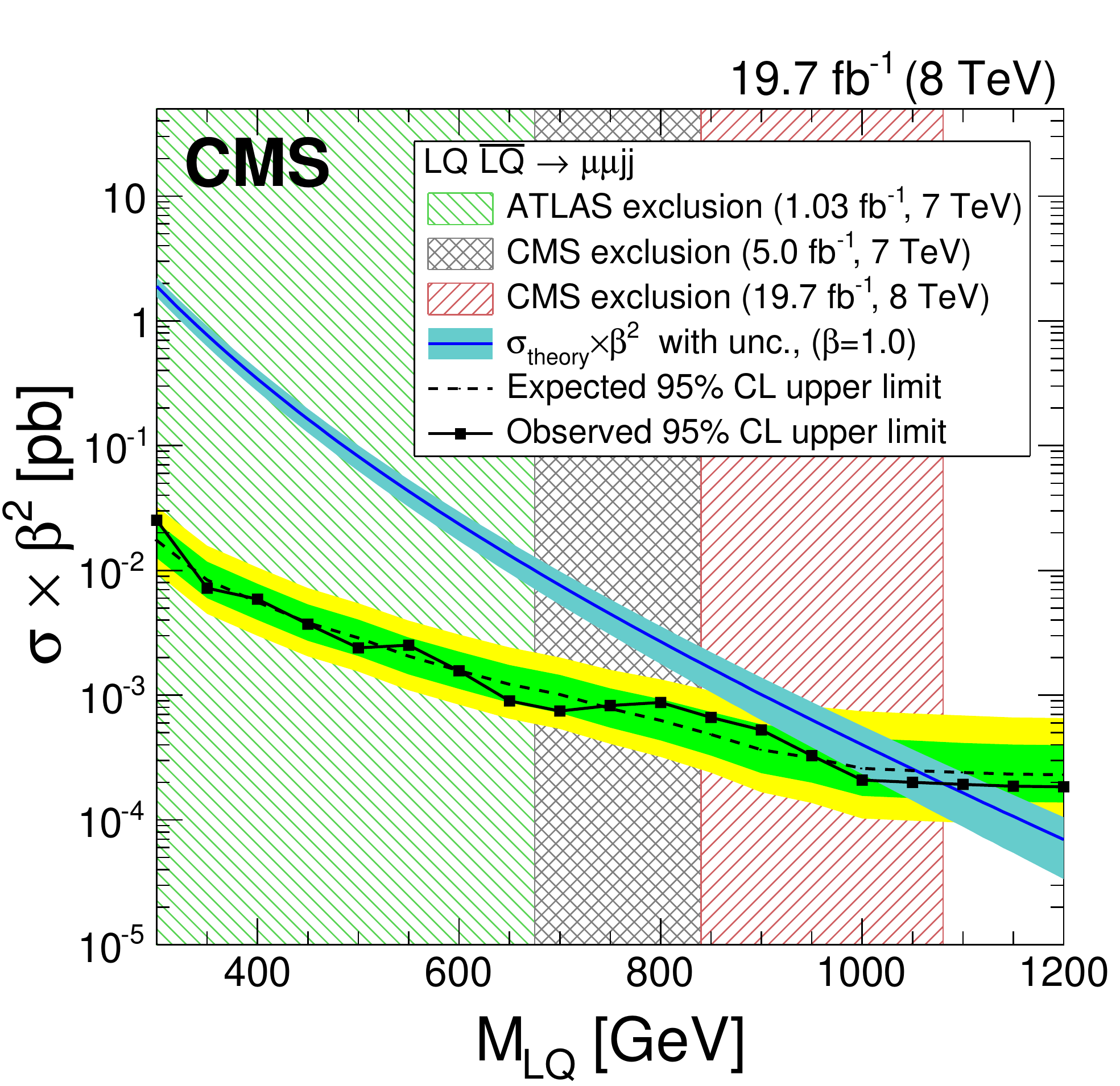}
  \includegraphics[width=0.45\textwidth]{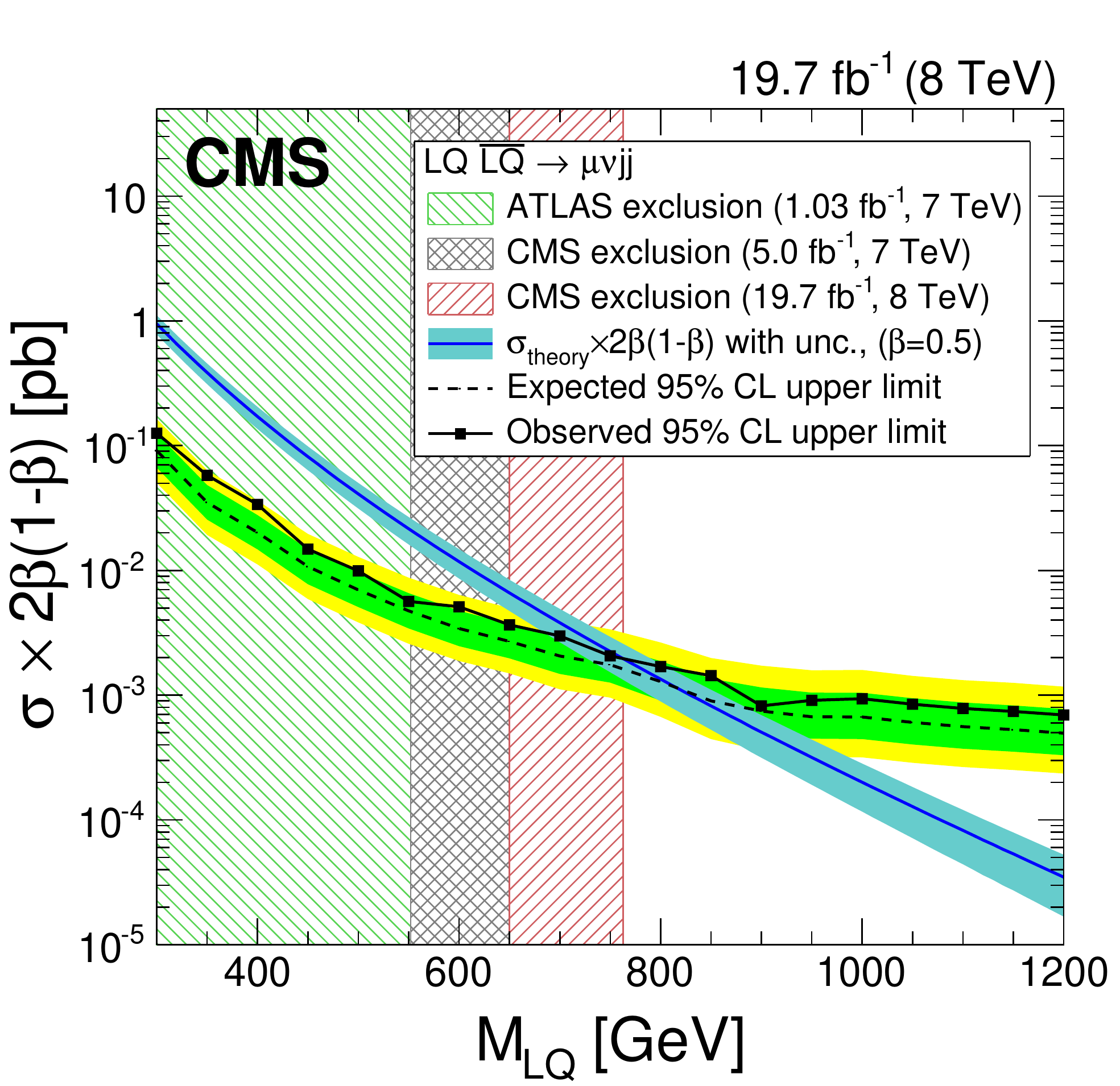}
   \caption{
    Frame on \cmsLeft\ (\cmsRight): the expected and observed upper limits at $95\%$~\CL on the LQ pair production
    cross section times $\beta^2$ ($2\beta(1-\beta)$) as a function of the second generation LQ mass obtained with the \mumujj~(\munujj) analysis. The expected limits and uncertainty bands represent the median expected limits and the 68\% and 95\% confidence intervals.
    The left shaded regions are excluded by Ref.~\cite{atlaslq2-2011} and the middle shaded regions are excluded by Ref.~\cite{cmslq-2011}. The right shaded region is excluded by the analysis presented in this paper.
    The $\sigma_{\rm theory}$ curves and their bands represent, respectively, the theoretical scalar LQ pair production cross section
    and the uncertainties due to the choice of PDF and renormalization/factorization scales.
    }
  \label{fig:limit_plots_mu}
\end{figure}

The 95\%~confidence level (\CL) upper limits on $\sigma \times \beta^2$ or $\sigma \times 2\beta(1-\beta)$ as a function of LQ mass are shown
together with the NLO predictions for the scalar LQ pair production cross section in Fig.~\ref{fig:limits_1d} for the \eejj~and \enujj~channels, and in Fig.~\ref{fig:limit_plots_mu} for the \mumujj~and \munujj~channels.
The theoretical cross sections are represented as the central values with a band indicating the sum in quadrature of the PDF uncertainty and the uncertainty associated with the choice of factorization/renormalization scale. The latter is estimated from the observed effect of varying the scale between half and twice the LQ mass.

By comparing the observed upper limit with the theoretical cross section values, first generation scalar LQ with masses less than 1010 (850)\GeV are excluded with the assumption that $\beta = 1$~$(0.5)$. This is to be compared with median expected limits of 1030 (890)\GeV.
Similarly, second generation scalar LQ with masses less than 1080 (760)\GeV are excluded with the same assumptions on $\beta$, to be compared with median expected limits of 1050 (820)\GeV.

\begin{figure}[htbp]
  \begin{center}
    {\includegraphics[width=.45\textwidth]{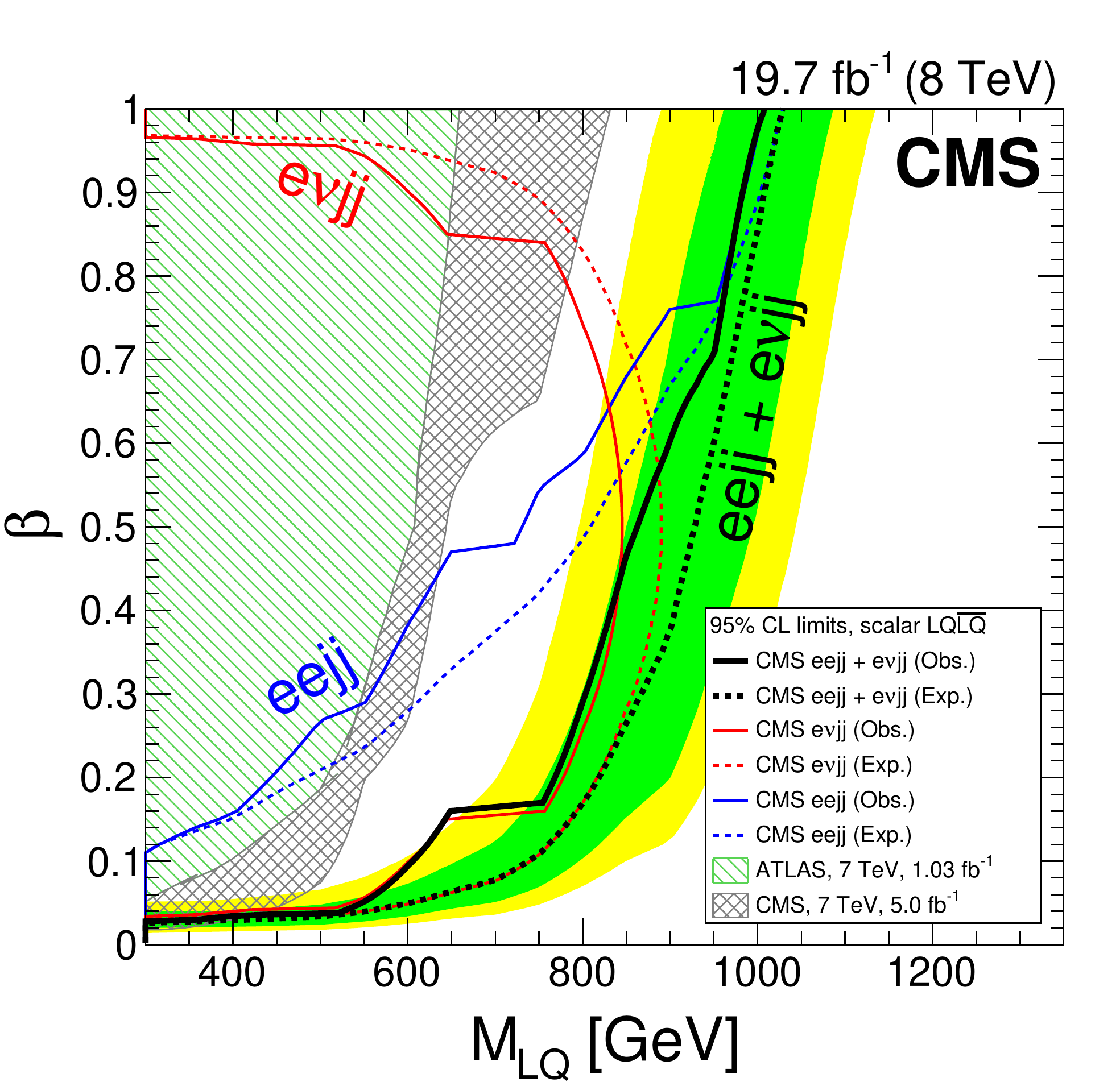}}
    {\includegraphics[width=.45\textwidth]{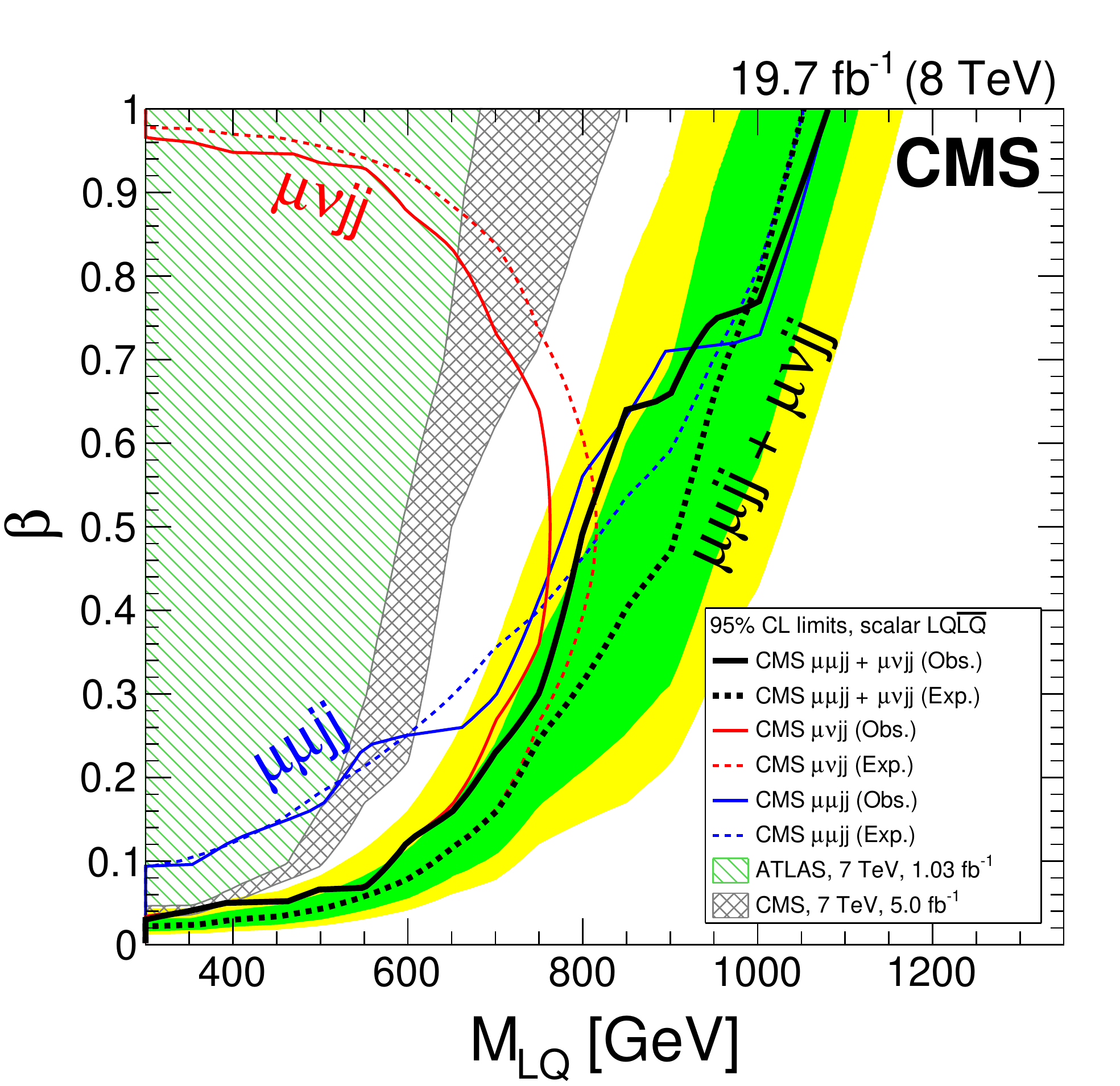}}
    \caption{
      The expected and observed exclusion limits at $95\%$~\CL on the
      first (\cmsLeft) and second (\cmsRight) generation scalar LQ hypothesis in the $\beta$ versus
      LQ mass plane using the central value of signal cross section
      for the individual $\lljj$ and $\lnujj$ channels and their
      combination.  The expected limits and uncertainty bands represent the median expected limits and the 68\% and 95\% confidence intervals. Solid
      lines represent the observed limits in each channel, and dashed
      lines represent the expected limits.
    }
    \label{fig:limits_2d}
  \end{center}
\end{figure}

The combination of the \lljj~and \lnujj~channels, shown in Fig.~\ref{fig:limits_2d}, excludes LQ masses as a function of $\beta$ using the intersection of the theoretical cross section central value and the excluded cross section. The combination can improve the mass exclusion reach for values of $\beta < 1$. Using the combined channels, second generation scalar LQ with masses less than 800\GeV are excluded for $\beta = 0.5$, compared with an expected limit of 910\GeV. In the case of first generation LQ, the combination does not lead to a change in the observed limit for $\beta = 0.5$.

The broad excess in the \eejj~and \enujj~channels is most significant for the final selection optimized for a LQ mass of 650\GeV, but has kinematic distributions that do not match those expected for a LQ hypothesis of that mass.
Figure~\ref{fig:limits_2d} shows that the presence of the excess does reduce the exclusion power of the analysis at small values of $\beta$ ($\lesssim 0.15$) for the selections optimized for LQ masses around 650\GeV. The exclusion limit for this region of the parameter space is dominated by the \enujj~channel.

\subsection{Additional interpretations}

Vector LQ signal samples were simulated with {\sc CalcHep} at the values of LQ mass detailed in Section~\ref{samples} for the four scenarios of anomalous couplings described in Section~\ref{introduction}. The cross sections for pair production of vector LQs are larger than the ones for the pair production of scalar LQs, therefore we expect a higher reach in the $M_{\mathrm{LQ}}$ exclusion limits. The cross sections for vector LQs have been calculated only at the LO level. We assume that the ratios of NLO to LO cross sections for the case of vector LQs are the same as the corresponding ratios for scalar LQs, which vary from 1.62--4.03 over the 300--1800\GeV mass range~\cite{kramer}.  In fact, the ratios of the NLO K-factors for scalar LQ pair production vs. vector LQ pair production are expected to be very similar to the analogous ratios for single LQ production, which have recently been published~\cite{VlqKfactors}.  Therefore, the limits we obtain by applying the scalar LQ K-factors to the vector LQ LO theoretical curves to obtain predictions for the NLO cross sections are expected to be conservative.  The distributions of the kinematic variables for scalar and vector LQs are sufficiently similar that the same event selections and final optimization thresholds can be used for both analyses. It is found that the cross section limits determined using the MC scenario agree within uncertainties with the YM, MM, and AM coupling scenarios. Thus, it is sufficient to overlay the theoretical cross section curves for all vector LQ scenarios with the limit curve calculated using the MC scenario.

\begin{figure}[htbp]
  \centering
  \includegraphics[width=0.45\textwidth]{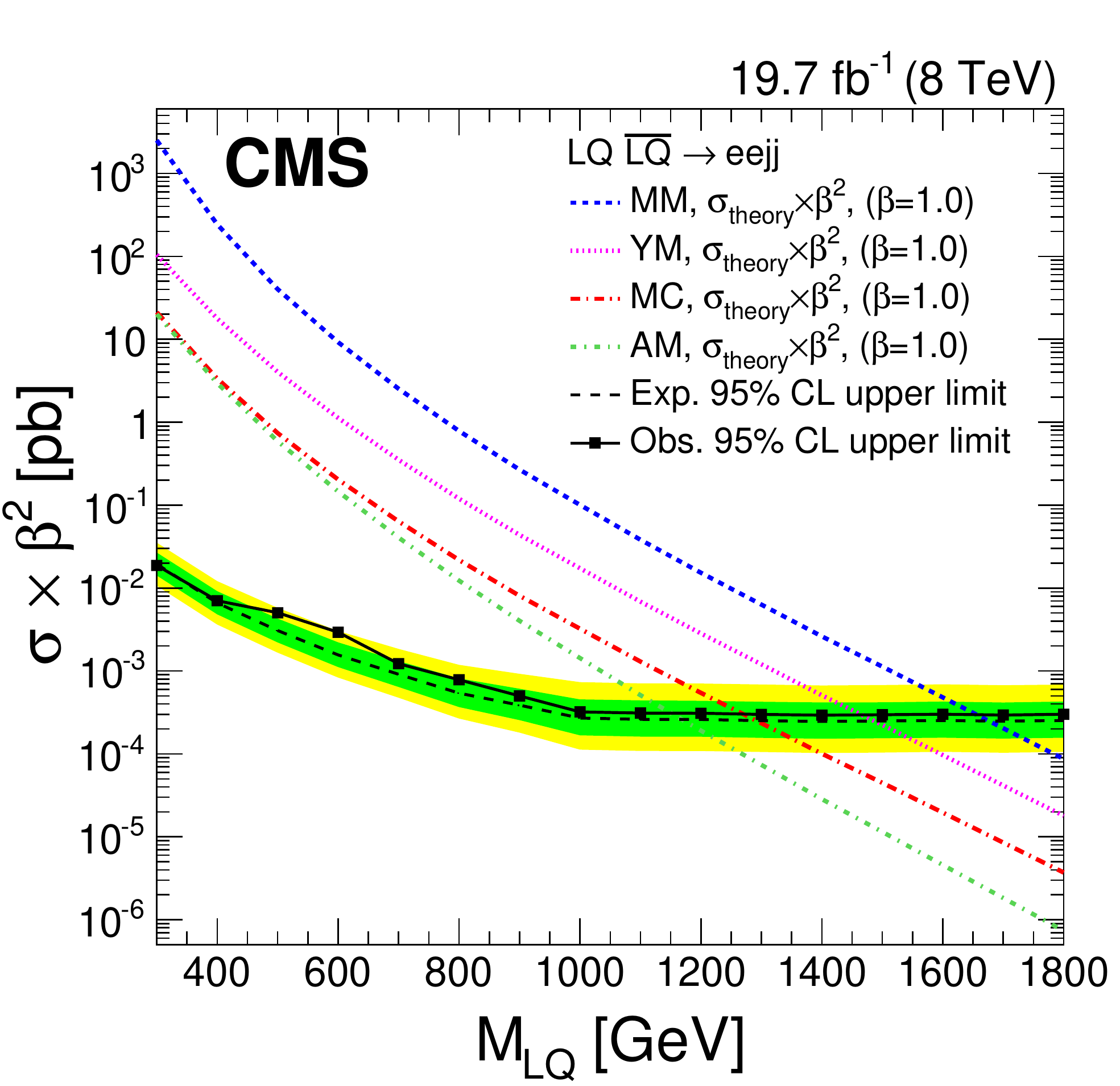}
  \includegraphics[width=0.45\textwidth]{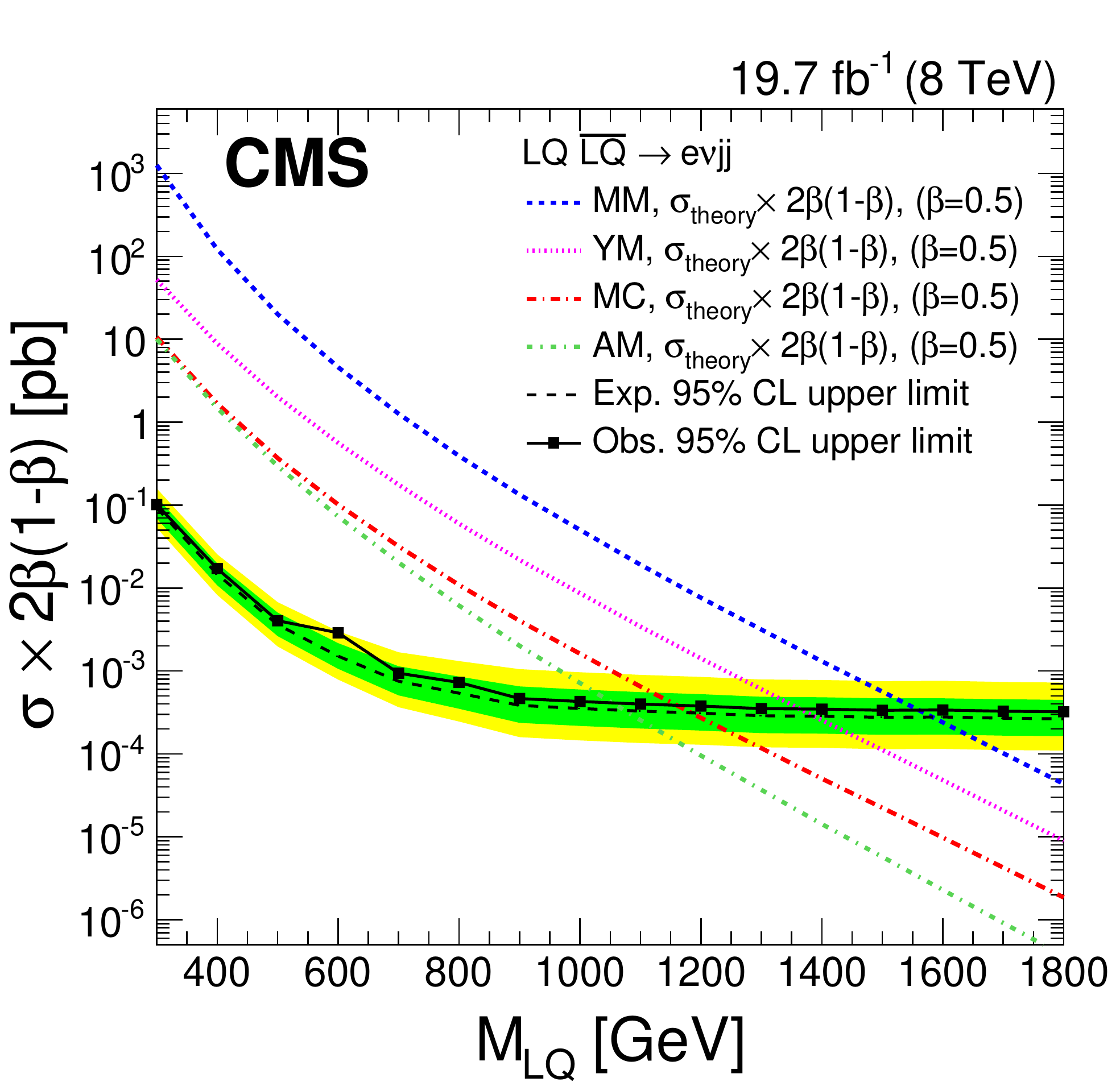}
   \caption{
    Frame on \cmsLeft\ (\cmsRight): the expected and observed upper limits at $95\%$~\CL on the vector leptoquark pair production
    cross section times $\beta^2$ ($2\beta(1-\beta)$) as a function of the first generation vector leptoquark mass, obtained with the \eejj~(\enujj) analysis for the four coupling scenarios (MC, YM, MM, and AM). The expected limits and uncertainty bands represent the median expected limits and the 68\% and 95\% confidence intervals using the MC scenario. Because of the kinematic similarity between the MC scenario and the other coupling scenarios, cross section limits are found to be the same within the uncertainties.
    }
  \label{fig:limit_plots_evec}
\end{figure}

Figure \ref{fig:limit_plots_evec} shows the experimental limits along with the four theoretical vector LQ cross sections for the \eejj~(\enujj) channel for $\beta = 1$~$(0.5)$. The experimental results yield a $95\%$~\CL upper limit exclusion of masses less than 1470 (1360)\GeV assuming YM couplings, 1270 (1160)\GeV for the MC couplings scenario, 1660 (1560)\GeV for the MM couplings scenario, and 1150 (1050)\GeV for the AM scenario. The increased energy and luminosity of the LHC results in considerably improved limits compared to the ones determined by the \DZERO~experiment at the Tevatron~\cite{dzero_vlq1}, which excluded leptoquark masses less than 340 (315)\GeV for the case of YM couplings.

\begin{figure}[htbp]
  \centering
  \includegraphics[width=0.45\textwidth]{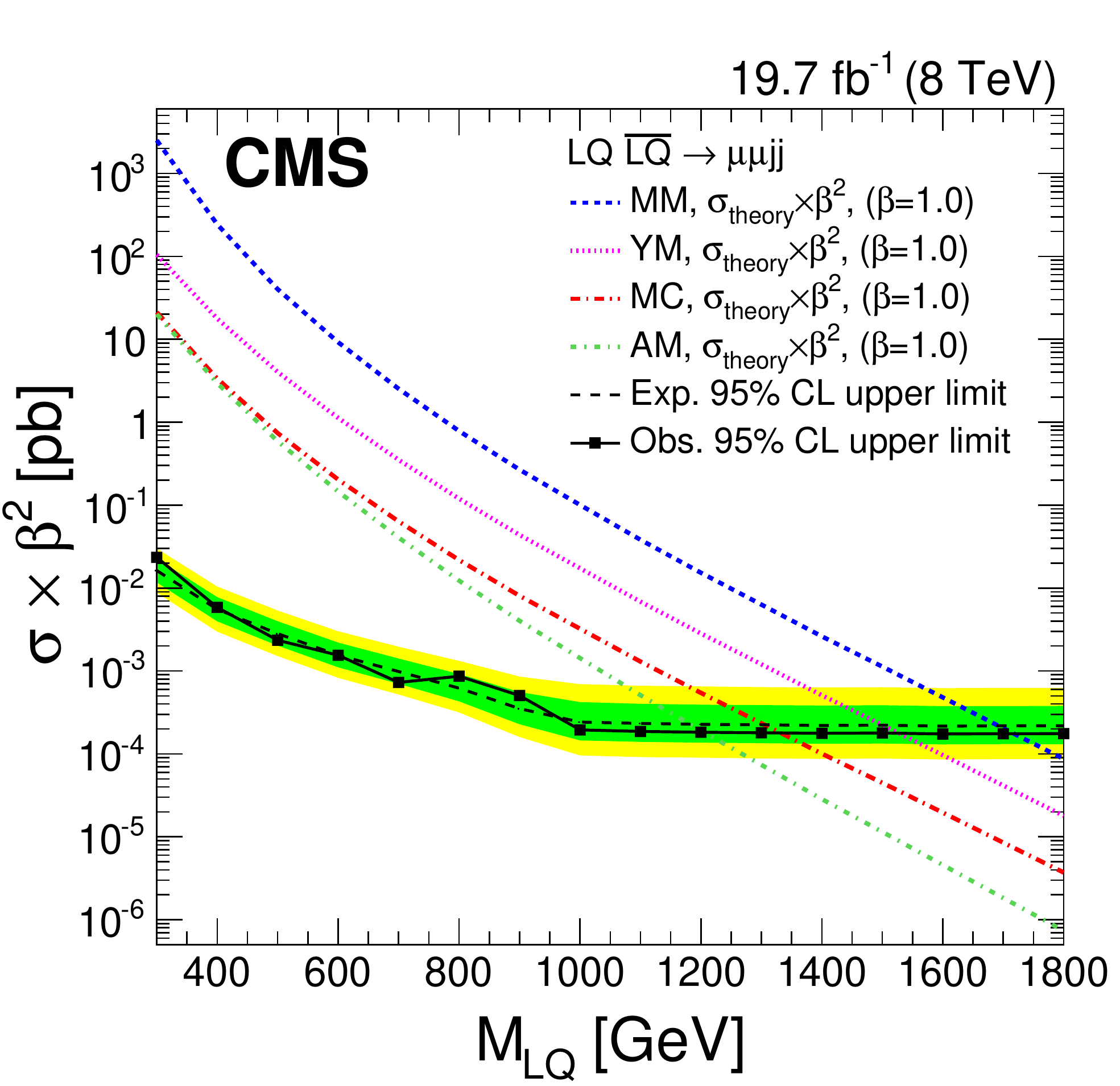}
  \includegraphics[width=0.45\textwidth]{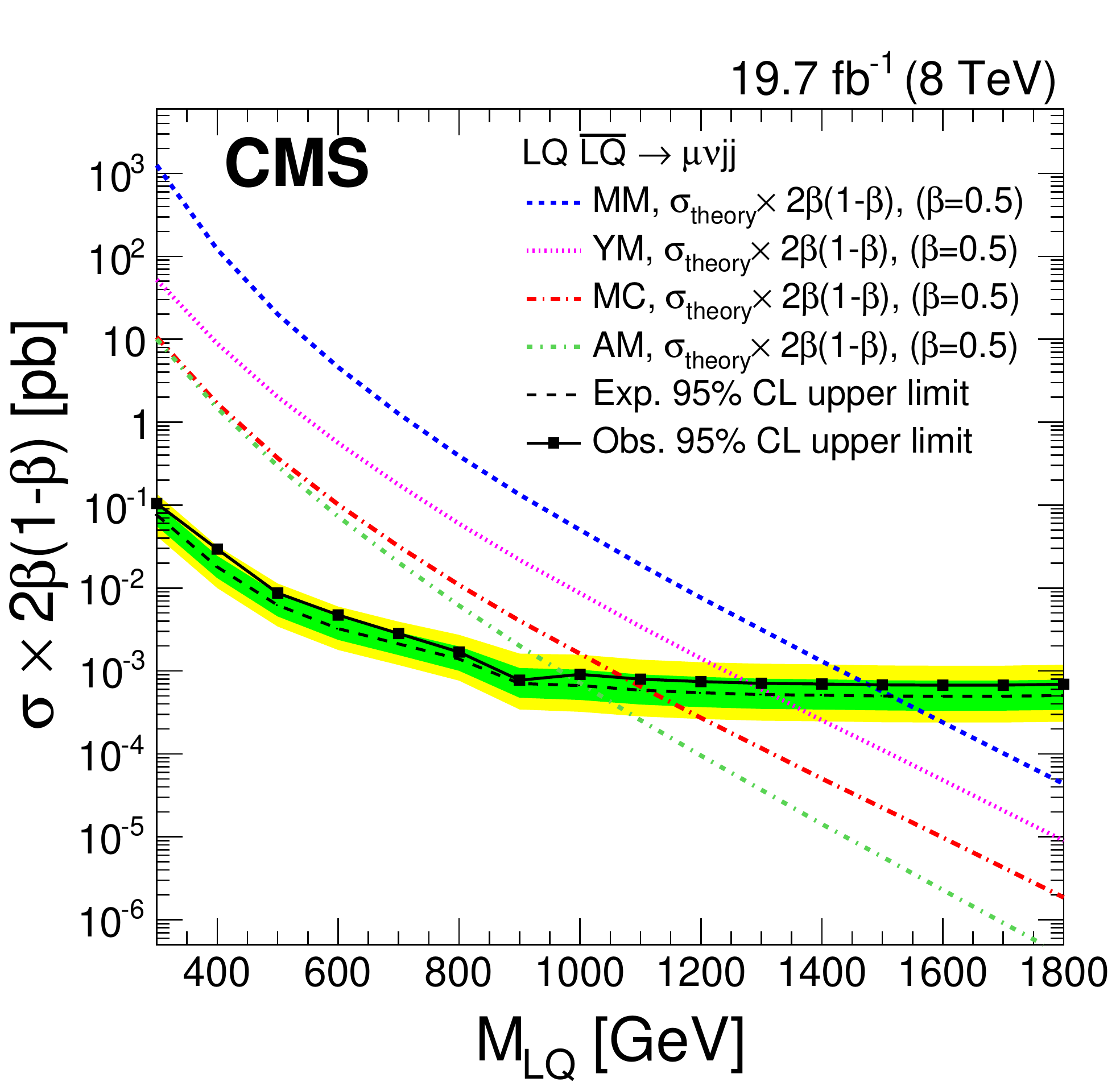}
   \caption{
    Frame on \cmsLeft\ (\cmsRight): the expected and observed upper limits at $95\%$~\CL on the vector leptoquark pair production
    cross section times $\beta^2$ ($2\beta(1-\beta)$) as a function of the second generation vector leptoquark mass, obtained with the \mumujj~(\munujj) analysis for the four coupling scenarios (MC, YM, MM, and AM). The expected limits and uncertainty bands represent the median expected limits and the 68\% and 95\% confidence intervals using the MC scenario. Because of the kinematic similarity between the MC scenario and the other coupling scenarios, cross section limits are found to be the same within the uncertainties.
    }
  \label{fig:limit_plots_muvec}
\end{figure}

Experimental limits along with the four theoretical vector LQ cross sections for the \mumujj~(\munujj) channel for $\beta = 1$~$(0.5)$ are shown in Fig.~\ref{fig:limit_plots_muvec} on the left (right).  In the \mumujj~(\munujj) channel, the experimental results yield a $95\%$~\CL upper limit exclusion of masses less than 1530 (1280)\GeV assuming YM couplings, 1330 (1070)\GeV for the MC scenario, 1720 (1480)\GeV for the MM couplings scenario, and 1200 (980)\GeV for the AM couplings scenario. These are the most stringent limits to date on second-generation vector LQ production.

The data have also been compared with an RPV SUSY model described in Ref.~\cite{rpvSusy2013}.  This model predicts light top squarks that decay to a lepton and quark through an R-parity violating top squark-lepton-quark vertex ($\lambda^{\prime}$) operator.  The $\lambda^{\prime}_{132}$ ($\lambda^{\prime}_{232}$) operator refers to top squark decay to one electron (muon) and one light-flavor quark.  In the case of direct top squark decay, this model is kinematically similar to LQ production, and the limits already described for $\beta=1$~scalar LQs can be applied simply by scaling for the small difference in production cross sections between top squarks and LQs.

\begin{figure}[!htb]
  \centering
  \includegraphics[width=0.45\textwidth]{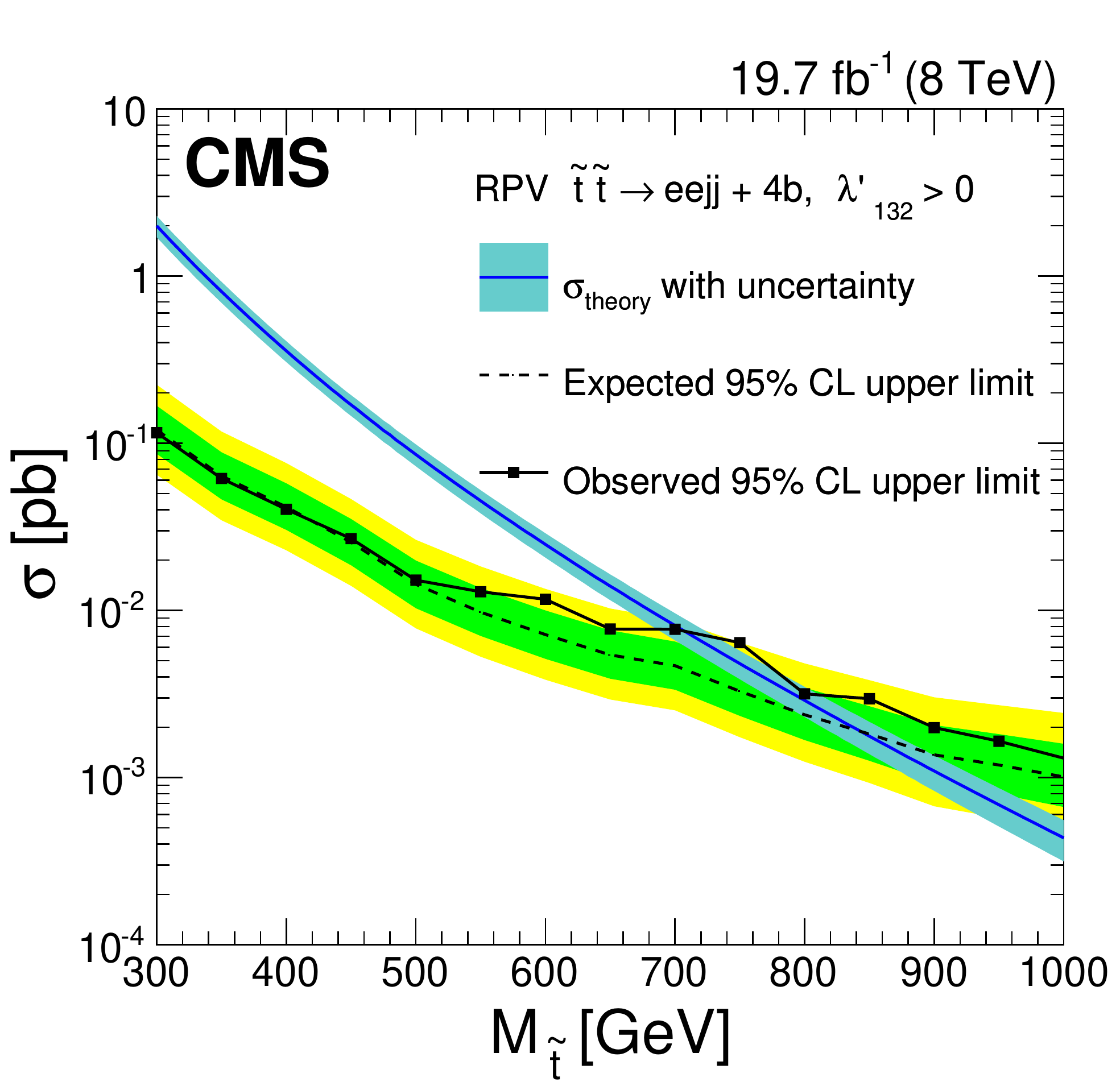}
  \includegraphics[width=0.45\textwidth]{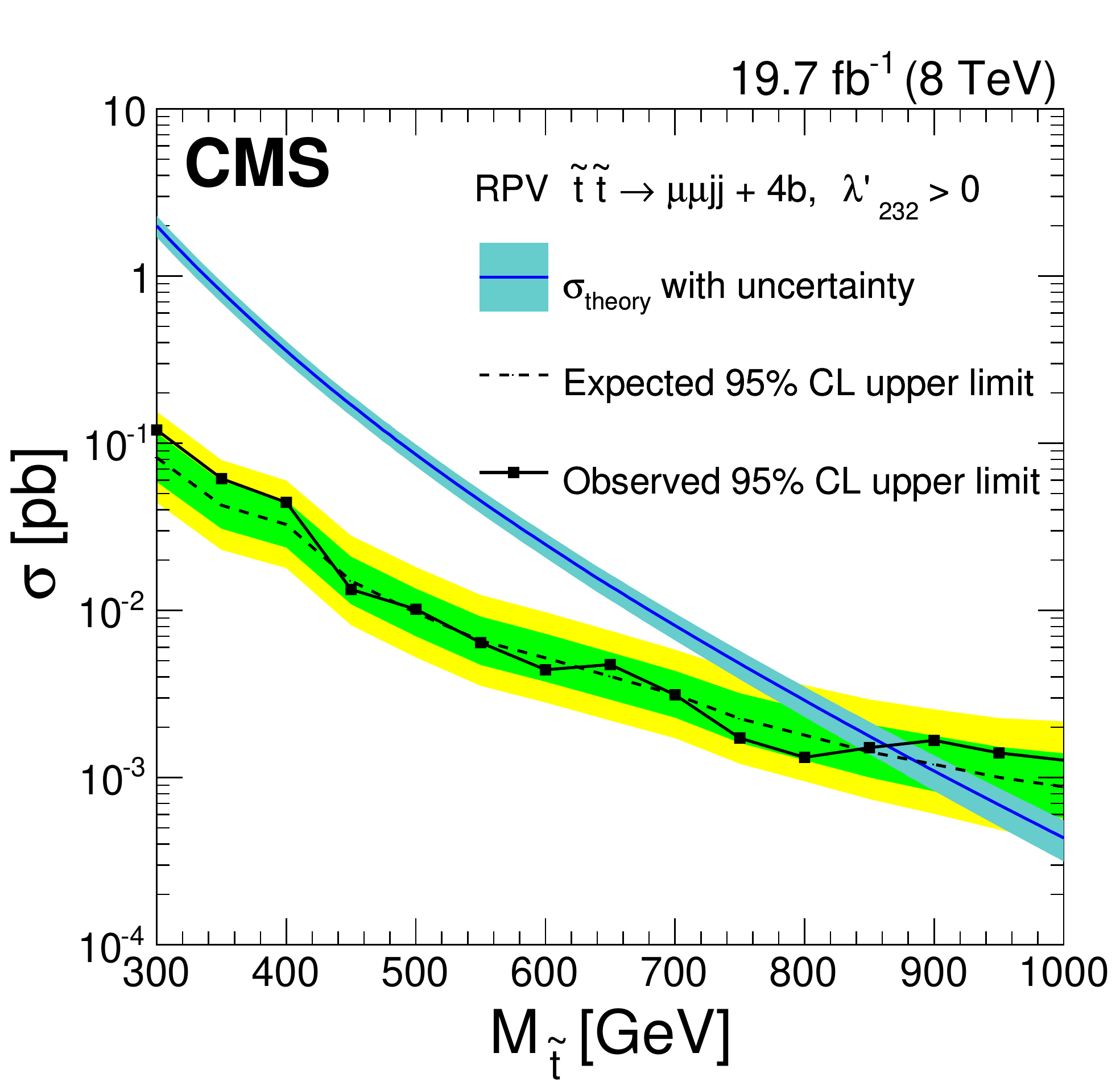}
   \caption{
    Frame on \cmsLeft\ (\cmsRight): the expected and observed upper limits at $95\%$~\CL on the top squark pair-production
    cross section for a Higgsino-mediated RPV SUSY model in the \eejj~(\mumujj) + 4 b quark final state as a function of the top squark mass, obtained with the \eejj~(\mumujj) analysis. The expected limits and uncertainty bands represent the median expected limits and the 68\% and 95\% confidence intervals.
    The $\sigma_{\rm theory}$ curves and their bands represent, respectively, the theoretical top squark pair production cross section
    and the uncertainties due to the choice of PDF and renormalization/factorization scales.
    }
  \label{fig:limit_plots_RPV}
\end{figure}

It is interesting to consider the case where top squark decay is mediated by a Higgsino with a mass $M_{\sHig}=M_{\PSQt}-100\GeV$ with a 100$\%$ branching fraction, as shown in Fig.~\ref{rpvFig}.  Because of higher jet multiplicity and hence softer kinematic spectra, the optimization selections described in Section~\ref{sec:lljjSelection} are shifted such that for a given top squark mass, the selections used correspond to a LQ mass lower by 100\GeV, determined by optimizing the expected limits.  The experimental limits along with the theoretical top squark pair production cross sections for the \eejj~(\mumujj) channel  are shown in Fig.~\ref{fig:limit_plots_RPV} on the left (right). Assuming this model, the experimental results yield a $95\%$~\CL observed upper limit exclusion of top squark masses less than 710\GeV in the first generation $\lambda^{\prime}_{132}$ model, compared with a median expected limit of 840\GeV.  The second generation $\lambda^{\prime}_{232}$ model yields an observed exclusion of top squark masses less than 860\GeV, compared with a median expected limit of 880\GeV.  These are the first experimental limits to date on $\lambda^{\prime}_{132}$ and $\lambda^{\prime}_{232}$ RPV SUSY top squark decays.

\section{Summary}
\label{summary}
A search has been conducted for pair production of first- and second-generation scalar leptoquarks in final states with either two electrons (or two muons) and two jets, or with one electron (or muon), significant missing transverse energy, and
two jets, using 8\TeV proton-proton collision data corresponding to an integrated luminosity of 19.7~\fbinv. The results are also interpreted in the context of models of vector leptoquark pair production and of R-parity violating supersymmetric models with similar final state signatures.

The selection criteria used for all the searches are optimized for each scalar leptoquark signal mass hypothesis.  In the first generation \eejj~(\enujj) channel, a broad 2.3 (2.6) standard deviation excess is observed in
the final selection optimized for leptoquarks with a mass of 650\GeV.
The excess does not peak in the $M_{\mathrm{\Pe\cPj}}$ distributions, as a leptoquark signal would, but does weaken the upper limit that can be set on the production cross section for leptoquark masses of about 650\GeV and values of $\beta \lesssim 0.15$.
Limits are placed with 95\%~\CL
on first-generation scalar leptoquarks with masses less than 1010 (850)\GeV, assuming $\beta = 1.0$~$(0.5)$.
This is to be compared with the expected 95\%~\CL exclusions of 1030 (890)\GeV.  In the secondgeneration leptoquark search the number of observed candidates for each mass hypothesis agrees within uncertainties with the number of expected standard model background events. Second-generation scalar leptoquarks are excluded at 95\%~\CL with masses below 1080 (800)\GeV for $\beta = 1.0$~$(0.5)$. This is to be compared with a median expected limit of 1050 (910)\GeV.  These results for pair production of scalar leptoquarks are closely comparable to those of Ref.~\cite{atlas_lq_8tev}.

Limits are set on four coupling scenarios for vector leptoquarks, and for the \eejj~(\enujj) channel yield $95\%$~\CL upper limit exclusions of masses in the range of $1150-1660$~($1050-1560$)\GeV.   In the \mumujj~(\munujj) channel, the experimental results yield $95\%$~\CL upper limit exclusions of masses in the range of $1200-1720$~($980-1480$)\GeV. These represent the most stringent limits on vector LQ production to date.

Limits are also set for top squark production in an R-parity violating supersymmetric model via the $\lambda^{\prime}_{132}$ or $\lambda^{\prime}_{232}$ operators.  For direct top squark decay, the scalar LQ limits can be applied directly.  Interpretation is also made in Higgsino-mediated top squark decay, where the experimental results yield a $95\%$~\CL observed upper limit exclusion of top squark masses less than 710\GeV in the first generation $\lambda^{\prime}_{132}$ model, compared with a median expected limit of 840\GeV.  The second generation $\lambda^{\prime}_{232}$ model yields an observed exclusion of top squark masses less than 860\GeV, compared with a median expected limit of 880\GeV.  These represent the most stringent experimental limits to date on $\lambda^{\prime}_{132}$ and $\lambda^{\prime}_{232}$ RPV SUSY top squark decays and the first experimental limits on the Higgsino-mediated decays.

\begin{acknowledgments}
\hyphenation{Bundes-ministerium Forschungs-gemeinschaft Forschungs-zentren} We congratulate our colleagues in the CERN accelerator departments for the excellent performance of the LHC and thank the technical and administrative staffs at CERN and at other CMS institutes for their contributions to the success of the CMS effort. In addition, we gratefully acknowledge the computing centers and personnel of the Worldwide LHC Computing Grid for delivering so effectively the computing infrastructure essential to our analyses. Finally, we acknowledge the enduring support for the construction and operation of the LHC and the CMS detector provided by the following funding agencies: the Austrian Federal Ministry of Science, Research and Economy and the Austrian Science Fund; the Belgian Fonds de la Recherche Scientifique, and Fonds voor Wetenschappelijk Onderzoek; the Brazilian Funding Agencies (CNPq, CAPES, FAPERJ, and FAPESP); the Bulgarian Ministry of Education and Science; CERN; the Chinese Academy of Sciences, Ministry of Science and Technology, and National Natural Science Foundation of China; the Colombian Funding Agency (COLCIENCIAS); the Croatian Ministry of Science, Education and Sport, and the Croatian Science Foundation; the Research Promotion Foundation, Cyprus; the Ministry of Education and Research, Estonian Research Council via IUT23-4 and IUT23-6 and European Regional Development Fund, Estonia; the Academy of Finland, Finnish Ministry of Education and Culture, and Helsinki Institute of Physics; the Institut National de Physique Nucl\'eaire et de Physique des Particules~/~CNRS, and Commissariat \`a l'\'Energie Atomique et aux \'Energies Alternatives~/~CEA, France; the Bundesministerium f\"ur Bildung und Forschung, Deutsche Forschungsgemeinschaft, and Helmholtz-Gemeinschaft Deutscher Forschungszentren, Germany; the General Secretariat for Research and Technology, Greece; the National Scientific Research Foundation, and National Innovation Office, Hungary; the Department of Atomic Energy and the Department of Science and Technology, India; the Institute for Studies in Theoretical Physics and Mathematics, Iran; the Science Foundation, Ireland; the Istituto Nazionale di Fisica Nucleare, Italy; the Ministry of Science, ICT and Future Planning, and National Research Foundation (NRF), Republic of Korea; the Lithuanian Academy of Sciences; the Ministry of Education, and University of Malaya (Malaysia); the Mexican Funding Agencies (CINVESTAV, CONACYT, SEP, and UASLP-FAI); the Ministry of Business, Innovation and Employment, New Zealand; the Pakistan Atomic Energy Commission; the Ministry of Science and Higher Education and the National Science Centre, Poland; the Funda\c{c}\~ao para a Ci\^encia e a Tecnologia, Portugal; JINR, Dubna; the Ministry of Education and Science of the Russian Federation, the Federal Agency of Atomic Energy of the Russian Federation, Russian Academy of Sciences, and the Russian Foundation for Basic Research; the Ministry of Education, Science and Technological Development of Serbia; the Secretar\'{\i}a de Estado de Investigaci\'on, Desarrollo e Innovaci\'on and Programa Consolider-Ingenio 2010, Spain; the Swiss Funding Agencies (ETH Board, ETH Zurich, PSI, SNF, UniZH, Canton Zurich, and SER); the Ministry of Science and Technology, Taipei; the Thailand Center of Excellence in Physics, the Institute for the Promotion of Teaching Science and Technology of Thailand, Special Task Force for Activating Research and the National Science and Technology Development Agency of Thailand; the Scientific and Technical Research Council of Turkey, and Turkish Atomic Energy Authority; the National Academy of Sciences of Ukraine, and State Fund for Fundamental Researches, Ukraine; the Science and Technology Facilities Council, UK; the US Department of Energy, and the US National Science Foundation.

Individuals have received support from the Marie-Curie program and the European Research Council and EPLANET (European Union); the Leventis Foundation; the A. P. Sloan Foundation; the Alexander von Humboldt Foundation; the Belgian Federal Science Policy Office; the Fonds pour la Formation \`a la Recherche dans l'Industrie et dans l'Agriculture (FRIA-Belgium); the Agentschap voor Innovatie door Wetenschap en Technologie (IWT-Belgium); the Ministry of Education, Youth and Sports (MEYS) of the Czech Republic; the Council of Science and Industrial Research, India; the HOMING PLUS program of the Foundation for Polish Science, cofinanced from European Union, Regional Development Fund; the OPUS program of the National Science Center (Poland); the Compagnia di San Paolo (Torino); the Consorzio per la Fisica (Trieste); MIUR project 20108T4XTM (Italy); the Thalis and Aristeia programmes cofinanced by EU-ESF and the Greek NSRF; the National Priorities Research Program by Qatar National Research Fund; the Rachadapisek Sompot Fund for Postdoctoral Fellowship, Chulalongkorn University (Thailand); and the Welch Foundation, contract C-1845.
\end{acknowledgments}

\bibliography{auto_generated}
\cleardoublepage \appendix\section{The CMS Collaboration \label{app:collab}}\begin{sloppypar}\hyphenpenalty=5000\widowpenalty=500\clubpenalty=5000\textbf{Yerevan Physics Institute,  Yerevan,  Armenia}\\*[0pt]
V.~Khachatryan, A.M.~Sirunyan, A.~Tumasyan
\vskip\cmsinstskip
\textbf{Institut f\"{u}r Hochenergiephysik der OeAW,  Wien,  Austria}\\*[0pt]
W.~Adam, E.~Asilar, T.~Bergauer, J.~Brandstetter, E.~Brondolin, M.~Dragicevic, J.~Er\"{o}, M.~Flechl, M.~Friedl, R.~Fr\"{u}hwirth\cmsAuthorMark{1}, V.M.~Ghete, C.~Hartl, N.~H\"{o}rmann, J.~Hrubec, M.~Jeitler\cmsAuthorMark{1}, V.~Kn\"{u}nz, A.~K\"{o}nig, M.~Krammer\cmsAuthorMark{1}, I.~Kr\"{a}tschmer, D.~Liko, T.~Matsushita, I.~Mikulec, D.~Rabady\cmsAuthorMark{2}, B.~Rahbaran, H.~Rohringer, J.~Schieck\cmsAuthorMark{1}, R.~Sch\"{o}fbeck, J.~Strauss, W.~Treberer-Treberspurg, W.~Waltenberger, C.-E.~Wulz\cmsAuthorMark{1}
% ...
\vskip\cmsinstskip
% ...
1:~~Also at Vienna University of Technology, Vienna, Austria\\
2:~~Also at CERN, European Organization for Nuclear Research, Geneva, Switzerland\\
% ...

\end{sloppypar}
\end{document}